%% file: SUS-13-002_temp.tex
\pdfoutput=1

\documentclass[11pt,twoside,a4paper,cmspaper,final,collab]{cms-tdr}
\def\svnVersion{257109}\def\cmsCernNo{2013-037}\def\cmsCernDate{\today}\def\cmsMessage{Published in Physical Review D as \href{http://dx.doi.org/10.1103/PhysRevD.90.032006}{\doi{10.1103/PhysRevD.90.032006.}}}

\begin{document}\cmsNoteHeader{SUS-13-002}

\hyphenation{had-ron-i-za-tion}
\hyphenation{cal-or-i-me-ter}
\hyphenation{de-vices}
\RCS$Revision: 253074 $
\RCS$HeadURL: svn+ssh://svn.cern.ch/reps/tdr2/papers/SUS-13-002/trunk/SUS-13-002.tex $
\RCS$Id: SUS-13-002.tex 253074 2014-07-26 07:18:38Z ecampana $

\newlength\cmsFigWidth
\ifthenelse{\boolean{cms@external}}{\setlength\cmsFigWidth{0.95\columnwidth}}{\setlength\cmsFigWidth{0.6\textwidth}}
\ifthenelse{\boolean{cms@external}}{\providecommand{\cmsLeft}{top}}{\providecommand{\cmsLeft}{left}}
\ifthenelse{\boolean{cms@external}}{\providecommand{\cmsRight}{bottom}}{\providecommand{\cmsRight}{right}}
\ifthenelse{\boolean{cms@external}}{\providecommand{\cmsUpperLeft}{top}}{\providecommand{\cmsUpperLeft}{upper left}}
\ifthenelse{\boolean{cms@external}}{\providecommand{\cmsUpperRight}{middle}}{\providecommand{\cmsUpperRight}{upper right}}
\ifthenelse{\boolean{cms@external}}{\providecommand{\NA}{\ensuremath{\cdots}}}{\providecommand{\NA}{\text{---}}}
\ifthenelse{\boolean{cms@external}}{\providecommand{\CL}{C.L.\xspace}}{\providecommand{\CL}{CL\xspace}}
\providecommand{\slepton}{\ensuremath{\widetilde{\ell}}\xspace}
\providecommand{\selectron}{\ensuremath{\widetilde{\cmsSymbolFace{e}}}\xspace}
\providecommand{\smuon}{\ensuremath{\widetilde{\mu}}\xspace}
\providecommand{\stau}{\ensuremath{\widetilde{\tau}}\xspace}

\providecommand{\squark}{\ensuremath{\widetilde{\cmsSymbolFace{q}}}\xspace}
\providecommand{\sbone}{\ensuremath{\widetilde{\cmsSymbolFace{b}}_{1}}\xspace}
\providecommand{\stop}{\ensuremath{\widetilde{\cmsSymbolFace{t}}}\xspace}

\providecommand{\nall}{\ensuremath{\widetilde{\chi}^{0}}\xspace}
\providecommand{\nalli}{\ensuremath{\widetilde{\chi}^{0}_{i}}\xspace}
\providecommand{\none}{\ensuremath{\widetilde{\chi}^{0}_{1}}\xspace}
\providecommand{\ntwo}{\ensuremath{\widetilde{\chi}^{0}_{2}}\xspace}
\providecommand{\nthree}{\ensuremath{\widetilde{\chi}^{0}_{3}}\xspace}

\providecommand{\call}{\ensuremath{\widetilde{\chi}^{\pm}}\xspace}
\providecommand{\cone}{\ensuremath{\widetilde{\chi}^{\pm}_{1}}\xspace}
\providecommand{\conem}{\ensuremath{\widetilde{\chi}^{-}_{1}}\xspace}
\providecommand{\conep}{\ensuremath{\widetilde{\chi}^{+}_{1}}\xspace}
\providecommand{\ctwo}{\ensuremath{\widetilde{\chi}^{\pm}_{2}}\xspace}

\providecommand{\gluino}{\ensuremath{\widetilde{g}}\xspace}

\providecommand{\goldstino}{\ensuremath{\widetilde{\cmsSymbolFace{G}}}\xspace}

\providecommand{\tauh}{\ensuremath{\Pgt_\textrm{h}}\xspace}

\renewcommand{\PASQt}{\ensuremath{\widetilde{\cmsSymbolFace{t}}^*}\xspace}

\cmsNoteHeader{SUS-13-002} % This is over-written in the CMS environment: useful as preprint no. for export versions
\title{Search for anomalous production of events with three or more leptons in \texorpdfstring{$\Pp\Pp$ collisions at $\sqrt{s} = 8$\TeV}{pp collisions at sqrt(s)=8 TeV}}

\date{\today}

\abstract{
A search for physics beyond the standard model in
events with at least three leptons is presented.
The data sample, corresponding to an integrated
luminosity of 19.5\fbinv of proton-proton collisions with
center-of-mass energy $\sqrt{s} = 8\TeV$, was collected by
the CMS experiment at the LHC during 2012.
The data are divided into exclusive categories based
on the number of leptons and their flavor, the presence or
absence of an opposite-sign, same-flavor lepton pair (OSSF),
the invariant mass of the OSSF pair, the presence or absence
of a tagged bottom-quark jet, the number of identified
hadronically decaying $\tau$ leptons, and the magnitude of
the missing transverse energy and of the scalar sum of jet transverse
momenta.
The numbers of observed events are found to be consistent
with the expected numbers from standard model processes, and
limits are placed on new-physics scenarios that yield
multilepton final states. In particular, scenarios that
predict Higgs boson production in the context of supersymmetric
decay chains are examined.
We also place a 95\% confidence level upper limit of 1.3\% on the branching fraction for the
decay of a top quark to a charm quark and a Higgs boson ($\cPqt \to \cPqc \PH$),
which translates to a bound on the left- and right-handed top-charm flavor-violating Higgs
Yukawa couplings, $\lambda_{\cPqt \cPqc}^{\PH}$ and $\lambda_{\cPqc \cPqt}^{\PH}$,
respectively, of $\sqrt{\abs{\lambda_{\cPqt \cPqc}^{\PH}}^2 + \abs{\lambda_{\cPqc \cPqt}^{\PH}}^2} < 0.21$.}

\hypersetup{%
pdfauthor={CMS Collaboration},%
pdftitle={Search for anomalous production of events with three or more leptons in pp collisions at sqrt(s)=8 TeV},%
pdfsubject={CMS},%
pdfkeywords={CMS, physics, software, computing}}

\maketitle %maketitle comes after all the front information has been supplied

\section{Introduction}
\label{sec:intro}

The recent discovery of a Higgs boson~\cite{Aad:2012tfa,Chatrchyan:2012ufa,Chatrchyan:2013lba}
at the relatively low mass of about 125\GeV implies that physics beyond the standard model (BSM)
may be observable at energy scales of around 1\TeV. Supersymmetry (SUSY) is a prominent candidate for BSM physics
because it provides a solution to the hierarchy problem, predicts gauge-coupling unification, and contains a ``natural" candidate
for dark matter~\cite{Nilles:1983ge,Haber:1984rc,deBoer:1994dg}.
Supersymmetry postulates the existence of fermionic superpartners
for each standard model (SM) boson, and of bosonic superpartners
for each SM fermion. For example, gluinos, squarks, and winos are the
superpartners of gluons, quarks, and $\PW$ bosons, respectively.
The superparter of a lepton is a slepton.
In $R$-parity~\cite{Farrar:1978xj} conserving SUSY models, supersymmetric particles
are created in pairs, and the lightest supersymmetric particle (LSP)
is stable. If the LSP interacts only weakly, as in the case of a dark matter candidate,
it escapes detection, leading to missing transverse energy ($\MET$).
Here, $R$-parity is defined by $R = (-1)^{3B+L+2s}$,
with $B$ and $L$ the baryon and lepton numbers, and $s$ the particle spin.
All SM particles have $R = +1$ while all superpartners have $R = -1$.

A wide range of BSM scenarios predict multilepton final states~\cite{Brooijmans:2010tn},
where by ``multilepton," we mean three or more charged leptons.
Since multilepton states are relatively rare in the SM, searches in the
multilepton channel have good potential to uncover BSM physics.

Given the rich SUSY particle spectrum, multilepton final states
in SUSY events take on multiple forms. For example, a cascade of
particles initiated by the decay of a heavy gluino can
proceed through intermediate squarks, winos, and sleptons to
produce a final state that is democratic in lepton flavor, \ie, equally
likely to contain electrons, muons, or $\tau$ leptons. Direct
pair production of the superpartners of the electron and
muon (selectron and smuon, respectively) can yield a multilepton
state dominated by $\tau$ leptons should the superpartner of
the $\tau$ lepton (stau) be substantially lighter than the selectron
and smuon, as is expected in some models. Another path to a multileptonic
final state arises from top-squark production in which the top squark decays
to leptonically decaying third-generation quarks and to a $\cPZ$ boson that
yields an opposite-sign same-flavor (OSSF) lepton pair. In these latter events,
bottom-quark jets ($\cPqb$ jets) might also be present. Similarly, many other
multileptonic signatures are possible.

Besides SUSY, other BSM scenarios can yield multileptonic final states,
such as $\cPqt \to \cPqc \PH$ transitions, with $\cPqt$ a top quark, $\cPqc$ a charm quark,
and $\PH$ a Higgs boson. The $\cPqt \to \cPqc \PH$ process is extremely rare in the SM
but can be enhanced through the production of new particles in loops~\cite{Craig:2012vj, Chen:2013qta}.
The top quark is the heaviest SM particle, and is thus the SM particle that is most strongly coupled to the
Higgs boson. Since the $\cPqt \to \cPqc \PH$ process directly probes the
flavor-violating couplings of the top quark to the Higgs boson, it provides a
powerful means to search for BSM physics regardless of the underlying new-physics mechanism.
The $\cPqt \to \cPqc \PH$ decay can give rise to a
multilepton signature when a top quark in a top quark-antiquark ($\ttbar$) pair decays
to the $\cPqc \PH$ state, followed by the decay of the Higgs boson to leptons
through, \eg, $\PH \to \cPZ \cPZ^*$ or $\PH \to \PW \PW^*$ decays,
in conjunction with the leptonic decay of the other top quark in the $\ttbar$ pair.

In this paper, we present a search for BSM physics in multilepton channels.
The search is based on a sample of proton-proton collision data collected at
$\sqrt{s} = 8\TeV$ with the Compact Muon Solenoid (CMS) detector at the CERN
Large Hadron Collider (LHC) in 2012, corresponding to an integrated luminosity of 19.5\fbinv.
The study is an extension of our earlier work~\cite{Chatrchyan:2012mea},
which was based on a data set of 5.0\fbinv collected at $\sqrt{s} = 7\TeV$.
A related search, presented in Ref.~\cite{Chatrchyan:2013xsw}, uses the 8\TeV data set
to investigate $R$-parity-violating SUSY scenarios. References~\cite{ATLAS:2012kr,Aad:2014dya,Aad:2014pda,Aad:2014mha,Aad:2014nua} contain recent, related results from the ATLAS Collaboration, and Refs.~\cite{Khachatryan:2014qwa,Khachatryan:2014doa,Chatrchyan:2014lfa,Chatrchyan:2013mya,Chatrchyan:2013fea,Chatrchyan:2013xna,Chatrchyan:2013xsw,Chatrchyan:2013wxa,Chatrchyan:2013lya,Chatrchyan:2012paa, Chatrchyan:2013iqa} contain the same from CMS.

Because of the wide range of possible BSM signatures, we have adopted a search
strategy that is sensitive to different kinematical and topological signatures,
rather than optimizing the analysis for a particular model. We retain all
observed multilepton candidate events and classify them into multiple
mutually exclusive categories based on the number of leptons, the lepton
flavor, the presence of $\cPqb$ jets, the presence of an OSSF pair indicative of a
$\cPZ$ boson, and kinematic characteristics such as $\MET$ and $\HT$, where $\HT$
is the scalar sum of jet transverse momentum ($\pt$) values.
We then confront a number of BSM scenarios that exhibit diverse characteristics
with respect to the population of these categories.

This paper is organized as follows. In Sec.~\ref{detector}, a brief summary of
the CMS detector and a description of the trigger is presented. Section~\ref{data} discusses the event
reconstruction procedures, event selection, and event simulation. The search strategy and the
background evaluation methods are outlined in Secs.~\ref{classif} and~\ref{background}.
Section~\ref{syst} contains a discussion of systematic uncertainties. The results are presented
in Sec.~\ref{results}. Sections~\ref{models} and~\ref{tch} present the interpretations of our
results for SUSY scenarios and for the $\cPqt \to \cPqc \PH$ process, respectively.
A summary is given in Sec.~\ref{summary}.

\section{Detector and trigger}
\label{detector}

The CMS detector has cylindrical symmetry around the direction of the beam axis.
The coordinate system is defined with the origin at the nominal collision point and
the $z$ axis along the direction of the counterclockwise proton beam. The $x$ axis
points toward the center of the LHC ring and the $y$ axis vertically upwards. The polar
angle $\theta$ is measured with respect to the $z$ axis. The azimuthal angle $\phi$ is
measured in the $x-y$ plane, relative to the $x$ axis. Both angles are
measured in radians. Pseudorapidity $\eta$ is defined as $\eta = -\ln [\tan(\theta/2)]$.
The central feature of the detector is a superconducting solenoidal magnet of field
strength 3.8\unit{T}. Within the field volume are a silicon pixel and strip tracker, a lead
tungstate crystal calorimeter, and a brass-and-scintillator hadron calorimeter. The
tracking detector covers the region $\abs{\eta} < 2.5$ and the calorimeters $\abs{\eta} < 3.0$.
Muon detectors based on gas-ionization detectors lie outside the solenoid,
covering $\abs{\eta} < 2.4$. A steel-and-quartz-fiber forward calorimeter covers $\abs{\eta} < 5.0$.
A detailed description of the detector can be found in Ref.~\cite{CMS:2008zzk}.

A double-lepton trigger ($\Pe \Pe$, $\Pgm \Pgm$, or $\Pe \Pgm$)
is used for data collection. At the trigger level, the leptons with the
highest and second-highest transverse momentum are
required to satisfy $\pt >17\GeV$ and $\pt > 8\GeV$, respectively.
The lepton trigger efficiency is determined using an independent data sample based on
minimum requirements for $\HT$~\cite{Chatrchyan:2012mea}. After application
of all selection requirements, the trigger efficiencies are found to be 95\%, 90\%,
and 93\%, respectively, for the $\Pe \Pe$, $\Pgm \Pgm$, and $\Pe \Pgm$ triggers.
Corrections are applied to account for the trigger inefficiencies.

\section{Event reconstruction, selection, and simulation}
\label{data}

The particle-flow (PF) method~\cite{PFT-10-002,CMS-PAS-PFT-09-001}
is used to reconstruct the physics objects used in this analysis: electrons,
muons, hadronically decaying $\tau$ leptons ($\tauh$), jets, and $\MET$.

Electrons and muons are reconstructed using measured quantities from
the tracker, calorimeter, and muon system. The candidate tracks must
satisfy quality requirements and spatially match energy deposits in the
electromagnetic calorimeter or tracks in the muon detectors, as appropriate.
Details of the reconstruction and identification procedures can be found in
Ref.~\cite{EGM-10-004} for electrons and in Ref.~\cite{JME-10-005}
for muons.

Hadronically decaying $\tau$ leptons predominantly yield either a
single charged track (one-prong decays) or three charged tracks (three-prong decays)
with or without additional electromagnetic energy from neutral-pion
decays. Both one-prong and three-prong $\tauh$ decays are reconstructed
using the hadron plus strips algorithm~\cite{CMS-PAPERS-JME-10-009}.

The event primary vertex is defined to be the reconstructed vertex
with the largest sum of charged-track $\pt^2$ value and is required
to lie within 24\unit{cm} of the origin in the direction along the $z$ axis
and 2\unit{cm} in the transverse plane.

Jets are formed from reconstructed PF objects using the
anti-\kt algorithm~\cite{Cacciari:2008gp,Cacciari:2011ma}
with a distance parameter of 0.5. Corrections are applied as a function
of jet $\pt$ and $\eta$ to account for nonuniform detector
response~\cite{CMS-PAPERS-JME-10-011}. Contributions to the
jet $\pt$ values due to overlapping $\Pp\Pp$ interactions from the
same or neighboring bunch crossing ("pileup") are subtracted using
the jet area method described in Ref.~\cite{Cacciari:2007fd}.

Finally, $\MET$ is the magnitude of the vector sum of the transverse momenta
of all PF objects.

We require the presence of at least three reconstructed leptons,
where by "lepton" we mean an electron, muon, or $\tauh$ candidate.
Electron and muon candidates must satisfy $\pt >10\GeV$ and $\abs{\eta} < 2.4$.
At least one electron or muon candidate must satisfy $\pt > 20\GeV$.
The $\tauh$ candidates must satisfy $\pt > 20\GeV$ and $\abs{\eta} < 2.3$.
Events are allowed to contain at most one $\tauh$ candidate. Leptonically
decaying $\tau$ leptons populate the electron and muon channels.

Leptons from BSM processes are typically isolated, \ie, separated
in $\DR \equiv \sqrt{(\Delta\eta)^2 + (\Delta\phi)^2}$ from other physics objects.
To reduce background from the semileptonic decays of heavy quark flavors,
which generally yield leptons within jets, we apply lepton isolation criteria.
For electrons and muons, we define the relative isolation $I_{\text{rel}}$
to be the sum of the $\pt$ values of all PF objects within a cone of
radius $\DR = 0.3$ around the lepton direction (excluding the lepton itself),
divided by the lepton $\pt$ value, and require $I_{\text{rel}} < 0.15$.
For $\tauh$ leptons, the sum of energy $E^{\tauh}_{\text{iso}}$ within a cone of radius $\DR = 0.5$
around the lepton direction is required to satisfy $E^{\tauh}_{\text{iso}} < 2\GeV$. In all cases,
we account for the effects of pileup interactions~\cite{Cacciari:2007fd}.

The signal scenarios contain prompt leptons, where by ``prompt"
we mean that the parent particles decay near the primary vertex.
To ensure that the electrons and muons are prompt, their distance of closest
approach to the primary vertex is required to be less than 2\unit{cm} in the direction
along the beam axis and 0.02\unit{cm} in the transverse plane.

We construct OSSF pairs from charged lepton  $\ell^{+} \ell^{-}$ combinations, with $\ell$ an electron or muon.
Events with an OSSF pair that satisfies $m_{\ell^{+}\ell^{-}} < 12\GeV$ are rejected to
eliminate background from low-mass Drell--Yan processes and $\JPsi$ and $\PgU$ decays.
If there is more than one OSSF pair in the event, this requirement is applied to each pair.
Events with an OSSF pair outside the $\cPZ$ boson mass region
(defined by $75 < m_{\ell^{+}\ell^{-}} < 105\GeV$)
but that satisfy $75 < m_{\ell^{+}\ell^{-}\ell^{(\prime)\pm}} < 105\GeV$,
where $\ell^{(\prime) \pm}$ is an electron or muon with the same
(different) flavor as the OSSF pair, are likely to arise from final-state photon
radiation from the \cPZ-boson decay products, followed by
conversion of the photon to a charged lepton pair.
Events that meet this condition are rejected if they also exhibit kinematic characteristics consistent
with background from events with a $\cPZ$ boson and jets (\cPZ+jets background).

Jets are required to satisfy $\pt > 30\GeV$ and $\abs{\eta} < 2.5$ and
are rejected if they lie within a distance $\DR = 0.3$ from a lepton that
satisfies our selection criteria. The identification of $\cPqb$ jets is performed using
the CMS combined secondary-vertex algorithm~\cite{Chatrchyan:2012jua}
at the medium working point. This working point yields a tagging efficiency of
roughly 70\% for jets with a \pt value of 80\GeV, with a misidentification rate
for light-flavor events of less than 2\% and for charm-quark jets of roughly 20\%.

Samples of simulated events are used to determine signal acceptance and to evaluate some
SM backgrounds. The simulation of SM events is based on the
\MADGRAPH (version 5.1.3.30)~\cite{Maltoni:2002qb} event generator with leading-order
CTEQ6L1~\cite{Nadolsky:2008zw} parton distribution functions (PDF), with the
\GEANTfour~\cite{Agostinelli:2002hh} package used to describe detector response.
The cross sections are normalized to next-to-leading (NLO)
order~\cite{MCFM,Campbell:2012dh,Garzelli:2012bn}.
The simulation of signal events is performed using both the \MADGRAPH
and \PYTHIA (version 6.420)~\cite{Sjostrand:2007gs} generators, with the description of
detector response based on the CMS fast simulation program~\cite{Orbaker:2010zz}.
Parton showering for all simulated events is described using \PYTHIA.
The simulated events are adjusted to account for the multiplicity of pileup interactions
observed in the data, as well as for differences between data and simulation for the
jet energy scale, rate of events with initial-state radiation (ISR)~\cite{Chatrchyan:2013xna},
and \cPqb-jet tagging efficiency~\cite{Chatrchyan:2012jua}.

\section{Multilepton event classification}
\label{classif}

Multilepton event candidates are separated into mutually exclusive search channels.
The level of the SM background varies considerably between the different categories.
The overall sensitivity to new physics is maximized by separating the
low- and high-background channels. Events with exactly three leptons generally suffer
from a higher background level than events with four or more leptons, as do events
with a $\tauh$ candidate. We therefore categorize events with three leptons separately
from those with four or more, and events with a $\tauh$ candidate separately from those
without such a candidate. Similarly, events with a tagged $\cPqb$ jet suffer higher background
from $\ttbar$ events, and so are categorized separately from events without a tagged $\cPqb$ jet.

We also define categories based on the number $n$ of OSSF dilepton pairs that can
be formed using each lepton candidate only once (OSSF$n$).
For example, both $\Pgmp \Pgmm \Pgmm$ and $\Pgmp \Pgmm \Pem$ events fall into
the OSSF1 category, while $\Pgmp \Pgmp \Pem$ and $\Pgmp \Pgmm \Pep \Pem$ events
fall into the OSSF0 and OSSF2 categories, respectively.
Events with an OSSF pair exhibit larger levels of background than do OSSF0 events.

We further classify events with at least one OSSF pair as being ``on-$\cPZ$''
if the reconstructed invariant mass $m_{\ell^{+}\ell^{-}}$ of any of the OSSF
dilepton pairings in the event lies in the $\cPZ$-boson mass region
$75 < m_{\ell^{+}\ell^{-}} < 105\GeV$.
Since there is considerably less SM background above the $\cPZ$-boson region than
below it, we also define ``above-$\cPZ$'' and ``below-$\cPZ$'' categories, but for three-lepton
events only, where for above-$\cPZ$ (below-$\cPZ$) events all possible OSSF pairs
satisfy $m_{\ell^{+}\ell^{-}} > 105\GeV$ ($m_{\ell^{+}\ell^{-}} < 75\GeV$).
Additionally, we classify events with four leptons as being ``off-$\cPZ$'' if all possible OSSF
pairs have $m_{\ell^{+}\ell^{-}}$ values outside the $\cPZ$-boson mass region.

Events with SUSY production of squarks and gluinos may be characterized by a high level of hadronic
activity compared to SM events. We therefore separate events according to whether $\HT$ is larger
or smaller than 200\GeV. Similarly, we subdivide events into five $\MET$ bins: four bins of
width 50\GeV from 0 to 200\GeV, and a fifth bin with $\MET > 200\GeV$. For the purposes of 
presentation in Tables~\ref{tab:resultsTable1} and~\ref{tab:resultsTable2}, a coarser $\MET$ binning 
has been used.

\section{Background estimation}
\label{background}

\subsection{Overview}
\label{overview}

The largest background category for trilepton events arises from $\cPZ$+jets events
in which the $\cPZ$ boson decays to a lepton pair while the third lepton candidate is
either a misidentified hadron or a genuine lepton from heavy-flavor decay.
This background dominates the low-$\MET$ and low-$\HT$ channels.
As described below (Secs.~\ref{subsec:NonPromptLeptons},
~\ref{subsec:FakeHadronicTaus}, and~\ref{subsec:AsymmeticInternalPhotonConversions}),
this background is evaluated from data.

Search channels with $\tauh$ candidates suffer from higher background compared
to those with only electrons and muons because sufficiently
narrow jets tend to mimic hadronically decaying $\tau$ leptons.
We measure the background due to misidentified $\tauh$ decays
from data (Sec.~\ref{subsec:FakeHadronicTaus}).

Background events containing three or more prompt genuine leptons and a significant
level of $\MET$ can arise from SM processes such as $\PW \cPZ$+jets or $\cPZ \cPZ$+jets
production if both electroweak bosons decay leptonically. This type of background is referred
to as ``irreducible'' because its characteristics are similar to the search signature. We use
simulation to estimate the irreducible background (Sec.~\ref{subsec:WZproduction}).
Comparison between data and simulation demonstrates that the $\MET$ distribution is well
modeled for processes with genuine $\MET$, viz., SM model processes
with neutrinos~\cite{CMS-PAPERS-MUO-10-004,JME-10-005}.

Another major source of background is $\ttbar$ production in which each top quark
produces a $\PW$ boson that decays leptonically, with a third lepton arising from
the semileptonic decay of the \cPqb-jet daughter of one of the two top quarks. The character of
this background differs significantly from the background due to $\cPZ$+jets events, in which
the jets are relatively soft. Simulation is used to evaluate the $\ttbar$
background (Sec.~\ref{sub:TTbarProduction}).

Two varieties of photon conversion are relevant to consider.
``External'' conversion of an on-shell photon in the detector
material predominantly results in an $\Pep \Pem$ pair, which is eliminated
using a collection of tracking and kinematic criteria appropriate to
the small opening angle of the pair. In contrast, the ``internal'' or ``Dalitz''
conversion of a virtual photon produces a $\Pgmp \Pgmm$ pair almost as often
as an $\Pep \Pem$ pair. When an internal conversion is also asymmetric, \ie, when one of
the leptons has a very low $\pt$ value, the low $\pt$ track can fail to be reconstructed or
to satisfy the selection criteria. Drell--Yan processes accompanied by the high-$\pt$
lepton from an asymmetric conversion constitute a significant source of background for trilepton
channels. We estimate this background from data
(Sec.~\ref{subsec:AsymmeticInternalPhotonConversions}).

Remaining backgrounds arise from rare SM
processes such as triple-boson production or $\ttbar$ production
in association with a vector boson and are estimated from simulation.

In the following subsections we describe the estimation of main SM backgrounds.

\subsection{Misidentified prompt and isolated electrons and muons}
\label{subsec:NonPromptLeptons}

Processes such as $\cPZ (\to 2\ell) + \text{jets}$ and
$\PWp \PWm (\to 2\ell) + \text{jets}$ predominantly generate dilepton final states.
However, rare fluctuations in the hadronization process of an accompanying jet can 
provide what appears as a third prompt and isolated lepton, contributing to the 
background in the trilepton event category. Simulation of rare fragmentation processes 
can be unreliable. Therefore, we use dilepton data to evaluate this 
background~\cite{Chatrchyan:2012mea,CMS:2012ra}.

Consider a dilepton data sample, such as an $\Pep \Pem$ sample, that shares
attributes such as the $\MET$ and $\HT$ values with a trilepton search
channel such as $\Pep \Pem \Pgm$. The number of background
events in the $\Pep \Pem \Pgm$ channel that originate from $\Pep \Pem$
dilepton events is given by the number of misidentified isolated muons
in the $\Pep \Pem$ sample. We estimate this number to be the product of
the observed number of isolated tracks in the dilepton sample and a
proportionality factor $f_{\Pgm}$ between isolated tracks and muons.
The factor $f_{\Pgm}$ depends on the selection requirements of the
search channel and, in particular, its heavy-flavor content. Since the impact
parameters of tracks are generally larger for heavy-flavor decays than
for light-flavor (pion and kaon) decays, the average
impact parameter value of nonisolated tracks is a good indicator of
the heavy-quark content. Therefore, we characterize the
variation of $f_{\Pgm}$ from sample to sample as a function of the
average impact parameter value of nonisolated tracks in the dilepton sample.

The factor $f_{\Pgm}$ is determined in a procedure~\cite{Chatrchyan:2012mea}
that considers the numbers of nonisolated muons and tracks in the dilepton samples.
We use the difference between cross-checks performed with $\Pe \Pe$
and $\Pgm \Pgm$ samples to evaluate a systematic uncertainty.
From a sample of $\cPZ(\to \Pep \Pem) + \text{jets}$ events,
we determine $f_{\Pgm} = (0.6 \pm 0.2)\%$, where the uncertainty is systematic.
Using an analogous procedure with a sample of $\cPZ(\to \Pgmp \Pgmm) + \text{jets}$
events, we find $f_{\Pe} = (0.7 \pm 0.2)\%$ for the background from misidentified electron candidates.

\subsection{Misidentified \texorpdfstring{$\tauh$}{hadronic tau} leptons}
\label{subsec:FakeHadronicTaus}

The probability to misidentify an isolated $\tauh$ lepton is determined by calculating an extrapolation
ratio $f_{\tau}$ defined by the number of $\tauh$ candidates in the isolation-variable signal
region $E^{\tauh}_{\text{iso}} < 2.0\GeV$ to the number in a sideband
region $6.0 < E^{\tauh}_{\text{iso}} < 15.0\GeV$ for an event sample in which
no genuine $\tauh$ leptons are expected, namely $\cPZ$+jets events with $\cPZ \to \Pep \Pem$
or $\Pgmp \Pgmm$. The extrapolation ratio is sensitive to the level of jet activity in an event.
We study the variation of this ratio with respect to $\HT$ and the number of jets, using a variety
of jet-triggered and dilepton samples, and assign a systematic uncertainty of 30\% based on the
observed variation. Using this procedure we obtain $f_{\tau} = (20 \pm 6)\%$.

To estimate the $\tauh$ background in a search channel, the number
of candidates in the isolation sideband region of the corresponding
dilepton sample is multiplied by the extrapolation ratio, analogously
to the procedure for $f_{\mu}$ described in Sec.~\ref{subsec:NonPromptLeptons}
for the background from misidentified electrons and muons.

\subsection{Irreducible background from \texorpdfstring{$\PW \cPZ$}{WZ} and \texorpdfstring{$\cPZ \cPZ$}{ZZ} production}
\label{subsec:WZproduction}

The irreducible background, from $\PW \cPZ$+jets and $\cPZ \cPZ$+jets events
where both electroweak bosons decay leptonically, is evaluated
using samples of simulated events corrected for the measured lepton reconstruction efficiency
and $\MET$ resolution.
The simulated $\PW \cPZ$ and $\cPZ \cPZ$ distributions are normalized to corresponding
measured results obtained from $\PW \cPZ$- and $\cPZ \cPZ$-dominated data control samples,
defined by selecting events with on-$\cPZ$, low-$\HT$, and $50 < \MET < 100\GeV$ requirements,
or two-on-$\cPZ$, low-$\HT$, and  $\MET < 50\GeV$ requirements, respectively. The normalization
factors have statistical uncertainties of 6\% and 12\%, again respectively.

The $\MET$ distribution is examined in individual two-dimensional bins of $\HT$ and the number of reconstructed
vertices in the event. In an individual bin, the $x$ and $y$ components of $\MET$ are found to be approximately Gaussian.
The $\MET$ resolution is adversely affected by both pileup and jet activity, but in different ways. The effects
of pileup are stochastic, affecting the Gaussian widths of the distributions, while jet activity affects the tails.
We apply smearing factors to the Gaussian widths of the simulated events so that the $\MET$ resolution matches
that of the data. The corrections to the widths vary from a few percent to as high as around 25\% depending on
the bin. The effects of jet activity are accounted for in the evaluation of systematic uncertainties, which are determined
by varying the smearing factors and assessing the level of migration between different bins of $\MET$ and $\HT$.

\begin{figure}[!ht]
\centering
\includegraphics[width=\cmsFigWidth]{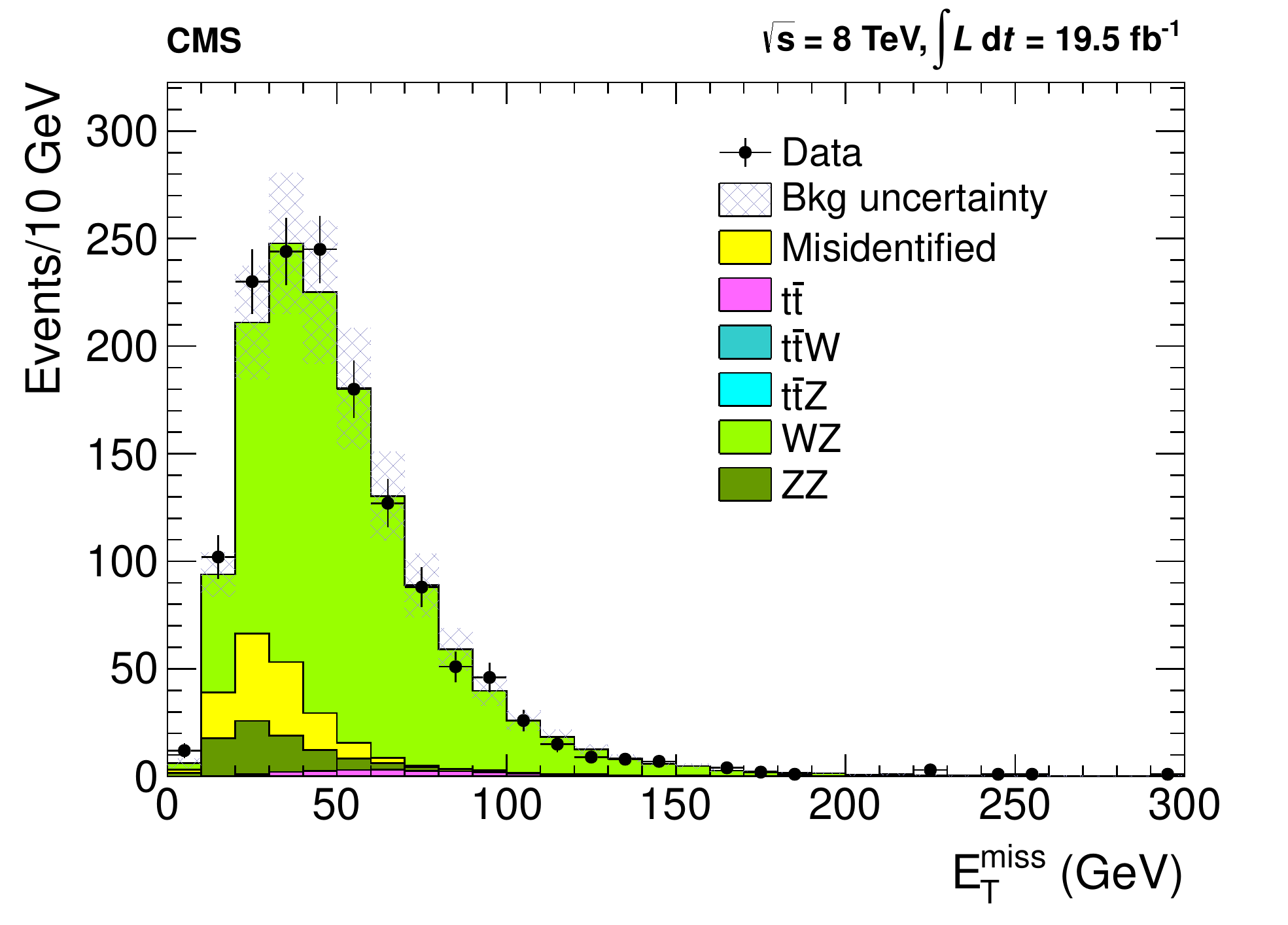}
\caption{Distribution of $\MET$ for a $\PW \cPZ$-enriched data control sample, in comparison to the result
from simulation. ``Misidentified" refers to SM background from Drell--Yan events, misidentified $\tauh$ decays,
and internal photon conversions. The simulation is normalized to a control region in data.
}
\label{fig:MET}
\end{figure}

For purposes of validation, Fig.~\ref{fig:MET} shows the distribution of $\MET$ for an on-$\cPZ$,
low-$\HT$, trilepton ($\Pe \Pe \Pe$, $\Pe \Pe \Pgm$, $\Pe \Pgm \Pgm$, and $\Pgm \Pgm \Pgm$),
$\PW \cPZ$-dominated data control sample defined by $75 < m_{\ell^+\ell^-} < 105\GeV$, $\HT < 200\GeV$,
and $50 < M_{\mathrm{T}} < 100\GeV$, where $M_{\mathrm{T}}$ is the transverse mass~\cite{CMS-PAPERS-TAU-11-001}
formed from the $\MET$ vector and the lepton not belonging to the OSSF pair. The results are shown
in comparison to simulated results that include the above-mentioned corrections.

\subsection{Background from \texorpdfstring{$\ttbar$}{TTbar} production}
\label{sub:TTbarProduction}

\begin{figure}[!ht]
\centering
\includegraphics[width=0.49\textwidth]{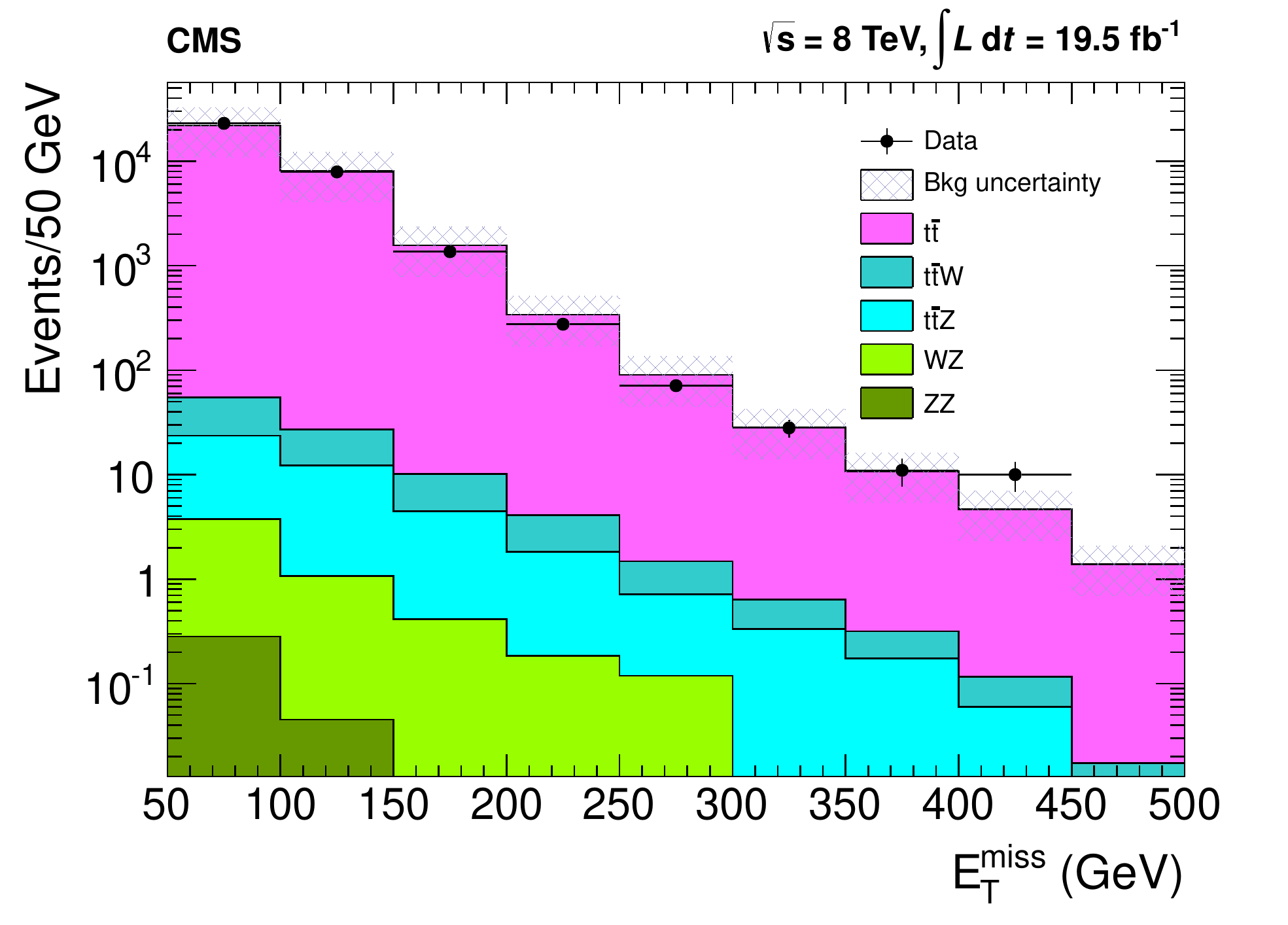}
\includegraphics[width=0.49\textwidth]{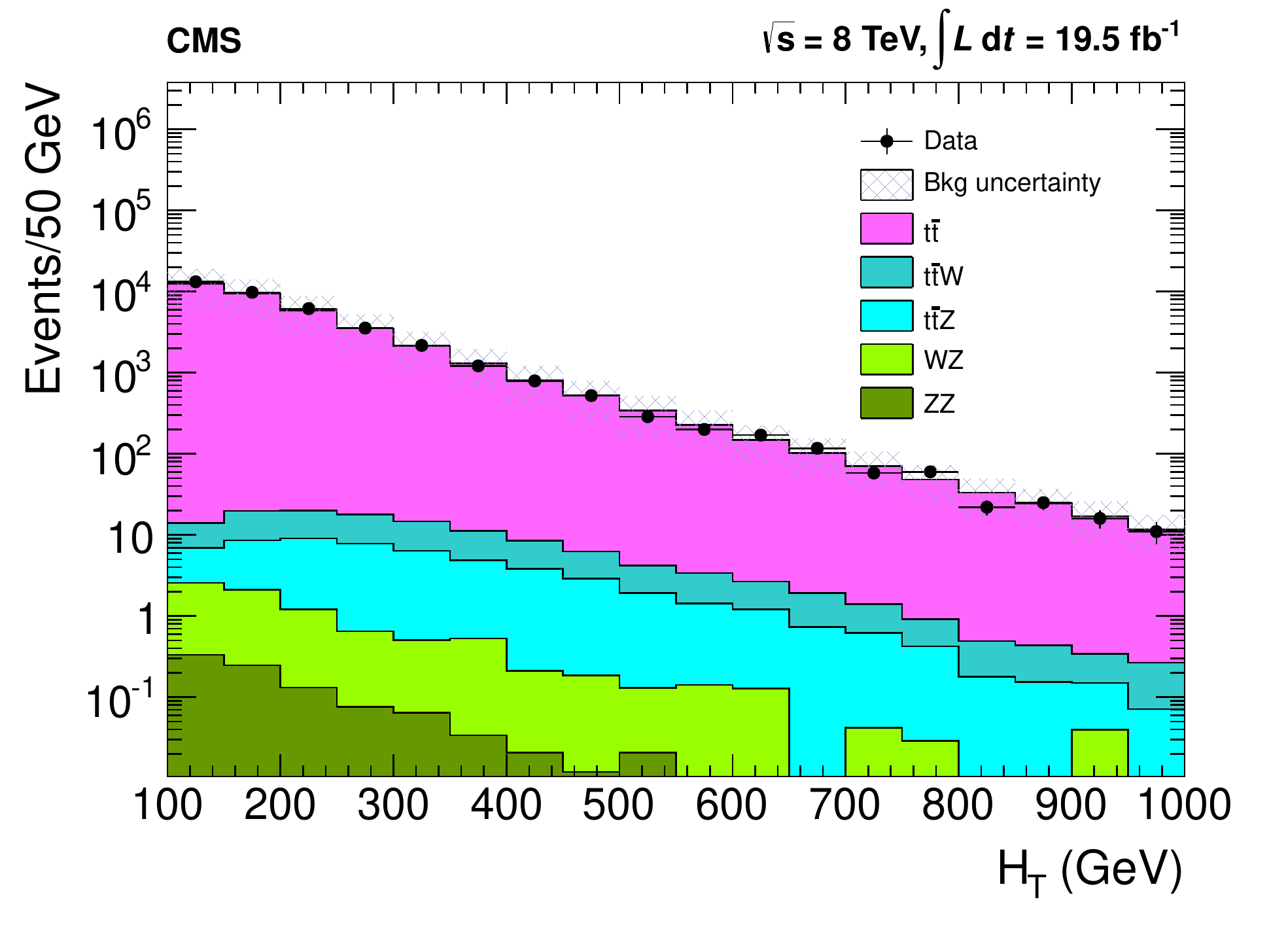}
\caption{Distribution of (\cmsLeft) $\MET$ and (\cmsRight) $\HT$ for a $\ttbar$-enriched data control sample, in
comparison to the result from simulation.
}
\label{fig:ttbar}
\end{figure}

The background from $\ttbar$ events is evaluated from simulation, with corrections applied for
lepton efficiencies and $\MET$ resolution as described in Sec.~\ref{subsec:WZproduction}.
Figure~\ref{fig:ttbar} shows the distributions of $\MET$ and $\HT$ for the data and corrected simulation
in a $\ttbar$-enriched control sample selected by requiring events to contain an
opposite-sign $\Pe \Pgm$ pair and at least one tagged $\cPqb$ jet.

\subsection{Backgrounds from asymmetric internal photon conversions}
\label{subsec:AsymmeticInternalPhotonConversions}

The background from photon conversions is evaluated from data by selecting a low-$\MET$, low-$\HT$
control region defined by $\MET < 30\GeV$ and $\HT < 200\GeV$ and measuring the ratio of the
number of events with $\abs{m_{\ell^+\ell^-\ell^{(\prime)\pm}}-m_{\cPZ}} < 15\GeV$ to those
with $\abs{m_{\ell^+\ell^-\gamma}-m{_\cPZ}} < 15\GeV$. We find a result of $(2.0 \pm 0.3)$\% for electrons
and $(0.7 \pm 0.1)$\% for muons, where the uncertainty is statistical. We multiply these factors by the
measured $\ell^+\ell^-\gamma$ rates in the signal regions to estimate the rate of photon-conversion
background events in these regions, with a systematic uncertainty of 50\%.

\section{Systematic uncertainties}
\label{syst}

The evaluation of systematic uncertainties for the SM background is partially discussed in the previous section.
In this section, we discuss additional sources of uncertainty, both for the background estimates
and the signal predictions.

Simulated signal and background samples are subject to uncertainties from the trigger, lepton-identification,
and isolation requirements. The latter two uncertainties are combined into a single term that is
approximately 1.5\% for leptons with $\pt > 20\GeV$.
The trigger efficiency uncertainties are approximately 5\%. Uncertainties associated with the
jet energy scale~\cite{CMS-PAPERS-JME-10-011}, \cPqb-jet tagging efficiency~\cite{Chatrchyan:2012jua},
$\MET$ resolution, and luminosity~\cite{CMS-PAS-LUM-13-001} affect signal efficiencies as well as background estimates
determined from simulation. The signal efficiencies are subject to an additional uncertainty,
from the ISR modeling~\cite{Chatrchyan:2013xna}.
Uncertainties in the cross section calculations affect the signal samples and simulation-derived background estimates,
with the exception of the background from $\PW \cPZ$ and $\cPZ \cPZ$ production, whose normalization is determined from data.

We assign a 50\% uncertainty to the estimate of the misidentified lepton background arising from $\ttbar$ production,
which is a combination of the uncertainty attributed to the cross section and an uncertainty derived from the level
of agreement between data and simulation for the distribution of the isolation variable.

The total systematic uncertainty per channel varies between 3\% and 40\%.
Table~\ref{tab:systematics} list representative values for some of the individual terms.

\begin{table}[!ht]
\centering
\topcaption{Typical values for systematic uncertainties.
}
\label{tab:systematics}
\begin{scotch}{lc}
Source of uncertainty & Magnitude (\%) \\
\hline
Luminosity & 2.6 \\
ISR modeling & 0--5 \\
$\MET$ resolution for $\PW \cPZ$ events & $\sim$ 4 \\
Jet energy scale ($\PW \cPZ$) &  0.5 \\
\cPqb-jet tagging & 0.1 ($\PW \cPZ$), 6 ($\ttbar$) \\
Muon ID/isolation at 10 (100)\GeV & 11 (0.2) \\
Electron ID/isolation at 10 (100)\GeV & 14 (0.6) \\
$\tauh$-lepton ID/isolation at 10 (100)\GeV & 2 (1.1)\\
Trigger efficiency & 5 \\
$\ttbar$ cross section/isolation variable & 50 \\
\end{scotch}
\end{table}

\section{Results}
\label{results}

Table~\ref{tab:resultsTable1} presents the results of the searches for events with four or more leptons,
and Table~\ref{tab:resultsTable2} the results for exactly three leptons. The observed numbers of events
are seen to be in overall agreement with the SM expectations.

Three excesses in the data relative to the SM estimates are worth noting in Table 2. All concern events
in the OSSF1, off-$\cPZ$ category with one $\tauh$-lepton candidate, no tagged $\cPqb$ jet, and $\HT < 200\GeV$.
Specifically, we observe 15, 4, and 3 events for $0 < \MET < 50\GeV$, $50 < \MET < 100\GeV$,
and $\MET > 100\GeV$, respectively, when only $7.5 \pm 2.0$, $2.1 \pm 0.5$, and $0.60 \pm 0.24$
SM events are expected, for an expectation of $10.1 \pm 2.4$ events in the combined $\MET$ range.
We determine the single-measurement probability to observe 22 or more events when the expected number
is $10.1 \pm 2.4$ events to be about 1\%. However, once trial factors are incorporated to account for the 64
independent channels of the analysis, the probability to observe such a fluctuation increases to about 50\%.
Alternatively, the joint probability to observe at least as large an excess for all three channels considered individually
is about 5\%. We account for systematic uncertainties and their correlations when evaluating these probabilities.

\begin{table*}[!ht]
\centering
\topcaption{Observed (Obs.) numbers of events with four or more leptons in comparison with the
expected (Exp.) numbers of SM background events. ``On-$\cPZ$" refers to events with
at least one $\Pep \Pem$ or $\Pgmp \Pgmm$ (OSSF) pair with dilepton mass
between 75 and 105\GeV, while ``Off-$\cPZ$" refers to events with one or two
OSSF pairs, none of which fall in this mass range. The OSSF$n$ designation refers to the number
of $\Pep \Pem$ and $\Pgmp \Pgmm$ pairs in the event, as explained in the text. Search channels
binned in $\MET$ have been combined into coarse $\MET$ bins for the purposes of presentation.
All uncertainties include both the statistical and systematic terms. The channel marked with an
asterisk is used for normalization purposes and is excluded from the search.
}
\label{tab:resultsTable1}
\resizebox{\textwidth}{!}{
\begin{scotch}{clc|cc|cc|cc|cc}
\multicolumn{1}{c}{$\geq$4 leptons} & $m_{\ell^+\ell^-}$ & $\MET$ & \multicolumn{2}{c|}{${N}_{\tauh} = 0$, ${N}_{\cPqb} = 0$} & \multicolumn{2}{c|}{${N}_{\tauh} = 1$, ${N}_{\cPqb} = 0$} & \multicolumn{2}{c|}{${N}_{\tauh} = 0$, ${N}_{\cPqb} \geq 1$} & \multicolumn{2}{c}{${N}_{\tauh} = 1$, ${N}_{\cPqb} \geq 1$} \\
$\HT > 200\GeV$ & & (\GeVns) & Obs. & Exp. & Obs. & Exp. & Obs. & Exp. & Obs. & Exp. \\
\hline
OSSF0 & \NA & (100, $\infty$) & 0 & $0.01^{+0.03}_{-0.01}$ & 0 & $0.01^{+0.06}_{-0.01}$ & 0 & $0.02^{+0.04}_{-0.02}$ & 0 & 0.11 $\pm$ 0.08 \\
OSSF0 & \NA & (50, 100) & 0 & $0.00^{+0.02}_{-0.00}$ & 0 & $0.01^{+0.06}_{-0.01}$ & 0 & $0.00^{+0.03}_{-0.00}$ & 0 & 0.12 $\pm$ 0.07 \\
OSSF0 & \NA & (0, 50) & 0 & $0.00^{+0.02}_{-0.00}$ & 0 & $0.07^{+0.10}_{-0.07}$ & 0 & $0.00^{+0.02}_{-0.00}$ & 0 & 0.02 $\pm$ 0.02 \\
OSSF1 & Off-$\cPZ$ & (100, $\infty$) & 0 & $0.01^{+0.02}_{-0.01}$ & 1 & 0.25 $\pm$ 0.11 & 0 & 0.13 $\pm$ 0.08 & 0 & 0.12 $\pm$ 0.12 \\
OSSF1 & On-$\cPZ$ & (100, $\infty$) & 1 & 0.10 $\pm$ 0.06 & 0 & 0.50 $\pm$ 0.27 & 0 & 0.42 $\pm$ 0.22 & 0 & 0.42 $\pm$ 0.19 \\
OSSF1 & Off-$\cPZ$ & (50, 100) & 0 & 0.07 $\pm$ 0.06 & 1 & 0.29 $\pm$ 0.13 & 0 & 0.04 $\pm$ 0.04 & 0 & 0.23 $\pm$ 0.13 \\
OSSF1 & On-$\cPZ$ & (50, 100) & 0 & 0.23 $\pm$ 0.11 & 1 & 0.70 $\pm$ 0.31 & 0 & 0.23 $\pm$ 0.13 & 1 & 0.34 $\pm$ 0.16 \\
OSSF1 & Off-$\cPZ$ & (0, 50) & 0 & $0.02^{+0.03}_{-0.02}$ & 0 & 0.27 $\pm$ 0.12 & 0 & $0.03^{+0.04}_{-0.03}$ & 0 & 0.31 $\pm$ 0.15 \\
OSSF1 & On-$\cPZ$ & (0, 50) & 0 & 0.20 $\pm$ 0.08 & 0 & 1.3 $\pm$ 0.5 & 0 & 0.06 $\pm$ 0.04 & 1 & 0.49 $\pm$ 0.19 \\
OSSF2 & Off-$\cPZ$ & (100, $\infty$) & 0 & $0.01^{+0.02}_{-0.01}$ & \NA & \NA & 0 & $0.01^{+0.06}_{-0.01}$ & \NA & \NA \\
OSSF2 & On-$\cPZ$ & (100, $\infty$) & 1 & $0.15^{+0.16}_{-0.15}$ & \NA & \NA & 0 & 0.34 $\pm$ 0.18 & \NA & \NA \\
OSSF2 & Off-$\cPZ$ & (50, 100) & 0 & 0.03 $\pm$ 0.02 & \NA & \NA & 0 & 0.13 $\pm$ 0.09 & \NA & \NA \\
OSSF2 & On-$\cPZ$ & (50, 100) & 0 & 0.80 $\pm$ 0.40 & \NA & \NA & 0 & 0.36 $\pm$ 0.19 & \NA & \NA \\
OSSF2 & Off-$\cPZ$ & (0, 50) & 1 & 0.27 $\pm$ 0.13 & \NA & \NA & 0 & 0.08 $\pm$ 0.05 & \NA & \NA \\
OSSF2 & On-$\cPZ$ & (0, 50) & 5 & 7.4 $\pm$ 3.5 & \NA & \NA & 2 & 0.80 $\pm$ 0.40 & \NA & \NA \\
\hline
\multicolumn{1}{c}{$\geq$4 leptons} & $m_{\ell^+\ell^-}$ & $\MET$ & \multicolumn{2}{c|}{${N}_{\tauh} = 0$, ${N}_{\cPqb} = 0$} & \multicolumn{2}{c|}{${N}_{\tauh} = 1$, ${N}_{\cPqb} = 0$} & \multicolumn{2}{c|}{${N}_{\tauh} = 0$, ${N}_{\cPqb} \geq 1$} & \multicolumn{2}{c}{${N}_{\tauh} = 1$, ${N}_{\cPqb} \geq 1$} \\
$\HT < 200\GeV$ & & (\GeVns) & Obs. & Exp. & Obs. & Exp. & Obs. & Exp. & Obs. & Exp. \\
\hline
OSSF0 & \NA & (100, $\infty$) & 0 & 0.11 $\pm$ 0.08 & 0 & 0.17 $\pm$ 0.10 & 0 & $0.03^{+0.04}_{-0.03}$& 0 & 0.04 $\pm$ 0.04 \\
OSSF0 & \NA & (50, 100) & 0 & $0.01^{+0.03}_{-0.01}$ & 2 & 0.70 $\pm$ 0.33 & 0 & $0.00^{+0.02}_{-0.00}$ & 0 & 0.28 $\pm$ 0.16 \\
OSSF0 & \NA & (0, 50) & 0 & $0.01^{+0.02}_{-0.01}$ & 1 & 0.7 $\pm$ 0.3 & 0 & $0.00^{+0.02}_{-0.00}$ & 0 & 0.13 $\pm$ 0.08 \\
OSSF1 & Off-$\cPZ$ & (100, $\infty$) & 0 & 0.06 $\pm$ 0.04 & 3 & 0.60 $\pm$ 0.24 & 0 & $0.02^{+0.04}_{-0.02}$ & 0 & 0.32 $\pm$ 0.20 \\
OSSF1 & On-$\cPZ$ & (100, $\infty$) & 1 & 0.50 $\pm$ 0.18 & 2 & 2.5 $\pm$ 0.5 & 1 & 0.38 $\pm$ 0.20 & 0 & 0.21 $\pm$ 0.10 \\
OSSF1 & Off-$\cPZ$ & (50, 100) & 0 & 0.18 $\pm$ 0.06 & 4 & 2.1 $\pm$ 0.5 & 0 & 0.16 $\pm$ 0.08 & 1 & 0.45 $\pm$ 0.24 \\
OSSF1 & On-$\cPZ$ & (50, 100) & 2 & 1.2 $\pm$ 0.3 & 9 & 9.6 $\pm$ 1.6 & 2 & 0.42 $\pm$ 0.23 & 0 & 0.50 $\pm$ 0.16 \\
OSSF1 & Off-$\cPZ$ & (0, 50) & 2 & 0.46 $\pm$ 0.18 & 15 & 7.5 $\pm$ 2.0 & 0 & 0.09 $\pm$ 0.06 & 0 & 0.70 $\pm$ 0.31 \\
OSSF1 & On-$\cPZ$ & (0, 50) & 4 & 3.0 $\pm$ 0.8 & 41 & 40 $\pm$ 10 & 1 & 0.31 $\pm$ 0.15 & 2 & 1.50 $\pm$ 0.47 \\
OSSF2 & Off-$\cPZ$ & (100, $\infty$) & 0 & 0.04 $\pm$ 0.03 & \NA & \NA & 0 & 0.05 $\pm$ 0.04 & \NA & \NA \\
OSSF2 & On-$\cPZ$ & (100, $\infty$) & 0 & 0.34 $\pm$ 0.15 & \NA & \NA & 0 & 0.46 $\pm$ 0.25 & \NA & \NA \\
OSSF2 & Off-$\cPZ$ & (50, 100) & 2 & 0.18 $\pm$ 0.13 & \NA & \NA & 0 &$0.02^{+0.03}_{-0.02}$ & \NA & \NA \\
OSSF2 & On-$\cPZ$ & (50, 100) & 4 & 3.9 $\pm$ 2.5 & \NA & \NA & 0 & 0.50 $\pm$ 0.21 & \NA & \NA \\
OSSF2 & Off-$\cPZ$ & (0, 50) & 7 & 8.9 $\pm$ 2.4 & \NA & \NA & 1 & 0.23 $\pm$ 0.09 & \NA & \NA \\
OSSF2 & On-$\cPZ$ & (0, 50) & *156 & 160 $\pm$ 34 & \NA & \NA & 4 & 2.9 $\pm$ 0.8 & \NA & \NA \\
\end{scotch}
}
\end{table*}

\begin{table*}[!ht]
\centering
\topcaption{Observed (Obs.) numbers of events with exactly three leptons in comparison with the
expected (Exp.) numbers of SM background events. ``On-$\cPZ$" refers to events with
an $\Pep \Pem$ or $\Pgmp \Pgmm$ (OSSF) pair with dilepton mass
between 75 and 105\GeV, while ``Above-$\cPZ$" and ``Below-$\cPZ$" refer to
events with an OSSF pair with mass above 105\GeV or below 75\GeV, respectively.
The OSSF$n$ designation refers to the number of $\Pep \Pem$ and $\Pgmp \Pgmm$ pairs in the event, as
explained in the text. Search channels binned in $\MET$ have been combined into coarse
$\MET$ bins for the purposes of presentation. All uncertainties include both the statistical
and systematic terms. The channels marked with an asterisk are used for normalization
purposes and are excluded from the search.
}
\label{tab:resultsTable2}
\resizebox{\textwidth}{!}{
\begin{scotch}{clc|cc|cc|cc|cc}
\multicolumn{1}{c}{$3$ leptons} & $m_{\ell^+\ell^-}$ & $\MET$ & \multicolumn{2}{c|}{${N}_{\tauh} = 0$, ${N}_{\cPqb} = 0$} & \multicolumn{2}{c|}{${N}_{\tauh} = 1$, ${N}_{\cPqb} = 0$} & \multicolumn{2}{c|}{${N}_{\tauh} = 0$, ${N}_{\cPqb} \geq 1$} & \multicolumn{2}{c}{${N}_{\tauh} = 1$, ${N}_{\cPqb} \geq 1$} \\
$\HT > 200\GeV$ & & (\GeVns) & Obs. & Exp. & Obs. & Exp. & Obs. & Exp. & Obs. & Exp. \\
\hline
OSSF0 & \NA & (100, $\infty$) & 5 & 3.7 $\pm$ 1.6 & 35 & 33 $\pm$ 14 & 1 & 5.5 $\pm$ 2.2 & 47 & 61 $\pm$ 30 \\
OSSF0 & \NA & (50, 100) & 3 & 3.5 $\pm$ 1.4 & 34 & 36 $\pm$ 16 & 8 & 7.7 $\pm$ 2.7 & 82 & 91 $\pm$ 46 \\
OSSF0 & \NA & (0, 50) & 4 & 2.1 $\pm$ 0.8 & 25 & 25 $\pm$ 10 & 1 & 3.6 $\pm$ 1.5 & 52 & 59 $\pm$ 29 \\
OSSF1 & Above-$\cPZ$ & (100, $\infty$) & 5 & 3.6 $\pm$ 1.2 & 2 & 10.0 $\pm$ 4.8 & 3 & 4.7 $\pm$ 1.6 & 19 & 22 $\pm$ 11 \\
OSSF1 & Below-$\cPZ$ & (100, $\infty$) & 7 & 9.7 $\pm$ 3.3 & 18 & 14.0 $\pm$ 6.4 & 8 & 9.1 $\pm$ 3.4 & 21 & 23 $\pm$ 11 \\
OSSF1 & On-$\cPZ$ & (100, $\infty$) & 39 & 61 $\pm$ 23 & 17 & 15.0 $\pm$ 4.9 & 9 & 14.0 $\pm$ 4.4 & 10 & 12.0 $\pm$ 5.8 \\
OSSF1 & Above-$\cPZ$ & (50, 100) & 4 & 5.0 $\pm$ 1.6 & 14 & 11.0 $\pm$ 5.2 & 6 & 6.8 $\pm$ 2.4 & 32 & 30 $\pm$ 15 \\
OSSF1 & Below-$\cPZ$ & (50, 100) & 10 & 11.0 $\pm$ 3.8 & 24 & 19.0 $\pm$ 6.4 & 10 & 9.9 $\pm$ 3.7 & 25 & 32 $\pm$ 16 \\
OSSF1 & On-$\cPZ$ & (50, 100) & 78 & 80 $\pm$ 32 & 70 & 50 $\pm$ 11 & 22 & 22.0 $\pm$ 6.3 & 36 & 24.0 $\pm$ 9.8 \\
OSSF1 & Above-$\cPZ$ & (0, 50) & 3 & 7.3 $\pm$ 2.0 & 41 & 33.0 $\pm$ 8.7 & 4 & 5.3 $\pm$ 1.5 & 15 & 23 $\pm$ 11 \\
OSSF1 & Below-$\cPZ$ & (0, 50) & 26 & 25.0 $\pm$ 6.8 & 110 & 86 $\pm$ 23 & 5 & 10.0 $\pm$ 2.5 & 24 & 26 $\pm$ 11 \\
OSSF1 & On-$\cPZ$ & (0, 50) & *135 & 130 $\pm$ 41 & 542 & 540 $\pm$ 160 & 31 & 32.0 $\pm$ 6.5 & 86 & 75 $\pm$ 19 \\
\hline
\multicolumn{1}{c}{$3$ leptons} & $m_{\ell^+\ell^-}$ & $\MET$ & \multicolumn{2}{c|}{${N}_{\tauh} = 0$, ${N}_{\cPqb} = 0$} & \multicolumn{2}{c|}{${N}_{\tauh} = 1$, ${N}_{\cPqb} = 0$} & \multicolumn{2}{c|}{${N}_{\tauh} = 0$, ${N}_{\cPqb} \geq 1$} & \multicolumn{2}{c}{${N}_{\tauh} = 1$, ${N}_{\cPqb} \geq 1$} \\
$\HT < 200\GeV$ & & (\GeVns) & Obs. & Exp. & Obs. & Exp. & Obs. & Exp. & Obs. & Exp. \\
\hline
OSSF0 & \NA & (100, $\infty$) & 7 & 11.0 $\pm$ 4.9 & 101 & 111 $\pm$ 54 & 13 & 10.0 $\pm$ 5.3 & 87 & 119 $\pm$ 61 \\
OSSF0 & \NA & (50, 100) & 35 & 38 $\pm$ 15 & 406 & 402 $\pm$ 152 & 29 & 26 $\pm$ 13 & 269 & 298 $\pm$ 151 \\
OSSF0 & \NA & (0, 50) & 53 & 51 $\pm$ 11 & 910 & 1035 $\pm$ 255 & 29 & 23 $\pm$ 10 & 237 & 240 $\pm$ 113 \\
OSSF1 & Above-$\cPZ$ & (100, $\infty$) & 18 & 13.0 $\pm$ 3.5 & 25 & 38 $\pm$ 18 & 10 & 6.5 $\pm$ 2.9 & 24 & 35 $\pm$ 18 \\
OSSF1 & Below-$\cPZ$ & (100, $\infty$) & 21 & 24 $\pm$ 9 & 41 & 50 $\pm$ 25 & 14 & 20 $\pm$ 10 & 42 & 54 $\pm$ 28 \\
OSSF1 & On-$\cPZ$ & (100, $\infty$) & 150 & 150 $\pm$ 26 & 39 & 48 $\pm$ 13 & 15 & 14.0 $\pm$ 4.8 & 19 & 23 $\pm$ 11 \\
OSSF1 & Above-$\cPZ$ & (50, 100) & 50 & 46.0 $\pm$ 9.7 & 169 & 140 $\pm$ 48 & 20 & 18 $\pm$ 8 & 85 & 93 $\pm$ 47 \\
OSSF1 & Below-$\cPZ$ & (50, 100) & 142 & 130 $\pm$ 27 & 353 & 360 $\pm$ 92 & 48 & 48 $\pm$ 23 & 140 & 133 $\pm$ 68 \\
OSSF1 & On-$\cPZ$ & (50, 100) & *773 & 780 $\pm$ 120 & 1276 & 1200 $\pm$ 310 & 56 & 47 $\pm$ 13 & 81 & 75 $\pm$ 32 \\
OSSF1 & Above-$\cPZ$ & (0, 50) & 178 & 200 $\pm$ 35 & 1676 & 1900 $\pm$ 540 & 17 & 18.0 $\pm$ 6.7 & 115 & 94 $\pm$ 42 \\
OSSF1 & Below-$\cPZ$ & (0, 50) & 510 & 560 $\pm$ 87 & 9939 & 9000 $\pm$ 2700 & 34 & 42 $\pm$ 11 & 226 & 228 $\pm$ 63 \\
OSSF1 & On-$\cPZ$ & (0, 50) & *3869 & 4100 $\pm$ 670 & *50188 & 50000 $\pm$ 15000 & *148 & 156 $\pm$ 24 & 906 & 925 $\pm$ 263 \\
\end{scotch}
}
\end{table*}

\section{Interpretation of results for supersymmetric scenarios}
\label{models}

We consider five new-physics scenarios that appear in the
framework of the minimal supersymmetric standard model (MSSM)~\cite{Nilles:1983ge,Haber:1984rc}.
They involve sleptons (including staus), bottom and top squarks, higgsinos, gravitinos, neutralinos,
and charginos, where higgsinos are the superpartners of the Higgs bosons, the gravitino $\goldstino$ is the superpartner
of the graviton, while neutralinos (charginos)
are mixtures of the superpartners of neutral (charged) electroweak vector and Higgs bosons. The first three scenarios
feature the gravitino as the LSP, while the lightest neutralino $\none$ is the LSP for the other two scenarios. The first
and last two scenarios proceed through the production of third-generation squarks, yielding final states rich in
heavy-flavor jets. Taken together, these five scenarios present a wide spectrum of multilepton signatures.

Our search results lack striking departures from the SM, and we set limits
on the production cross sections of the five scenarios. The limits are determined
using the observed numbers of events, the SM background estimates, and the
predicted event yields. For each scenario, we order the search channels by their
expected sensitivities and then combine channels, starting with the most sensitive
one. For ease of computation and with a negligible loss in overall sensitivity, we
do not consider channels once the number of signal events integrated over the retained
channels reaches 90\% of the total. The list of selected channels thus depends not only on
the scenario considered, but also on the assumed superpartner masses and branching fractions.

We set 95\% confidence level (\CL) upper limits on the signal parameters and cross sections using
the modified frequentest CL$_\mathrm{s}$ method with the LHC-style test statistic~\cite{ATLAS:1379837,Junk:1999kv,Read:2002hq}.
Lognormal nuisance-parameter distributions are used to account for uncertainties.

\subsection{Natural higgsino NLSP scenario}

\begin{figure*}[!ht]
\centering
\includegraphics[width=0.53\textwidth]{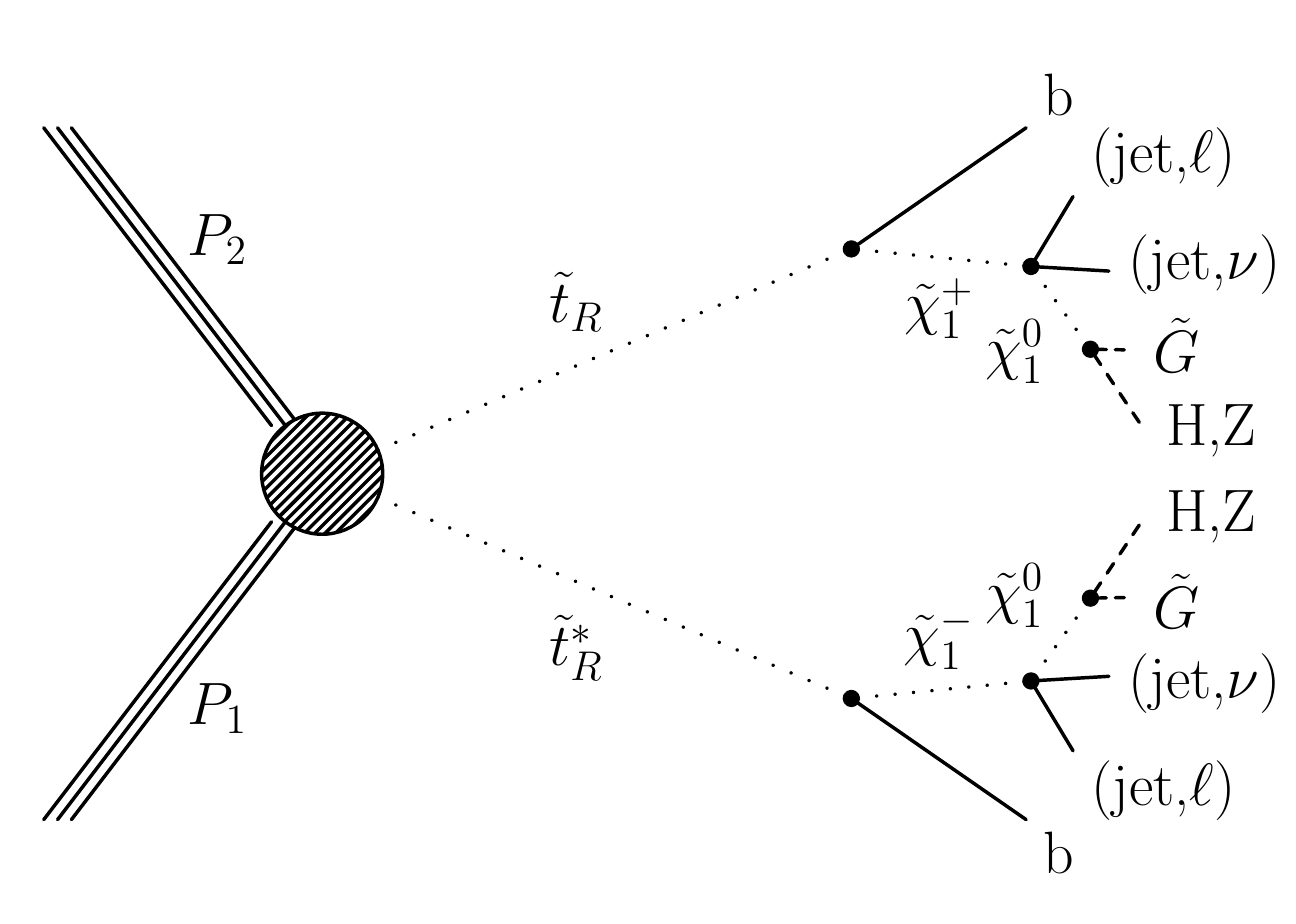}
\includegraphics[width=0.42\textwidth]{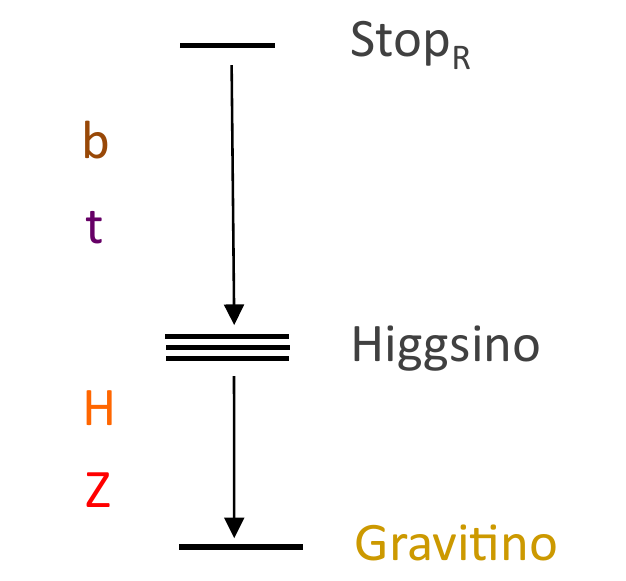}
\caption{Event diagram and a schematic superpartner mass spectrum for the GMSB natural higgsino NLSP scenario,
with $\cone$ ($\none$) the lightest chargino (neutralino), $\PH$ the lightest MSSM Higgs boson, and $\goldstino$ a gravitino.
Particles in parentheses in the event diagram have a soft $\pt$ spectrum.}
\label{fig:fd_spec_nh}
\end{figure*}

We first present a supersymmetric scenario
in which the $\none$ neutralino is a higgsino that forms the next-to-LSP (NLSP)
state~\cite{Chatrchyan:2013mya}. We refer to this scenario as
the ``natural higgsino NLSP" scenario. This scenario arises in gauge-mediated
SUSY-breaking (GMSB) models~\cite{PhysRevD.62.077702}.
Production proceeds through the right-handed top-antitop squark
pair $\PSQt_{R} \PASQt_{R}$, with the subsequent decays $\PSQt_{R} \to \cPqb \conep$
or $\PSQt_{R} \to \cPqt \nalli~(i = 1,2)$, where $\conep$ is the lightest chargino
and $\ntwo$ the second-lightest neutralino (both taken to be higgsinos), with the $\squark^{*}$
state the charge conjugate of the $\squark$ state. The $\conep$ and $\ntwo$ states each decay
to the $\none$ and SM particles. Figure~\ref{fig:fd_spec_nh} shows an event
diagram and a schematic mass spectrum. The last step in each of the two top-squark decay
chains is the decay $\none  \to \PH \goldstino$ or $\cPZ \goldstino$, yielding
an $\PH \PH$, $\PH \cPZ$, or $\cPZ \cPZ$ configuration, with $\MET$ from the undetected
gravitino. Note that we assume $\PH \goldstino$ and $\cPZ \goldstino$ to be the only two possible
decay modes for the $\none$ higgsino~\cite{PhysRevD.62.077702}.

Beyond the top-squark pair production diagram of Fig.~\ref{fig:fd_spec_nh},
the natural higgsino NLSP scenario also encompasses direct higgsino pair
production, in which the $\conep$ and $\conem$ states of Fig.~\ref{fig:fd_spec_nh}
(plus other di-higgsino states) are produced through electroweak interactions,
leading to the same $\PH \PH$, $\PH \cPZ$, and $\cPZ \cPZ$ configuration
as in Fig. 3, but with less jet activity~\cite{PhysRevD.62.077702}. Our search results
are also sensitive to this scenario.

Of the five new-physics scenarios we examine, the natural higgsino NLSP scenario
exhibits the largest range with respect to its population of the different search channels.
The channels with highest sensitivity are those that require $\cPqb$ jets, and,
for the decays through the  $\PH \cPZ$ and $\cPZ \cPZ$ states, the channels
with on-$\cPZ$  and off-$\cPZ$ requirements.

The natural higgsino NLSP scenario is complex because the higgsino can decay to either a $\cPZ$ or Higgs boson,
while the Higgs boson has many decay modes that lead to leptons. We consider seven decay channels for
the $\PH \PH$ configuration:  $\PW \PW^* \PW \PW^*$, $\cPZ \cPZ^* \cPZ \cPZ^*$, $\Pgt \Pgt \Pgt \Pgt$,
$\PW \PW^* \cPZ \cPZ^*$, $\PW \PW^* \Pgt \Pgt$, $\cPZ \cPZ^* \Pgt \Pgt$, and $\cPZ \cPZ^* \cPqb \cPqb$,
and three decay channels for the $\PH \cPZ$ configuration:  $\PW \PW^{*} \cPZ$, $\cPZ \cPZ^{*} \cPZ$,
and $\tau \tau \cPZ$, where $\PW^*$ and $\cPZ^*$ indicate off-shell vector bosons.

Signal events for the natural higgsino NLSP scenario are generated using \MADGRAPH,
as described in Sec.~\ref{data}. The $\none$ and $\ntwo$ higgsinos are assigned
masses 5\GeV below and above the mass of the $\cone$ higgsino, respectively,
while the gravitino is assumed to be massless. In the limit of no mixing between
higgsinos and gauginos, the light neutralinos and charginos become degenerate~\cite{PhysRevD.62.077702}. 
The 5 GeV splitting is representative of proximity to this limit. We generate signal events for a range
of $\PSQt_{R}$ and $\cone$ mass values. Cross sections for both the strong and
electroweak production processes are assigned an uncertainty of 20\%, which also 
accounts for the uncertainties associated with the PDFs and with the renormalization 
and factorization scales.

\begin{figure}[!ht]
\centering
\includegraphics[width=0.49\textwidth]{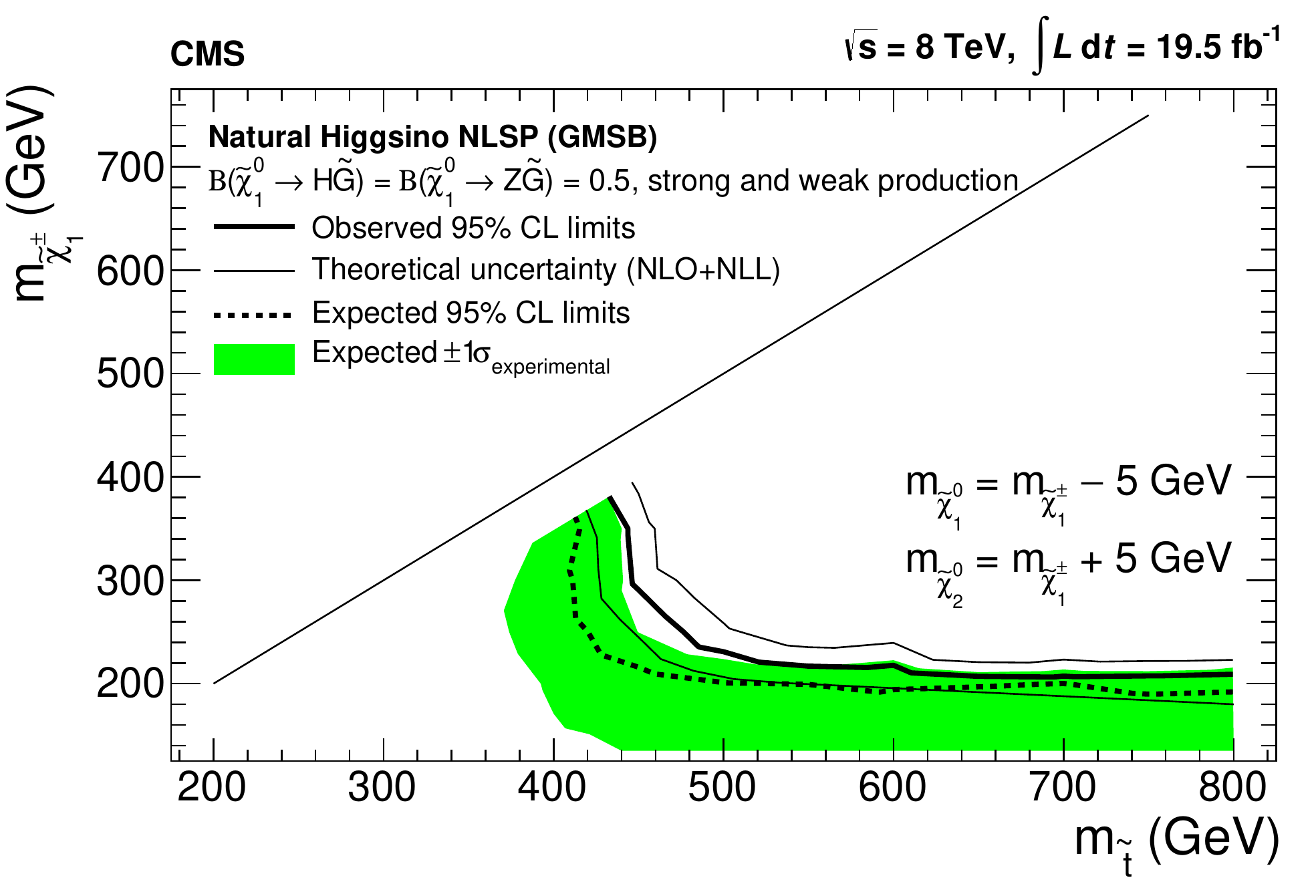}
\includegraphics[width=0.49\textwidth]{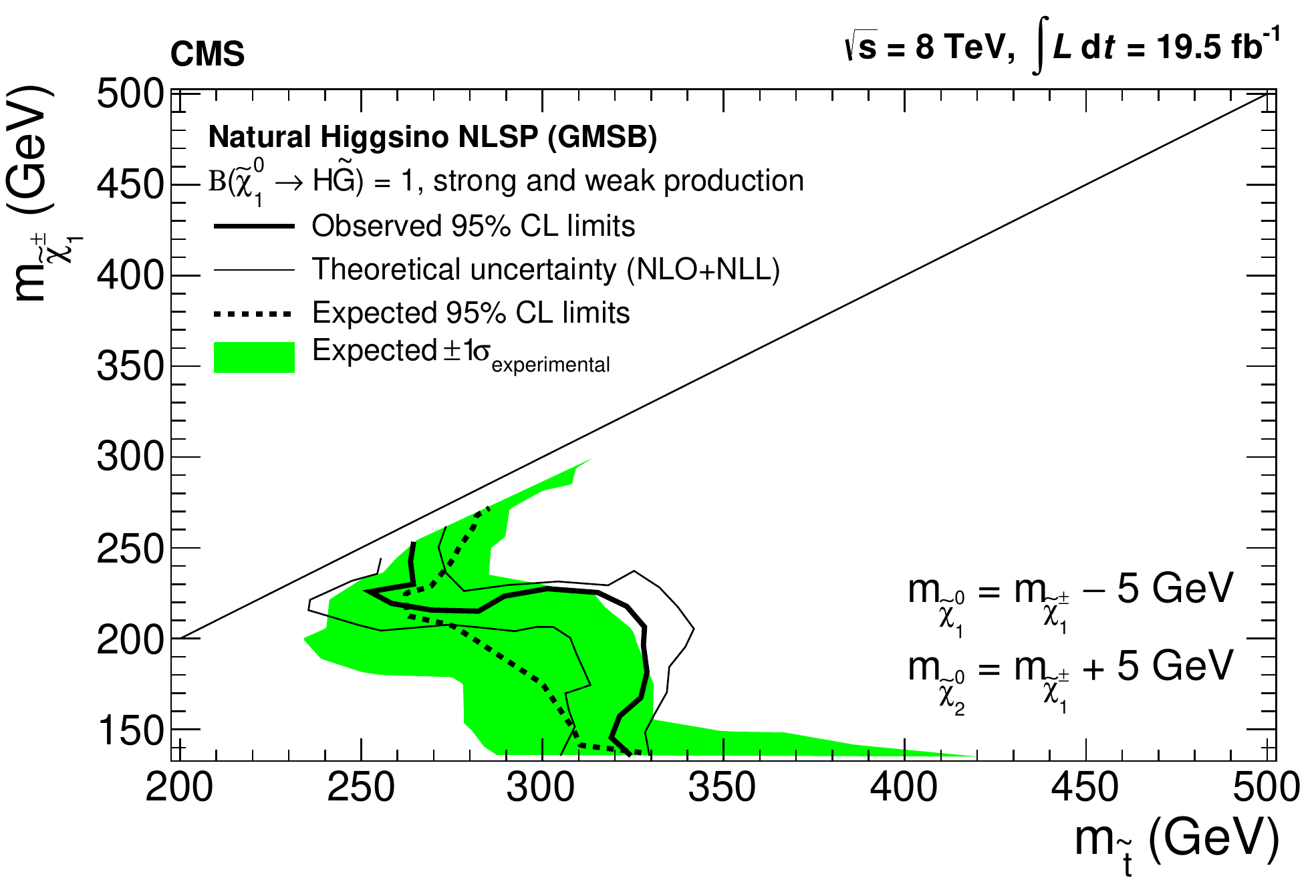}
\includegraphics[width=0.49\textwidth]{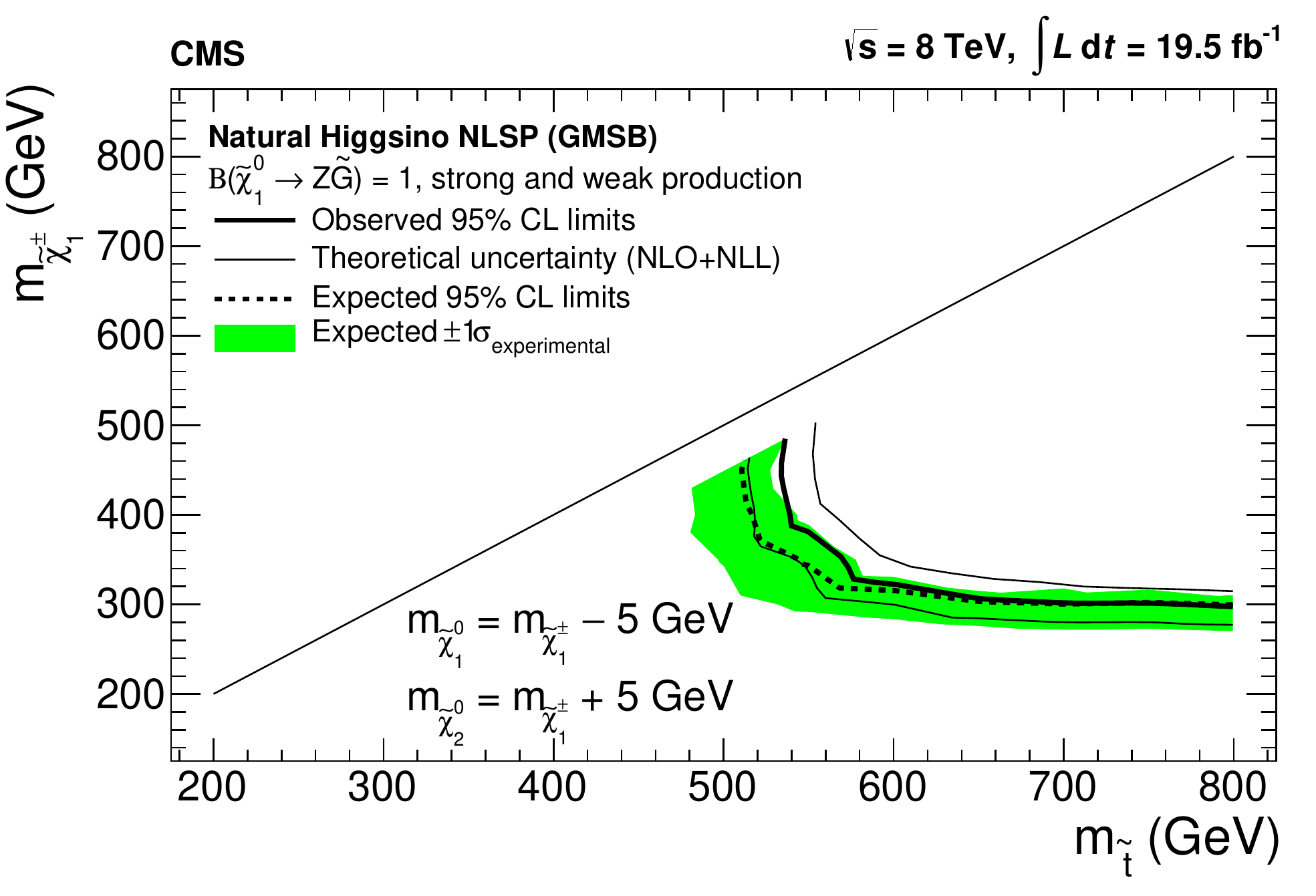}
\caption{
The 95\% confidence level upper limits in the top squark versus chargino mass plane, for the
natural higgsino NLSP scenario with the following $\none$ branching fractions:
$\mathcal{B}(\none \to \PH \goldstino) = \mathcal{B}(\none \to \cPZ \goldstino) = 0.5$ (\cmsUpperLeft),
$\mathcal{B}(\none \to \PH \goldstino) = 1.0$ (\cmsUpperRight),
and $\mathcal{B}(\none \to \cPZ \goldstino) = 1.0$ (bottom).
Both strong and electroweak production mechanisms are considered.
The region to the left and below the contours is excluded.
The region above the diagonal straight line is unphysical.
}
\label{fig:NaturalHiggsinoNLSPExclusion}
\end{figure}

\begin{figure*}[!ht]
\centering
\includegraphics[width=0.49\textwidth]{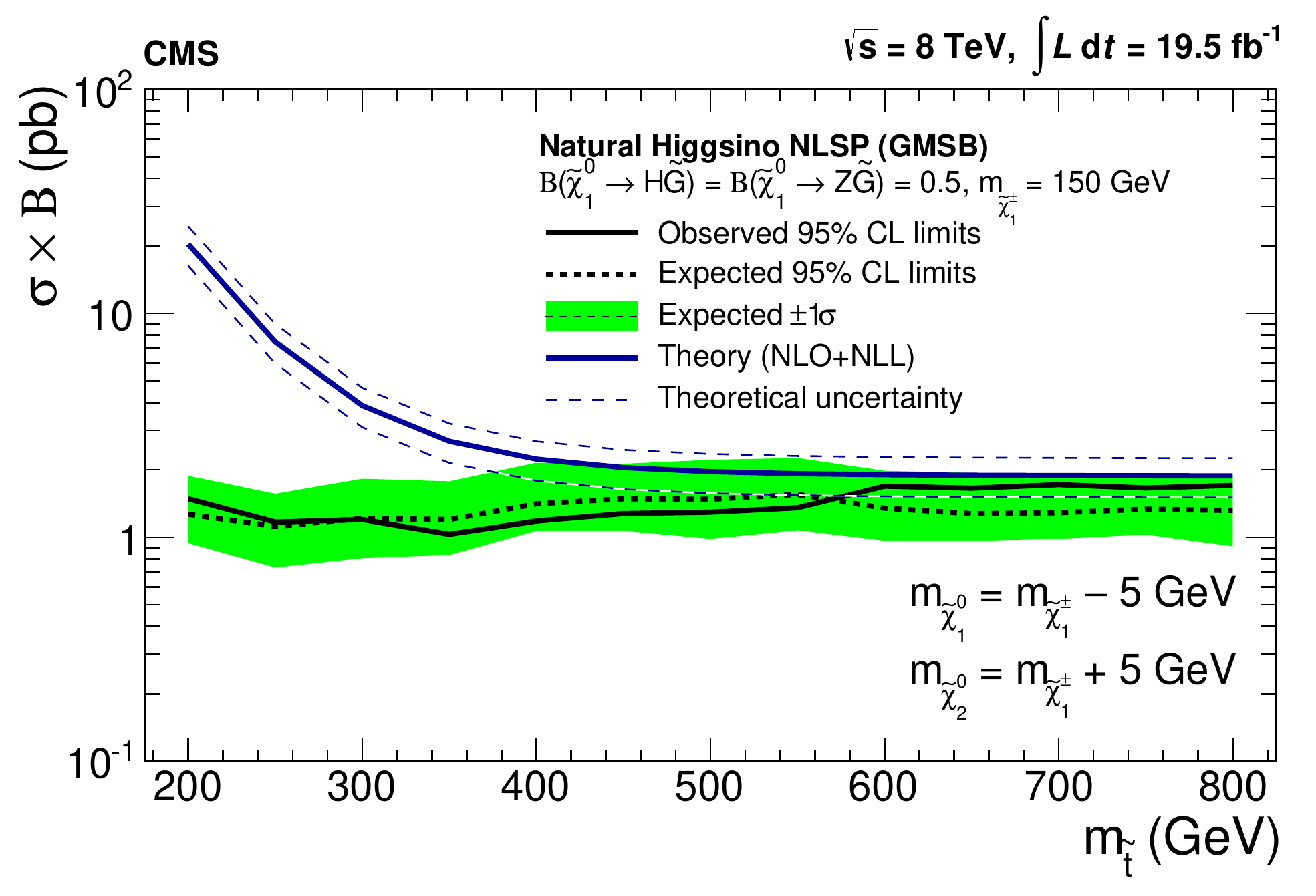}
\includegraphics[width=0.49\textwidth]{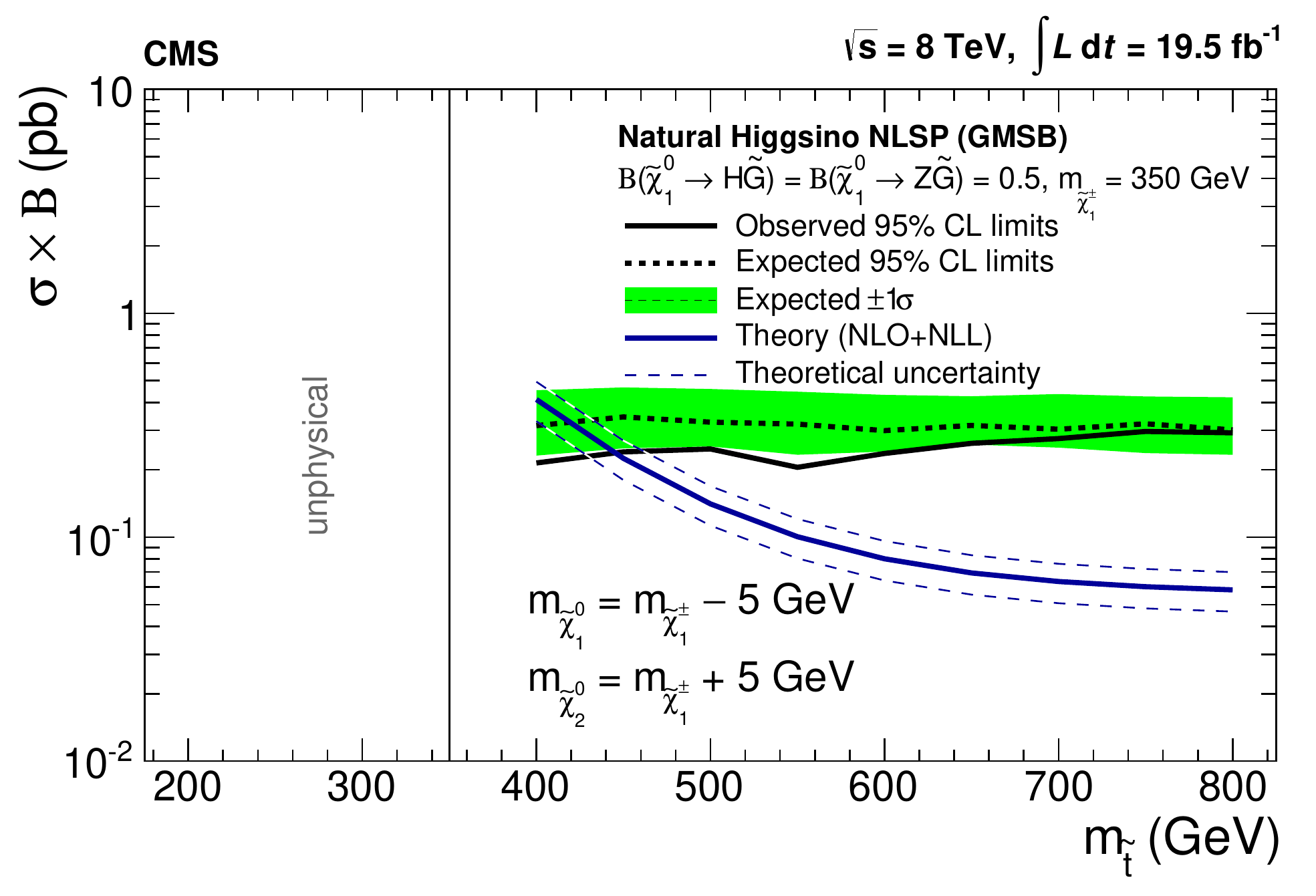}
\includegraphics[width=0.49\textwidth]{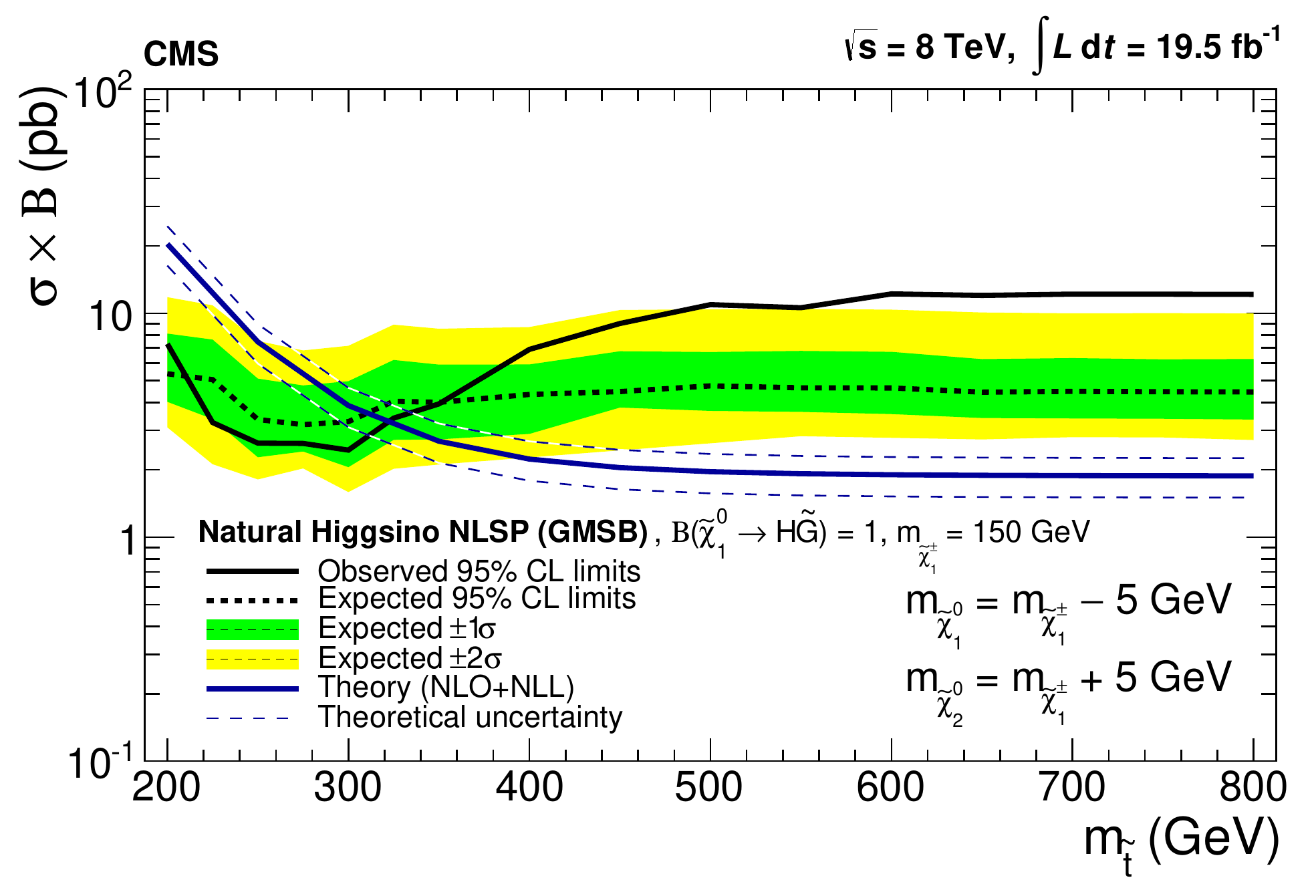}
\includegraphics[width=0.49\textwidth]{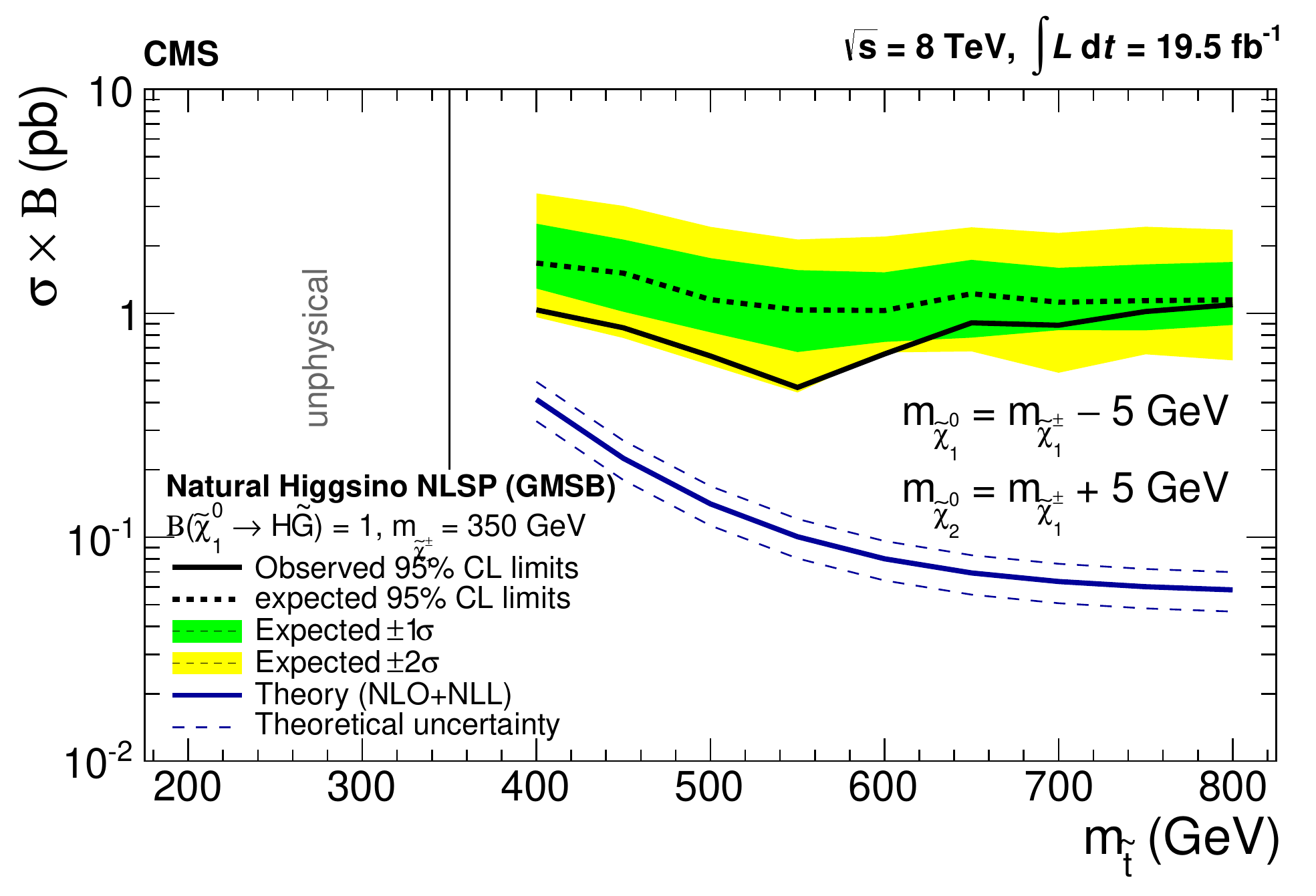}
\includegraphics[width=0.49\textwidth]{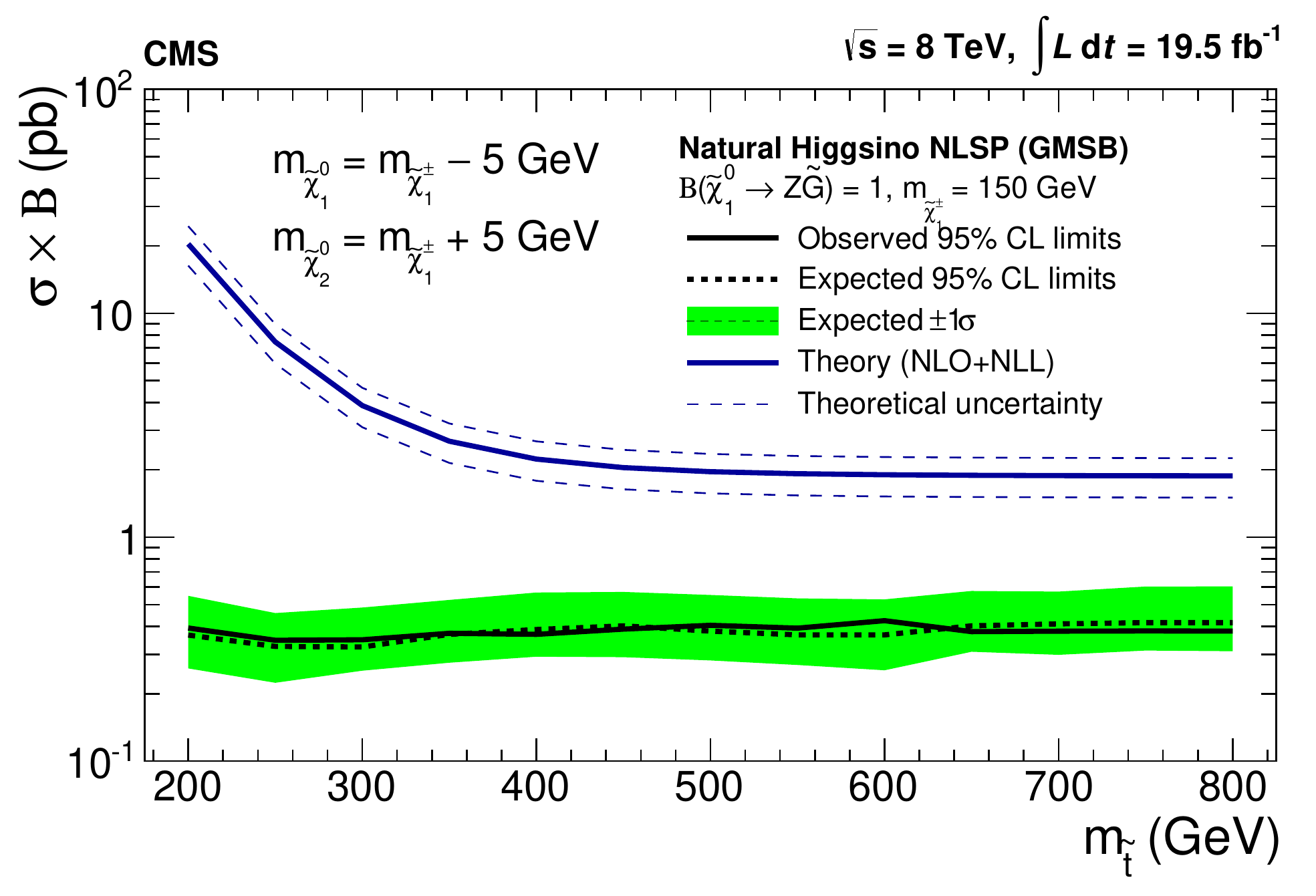}
\includegraphics[width=0.49\textwidth]{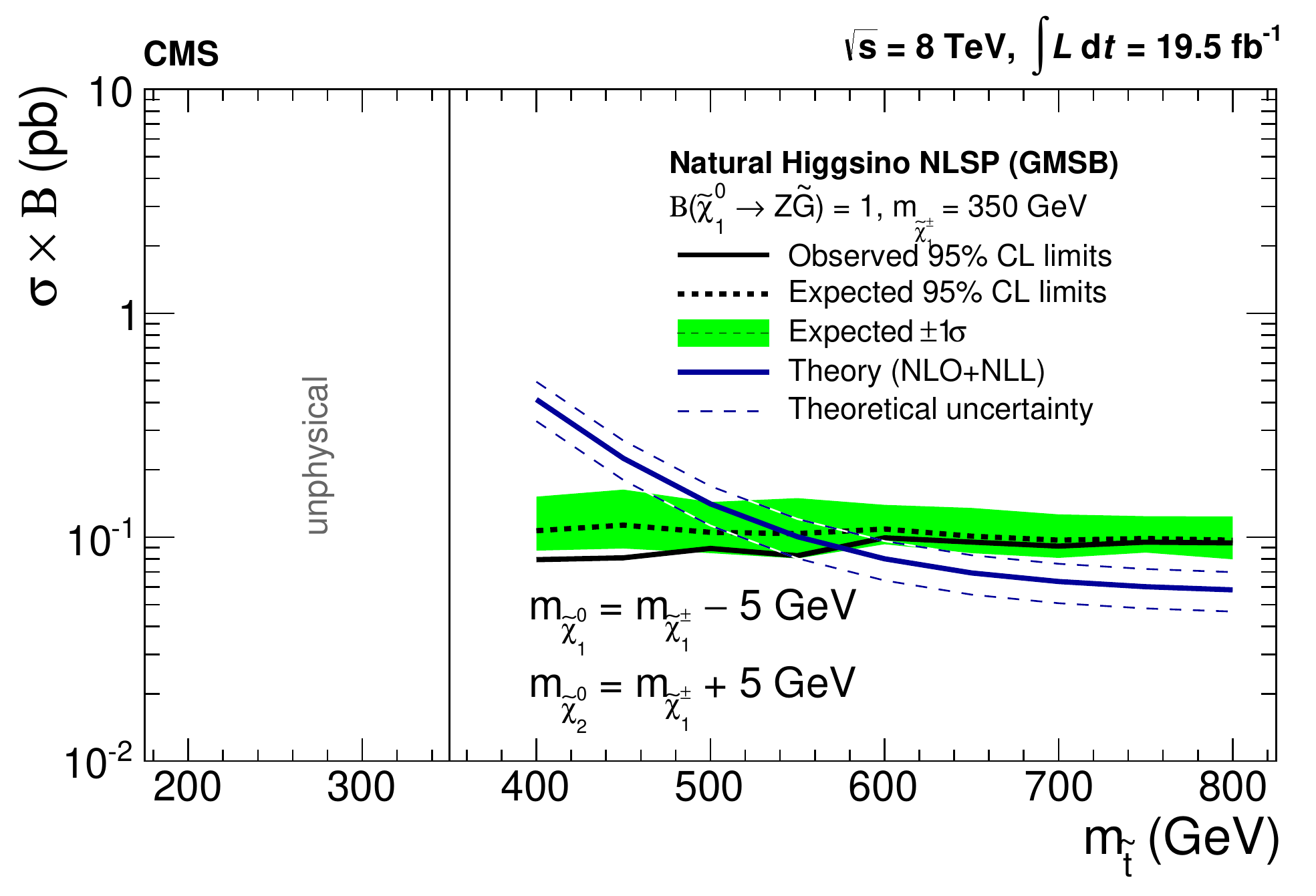}
\caption{
The 95\% confidence level upper limits on cross section times branching fraction $\mathcal{B}$,
for the natural higgsino NLSP scenario
for $\mathcal{B}(\none \to \PH \goldstino) = \mathcal{B}(\none \to \cPZ \goldstino) = 0.5$ (top),
$\mathcal{B}(\none \to \PH \goldstino) = 1.0$ (middle), and
$\mathcal{B}(\none \to \cPZ \goldstino) = 1.0$ (bottom). The
charged higgsino mass is fixed at 150\GeV (left) and 350\GeV (right).
The region to the left of the vertical line on the right plots is unphysical and limited by the charged higgsino mass.
}
\label{fig:NaturalHiggsinoNLSP_FixedHiggsino150_Exclusion}
\end{figure*}

\begin{figure}[!ht]
\centering
\includegraphics[width=0.49\textwidth]{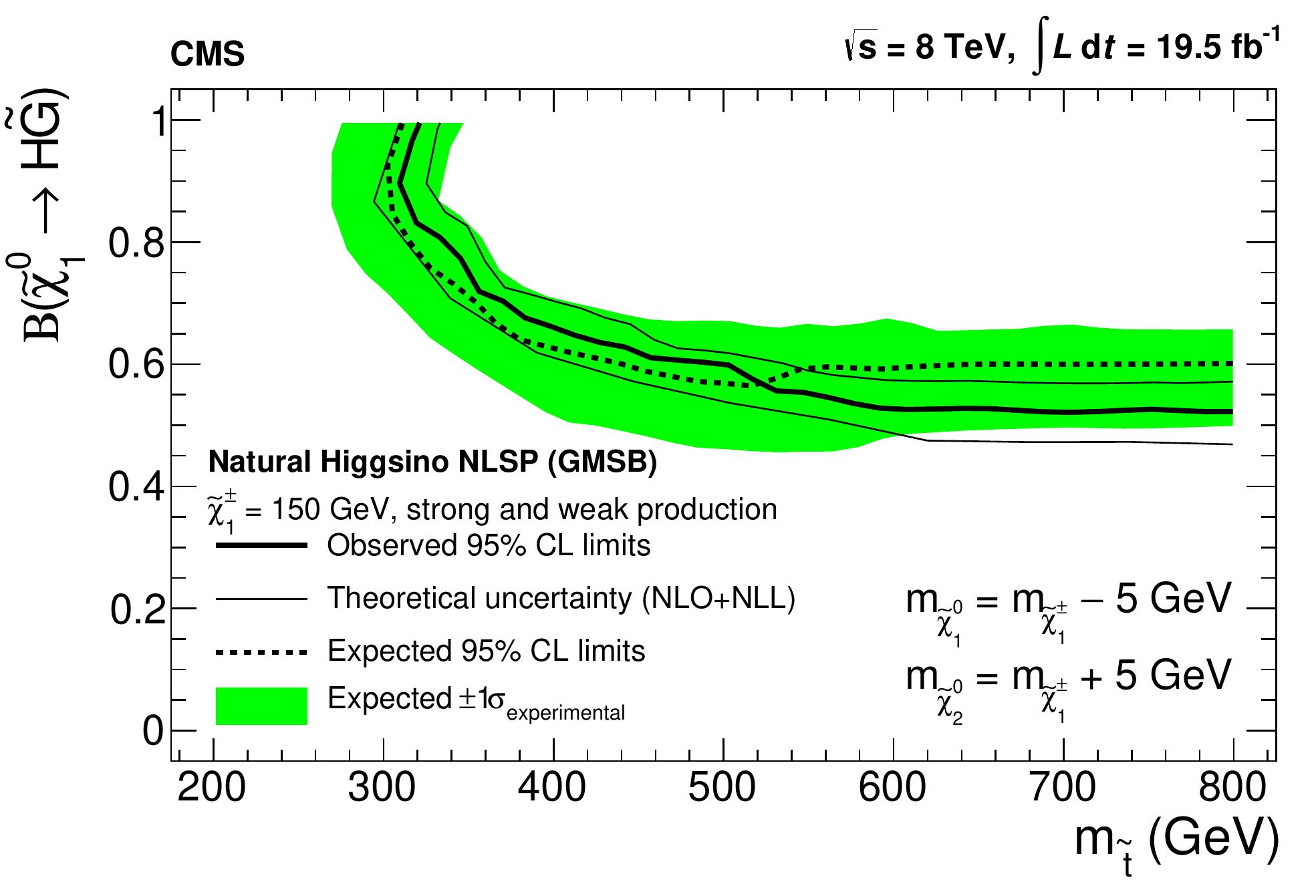}
\includegraphics[width=0.49\textwidth]{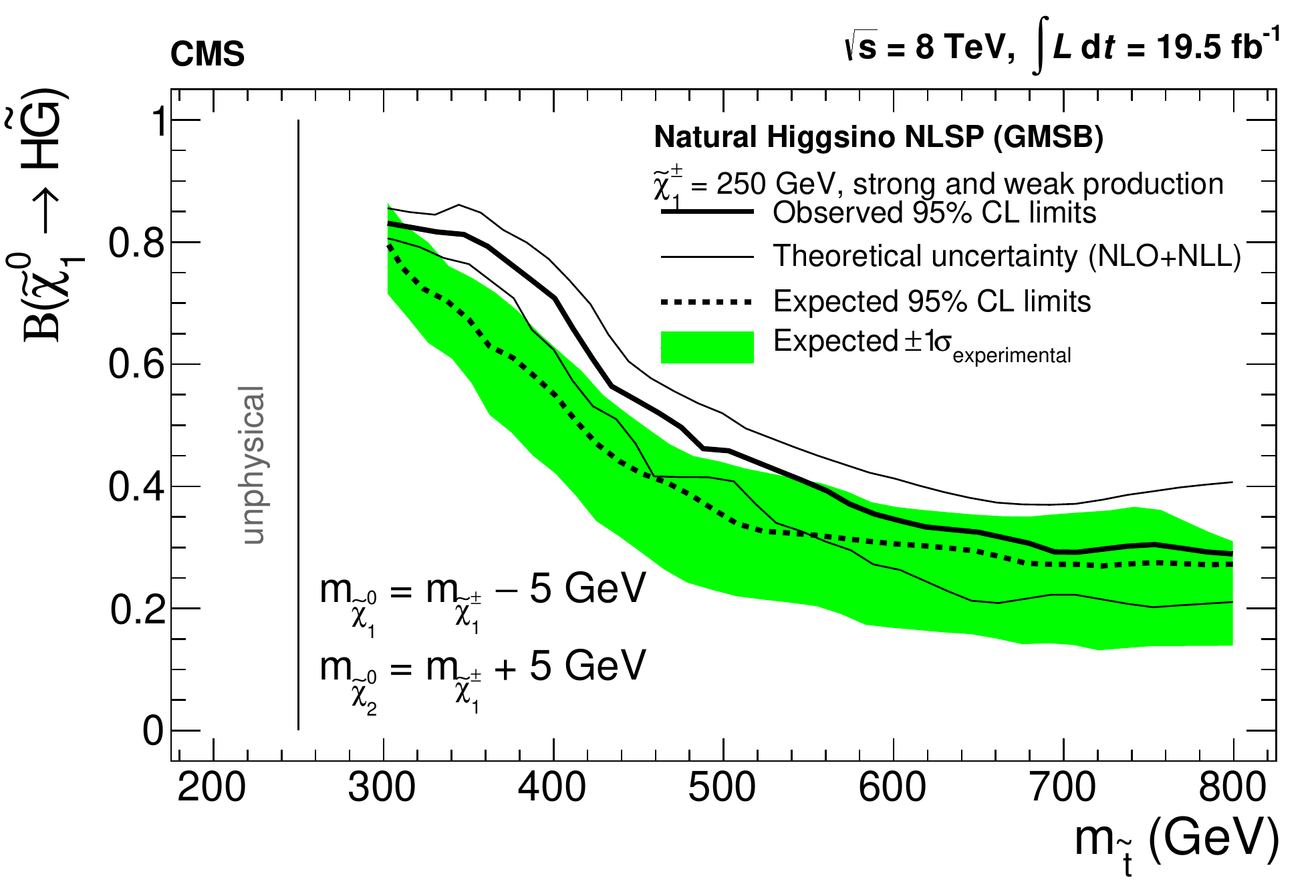}
\includegraphics[width=0.49\textwidth]{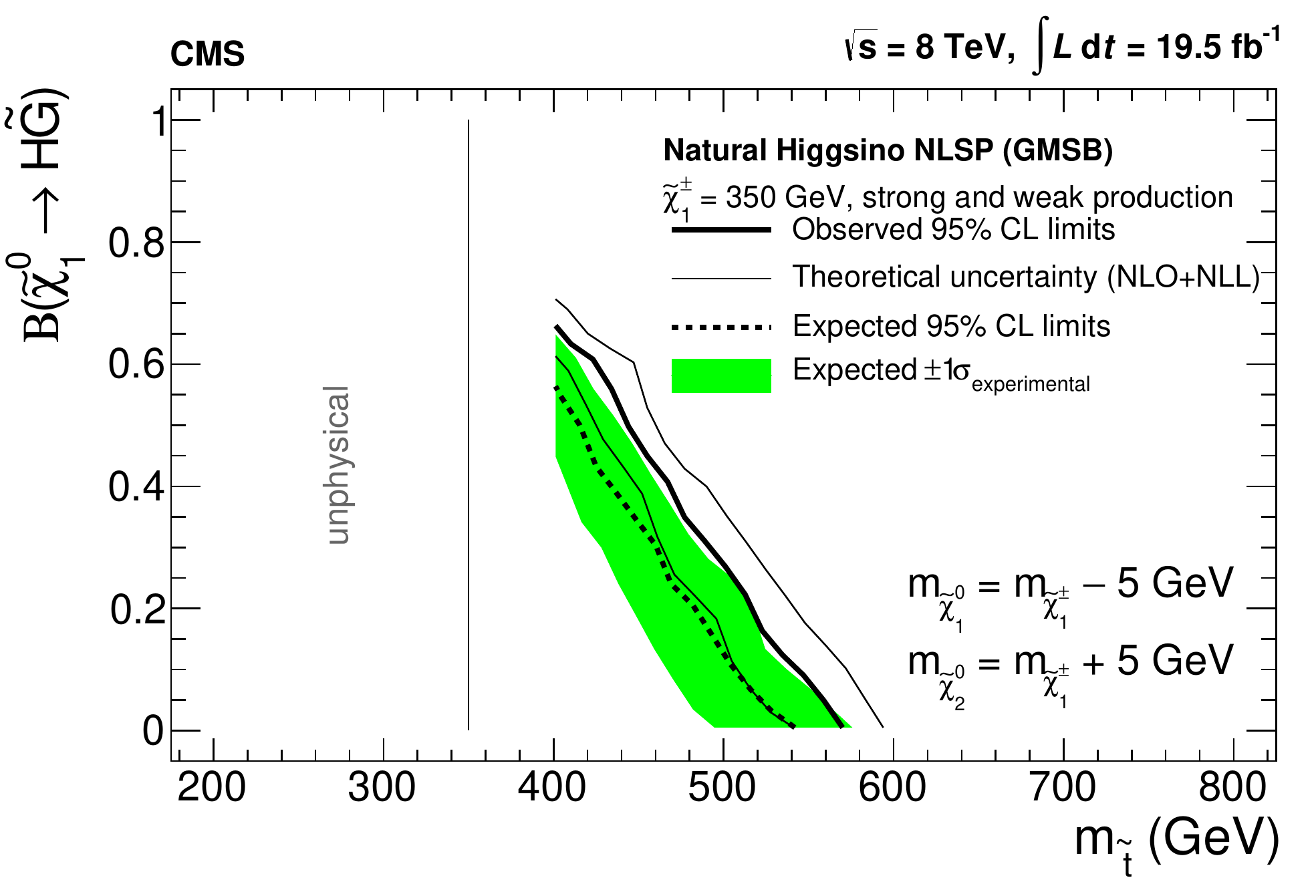}
\caption{The 95\% confidence level upper limits on the branching fraction $\mathcal{B}(\none \to \PH \goldstino)$
for the natural higgsino NLSP scenario with fixed charged higgsino mass of 150\GeV (\cmsUpperLeft), 250\GeV (\cmsUpperRight), and 350\GeV
(bottom) assuming $\mathcal{B}(\none \to \PH \goldstino) + \mathcal{B}(\none \to \cPZ \goldstino) = 1.0$.
The region to the left of the vertical line on the right plots is unphysical and limited by the charged higgsino mass.
}
\label{fig:NaturalHiggsinoNLSP_floatingBR_FixedHiggsino150_Exclusion}
\end{figure}

Figure~\ref{fig:NaturalHiggsinoNLSPExclusion} shows the excluded
regions in the plane of $m_{\cone}$ versus $m_{\PSQt}$. The results are shown
for several choices for the $\none \to \PH \goldstino$ branching fraction.
One-dimensional exclusion plots with fixed choices for the branching fraction
and chargino mass are shown in Fig.~\ref{fig:NaturalHiggsinoNLSP_FixedHiggsino150_Exclusion}.
The search sensitivity is larger
for lower chargino masses because of the larger cross section. There is
less sensitivity for the Higgs-boson-dominated mode in comparison with
the $\cPZ$-boson-dominated mode. Figure~\ref{fig:NaturalHiggsinoNLSP_floatingBR_FixedHiggsino150_Exclusion}
shows the results as a function of the $\none \to \PH \goldstino$
branching fraction and the top squark mass for different chargino masses.

\subsection{Slepton co-NLSP scenario}
\label{subsec:coNLSP}

\begin{figure*}[!ht]
\centering
\includegraphics[width=0.50\textwidth]{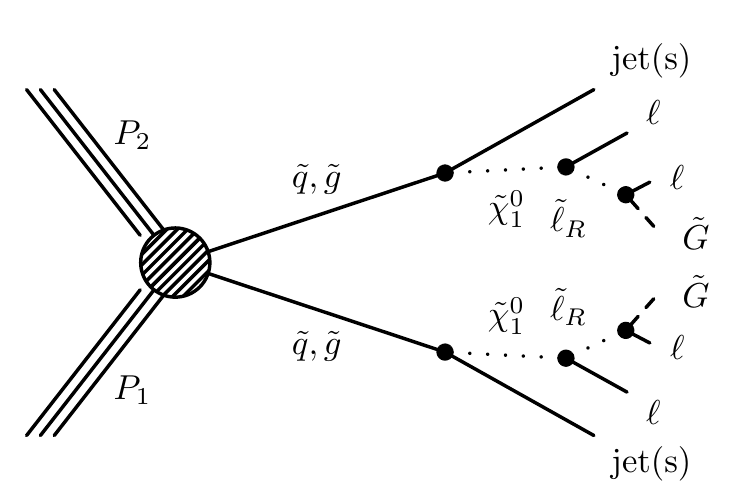}
\includegraphics[width=0.20\textwidth]{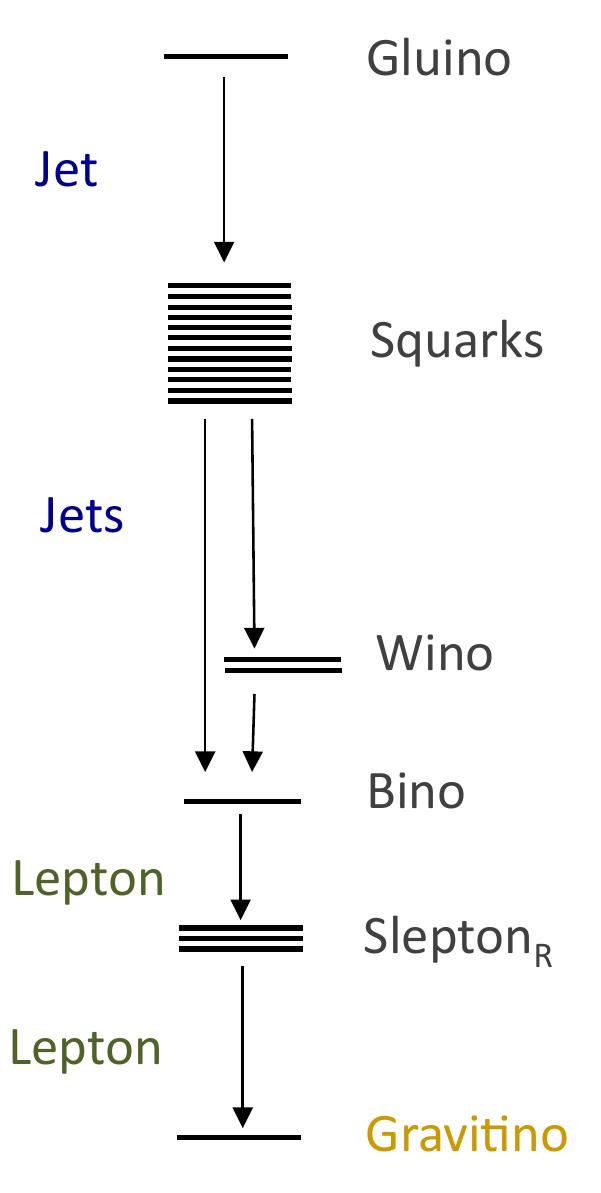}
\caption{Event diagram and a schematic superpartner mass spectrum for the GMSB slepton
co-NLSP scenario.}
\label{fig:fd_spec_slco}
\end{figure*}

We next consider the slepton co-NLSP scenario~\cite{Read:2002hq,Chatrchyan:2013mya}, in which mass-degenerate
right-handed sleptons $\slepton_{R}$ (selectron, smuon, stau) serve
together as the NLSP. This scenario arises in a broad class of GMSB models and can
lead to a multilepton final state~\cite{Dimopoulos:1996va,Culbertson:2000am,Ruderman:2010kj,Alves:2011wf}.
The process proceeds primarily through gluino~$\gluino$~and squark~$\squark$~pair production~\cite{Beenakker:1996ch}.
An event diagram and schematic mass spectrum are shown in Fig.~\ref{fig:fd_spec_slco}.
The $\none$ neutralino is taken to be a bino, the superpartner of the $\textrm{B}$ gauge boson.
The bino decays to a lepton and the NLSP, while the NLSP decays to the gravitino LSP and
an additional lepton. Depending on the mass spectrum, the events can have large $\HT$.
Channels with no tagged $\cPqb$ jets and off-$\cPZ$ OSSF pairs exhibit the largest sensitivity for this scenario.

Beyond production through squarks and gluinos, production through
chargino-neutralino or right-handed slepton pairs is possible.
The decay of each parent eventually leads to a bino $\none$,
which decays as shown in Fig.~\ref{fig:fd_spec_slco}, leading to a final state with
multileptons and $\MET$ as for the strong-production process. The relative importance
of the strong- and weak-production mechanisms depends on the values of the
superpartner masses.

Signal events for the slepton co-NLSP scenario are generated using the
\PYTHIA generator. The superpartner mass spectrum is parametrized in
terms of the masses of the $\cone$ chargino and the gluino.
The remaining superpartner masses are chosen to
be $m_{\slepton_R} = 0.3 m_{\cone}$,
$m_{\none}=0.5 m_{\cone}$, $m_{\slepton_L}=0.8 m_{\cone}$, and
$m_{\squark}=0.8 m_{\gluino}$,
with no mixing of the left- and right-handed slepton and squark components,
and with the higgsino masses so large that their contributions are negligible.
The cross sections are calculated  at NLO using K-factors
from \PROSPINO \cite{Beenakker:1996ed} and are assigned a 30\%
theoretical uncertainty, taking into account cross section, scale, and PDF uncertainties.

\begin{figure}[!ht]
\centering
\includegraphics[width=0.49\textwidth]{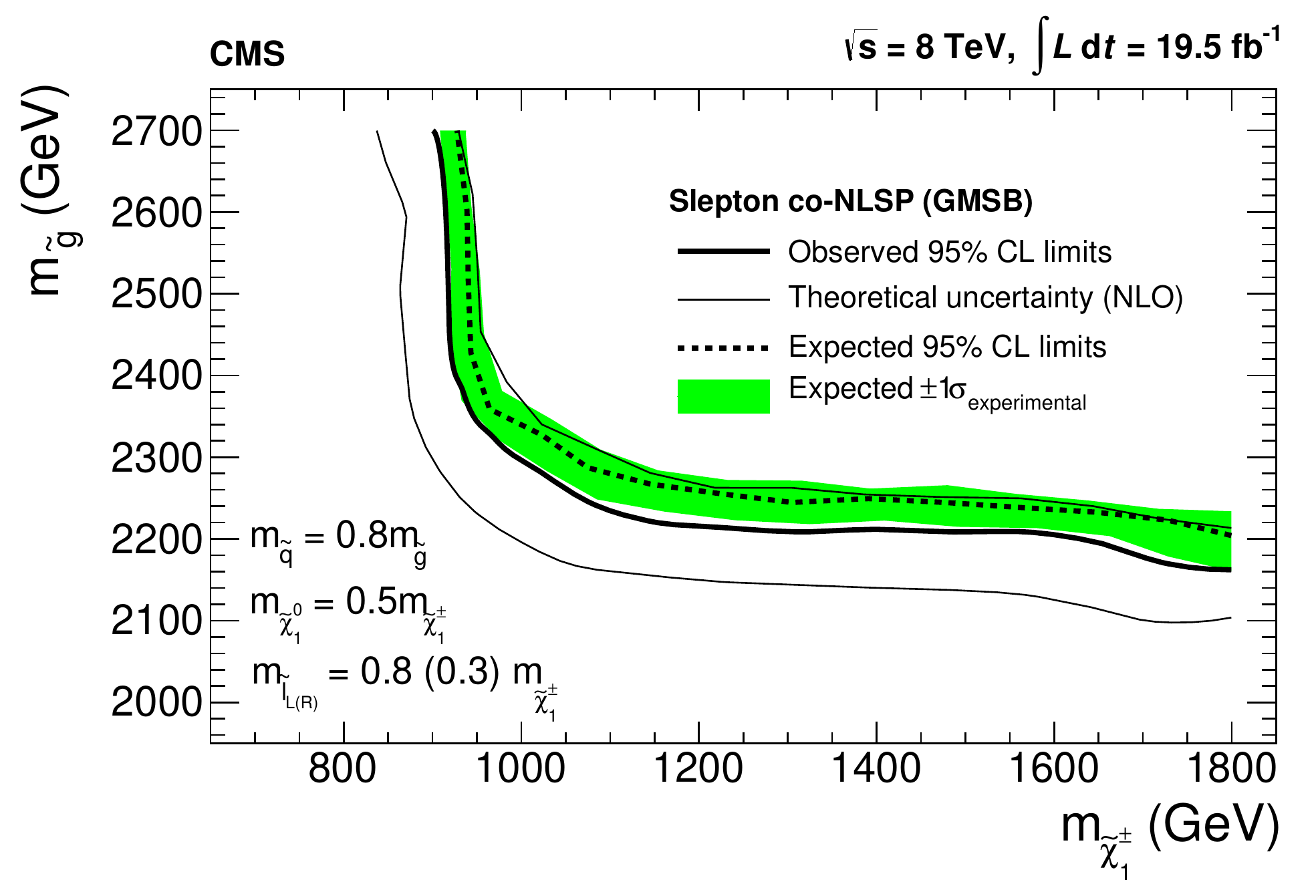}
\includegraphics[width=0.49\textwidth]{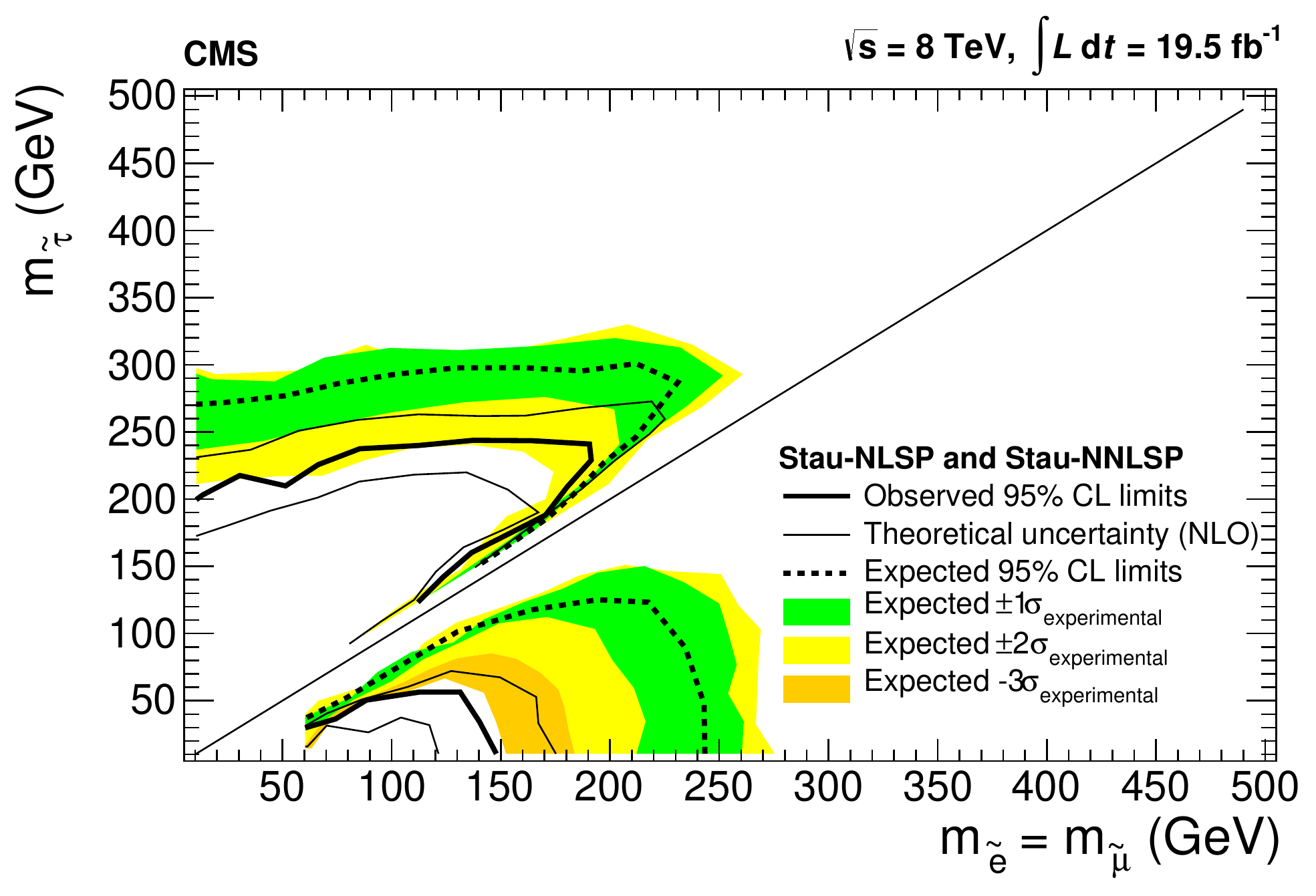}
\caption{
The 95\% confidence level upper limits for the slepton co-NLSP model in the
gluino versus chargino mass plane (\cmsLeft) and for the
stau-(N)NLSP scenarios in the stau versus degenerate-smuon and -selectron mass plane (\cmsRight).
The region to the left and below the contours is excluded.
}
\label{fig:GMSBExclusion}
\end{figure}

The 95\% \CL exclusions limits for the slepton co-NLSP scenario are shown
in Fig.~\ref{fig:GMSBExclusion} (\cmsLeft) as a function of the gluino and chargino masses.
In the region dominated by strong superpartner production, the exclusion curve asymptotically
approaches a horizontal plateau, while it tends towards a vertical line in the region dominated
by weak superpartner production.

\subsection{The stau-(N)NLSP scenario}

\begin{figure*}[!ht]
\centering
\includegraphics[width=0.50\textwidth]{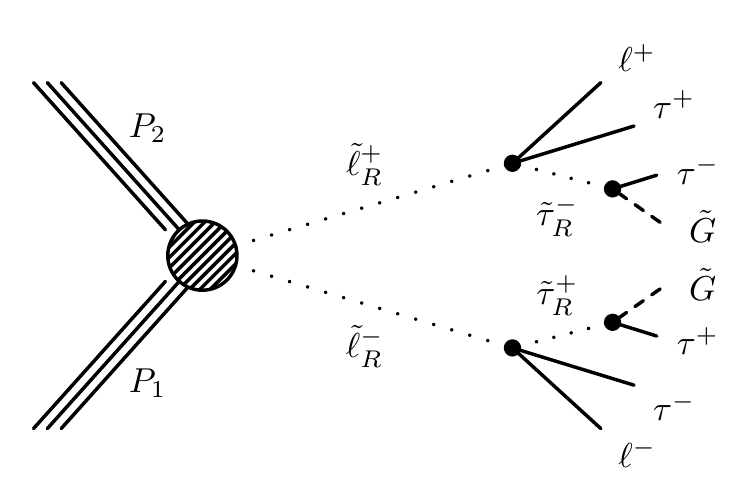}
\includegraphics[width=0.45\textwidth]{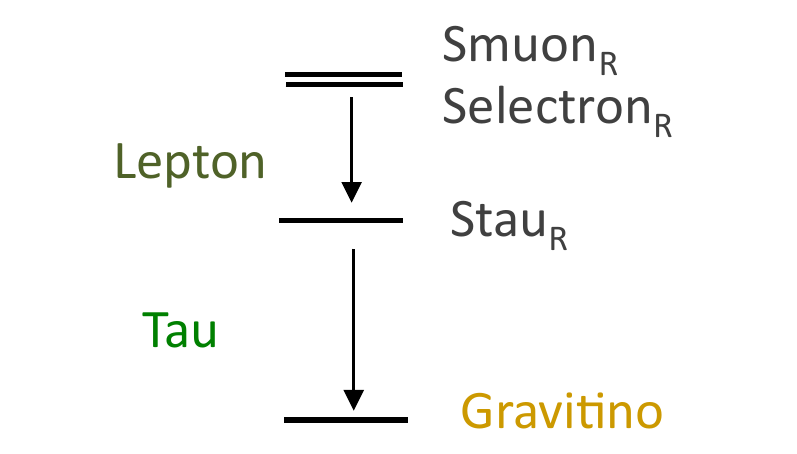}
\caption{Event diagram and a schematic superpartner mass spectrum for the GMSB stau-(N)NLSP scenario.}
\label{fig:fd_spec_stau}
\end{figure*}

In the stau-NLSP scenario, the right-handed stau lepton is the NLSP.
This scenario arises for moderate to large values of the MSSM
parameter $\tan \beta$~\cite{Nilles:1983ge,Haber:1984rc}.
Mass-degenerate right-handed selectrons and smuons decay to
the stau through the three-body processes $\selectron_{R} \to \stau_{R}  \Pgt  \Pe$
and $\smuon_{R} \to \stau_{R}  \Pgt  \Pgm$. The stau decays
as $\stau_{R} \to \goldstino  \Pgt$. Pair production of selectrons or
smuons leads to a multilepton final state dominated by $\tau$ leptons.
A diagram and schematic mass spectrum are shown in Fig.~\ref{fig:fd_spec_stau}.

Besides the stau-NLSP scenario, we also consider the stau-NNLSP scenario in which
mass-degenerate right-handed selectrons and smuons are co-NLSPs, while the
right-handed stau is the next-to-next-to-lightest SUSY particle (NNLSP). The process
proceeds via electroweak pair production of staus. The staus decay to the NLSP
and a $\tau$ lepton. The NLSPs decay to a $\tau$ lepton and gravitino.

The search channels most sensitive to the stau-(N)NLSP scenarios
contain $\tauh$ leptons, no tagged $\cPqb$ jets, off-$\cPZ$ OSSF pairs, and large $\MET$.
Signal events for the stau-(N)NLSP model are generated using \PYTHIA~\cite{Sjostrand:2007gs}.
The cross sections are normalized to NLO calculations using \PROSPINO \cite{Beenakker:1996ed}
and are assigned a 30\% theoretical uncertainty.

The 95\% \CL exclusion limits for the stau-(N)NLSP scenario are shown in
Fig.~\ref{fig:GMSBExclusion} (bottom). When the mass difference between
the stau and the other sleptons is small, the leptons are soft. This results
in low signal efficiency, which causes the exclusion contour to become nearly
parallel to the diagonal for points near the diagonal. The difference between the
expected and observed limits in the region below the diagonal is driven by the
excesses observed between the data and SM estimates
in the four-or-more lepton, OSSF1, off-$\cPZ$, $\tauh$ channels without $\cPqb$ jets, noted
in Sec.~\ref{results}.

\subsection{Third-generation SMS scenario T1tttt}

\begin{figure*}[!ht]
\centering
\includegraphics[width=0.55\textwidth]{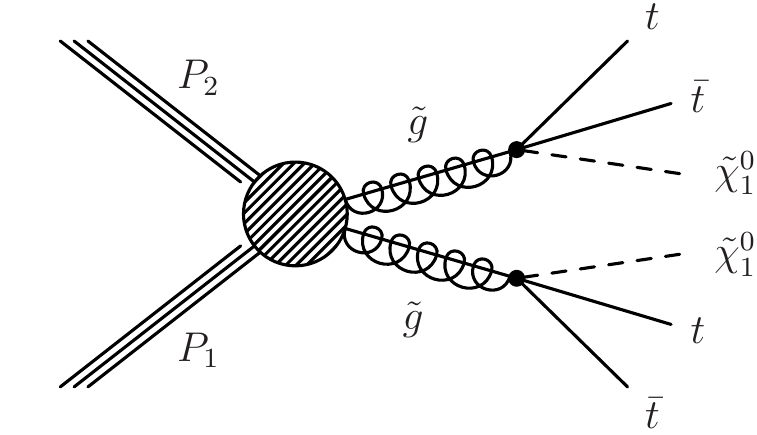}
\includegraphics[width=0.40\textwidth]{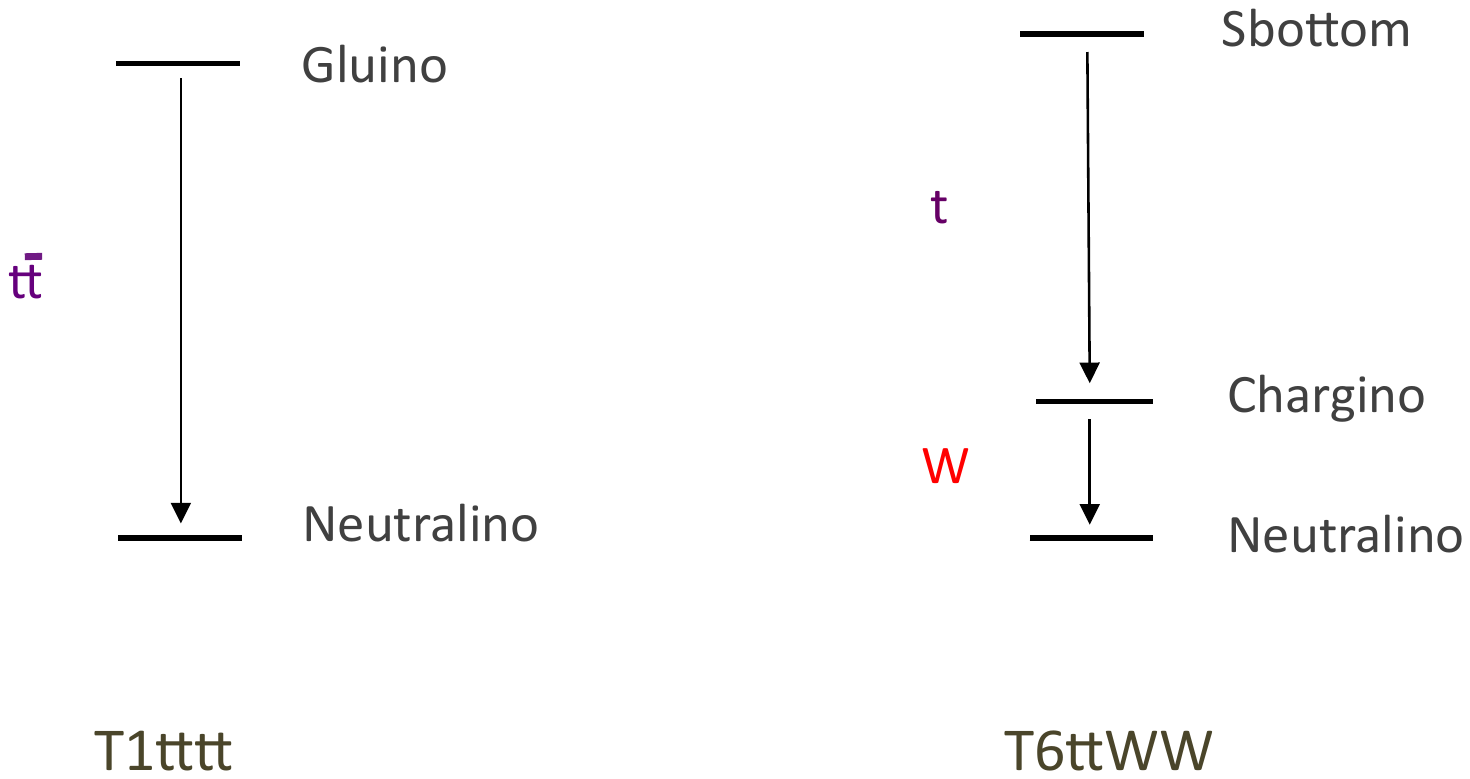}
\caption{Event diagram and a schematic superpartner mass spectrum for the SMS T1tttt scenario.}
\label{fig:fd_spec_T1}
\end{figure*}

In the T1tttt simplified model spectra (SMS) scenario~\cite{Alves:2011wf,Essig:2011qg,Chatrchyan:2013sza},
pair-produced gluinos each decay to a top quark and a virtual top squark. The virtual
top squark decays to a top quark and the LSP, where the LSP is the lightest neutralino.
Thus each gluino undergoes an effective three-body decay to two top quarks and the LSP,
yielding four top quarks in the final state. Each top quark can potentially yield a $\cPqb$ jet and a
leptonically decaying $\PW$ boson, leading to a multilepton final state with $\cPqb$ jets and $\MET$.
Because of the large number of jets, the $\HT$ value can be quite large. An event diagram
and schematic mass spectrum are shown in Fig.~\ref{fig:fd_spec_T1}.

The presence of four top quarks in the final state results in four $\cPqb$ quarks and four $\PW$ bosons.
The $\PW$-boson decays can produce up to four leptons with large $\MET$. The SM background is
significantly reduced by requiring the presence of a $\cPqb$ jet. This requirement represents an improvement
with respect to our analysis of the 7\TeV data~\cite{Chatrchyan:2012mea}.

Signal events for the T1tttt scenario are generated using \MADGRAPH.
The cross sections are calculated at the NLO plus next-to-leading-logarithm (NLL)
level~\cite{Beenakker:1996ch,Kulesza:2008jb,Kulesza:2009kq,Beenakker:2009ha,Beenakker:2011fu}
with uncertainties that vary between 23\% and 27\%~\cite{Kramer:2012bx}.

\begin{figure}[!ht]
\centering
\includegraphics[width=0.49\textwidth]{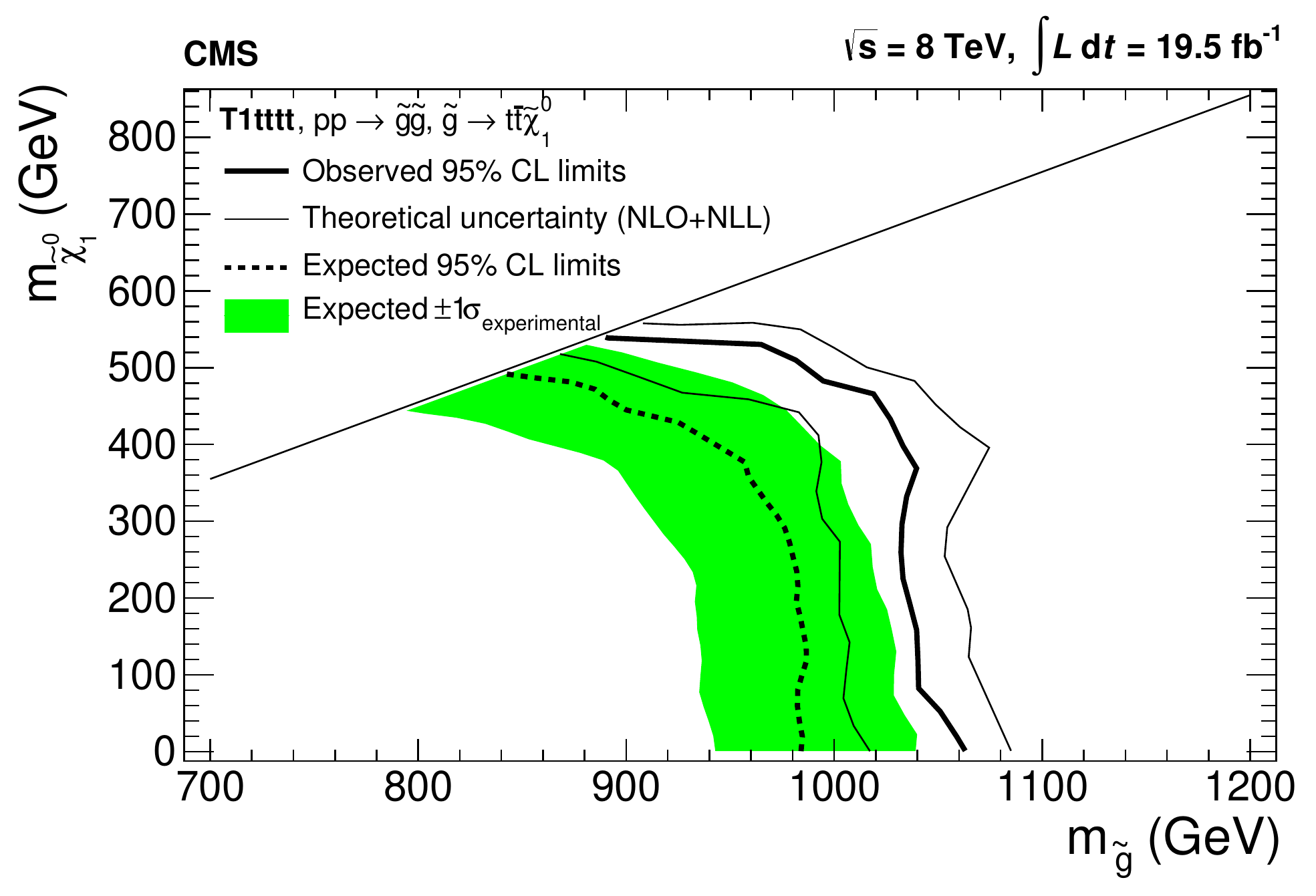}
\includegraphics[width=0.49\textwidth]{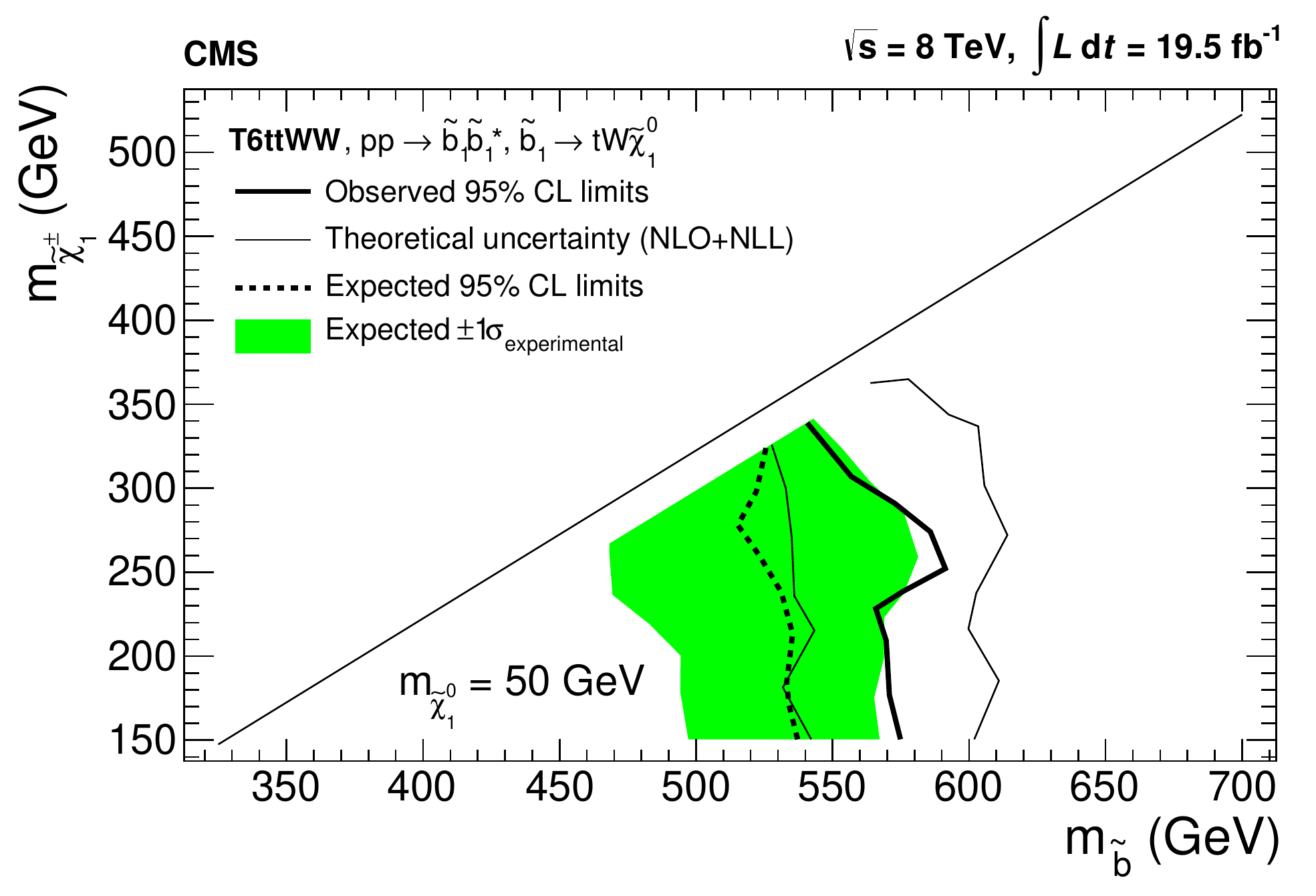}
\caption{95\% confidence level upper limits for the T1tttt scenario in the LSP versus gluino mass plane (\cmsLeft)
and for the T6ttWW scenario in the chargino versus bottom-squark mass plane (\cmsRight) are shown. Masses to the left
and below the contours are excluded.
}
\label{fig:SMSExclusion}
\end{figure}

The 95\% \CL exclusion limits in the gluino versus LSP mass plane are shown in Fig.~\ref{fig:SMSExclusion} (\cmsLeft).
We exclude gluinos with mass values below 1\TeV over much of this plane.

\subsection{Third-generation SMS scenario T6ttWW}

\begin{figure*}[!ht]
\centering
\includegraphics[width=0.55\textwidth]{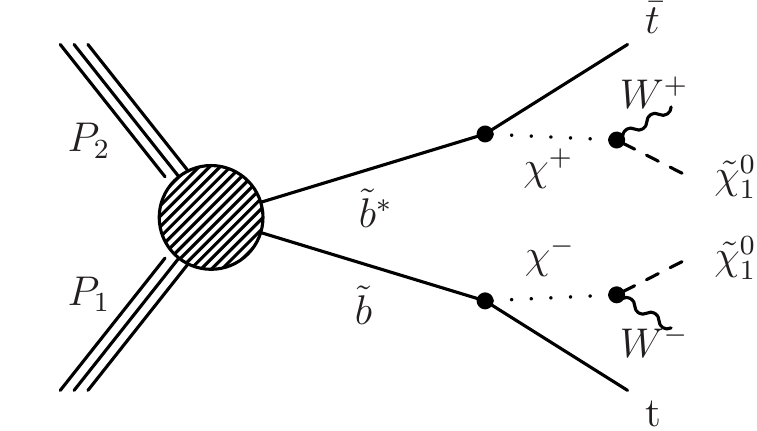}
\includegraphics[width=0.40\textwidth]{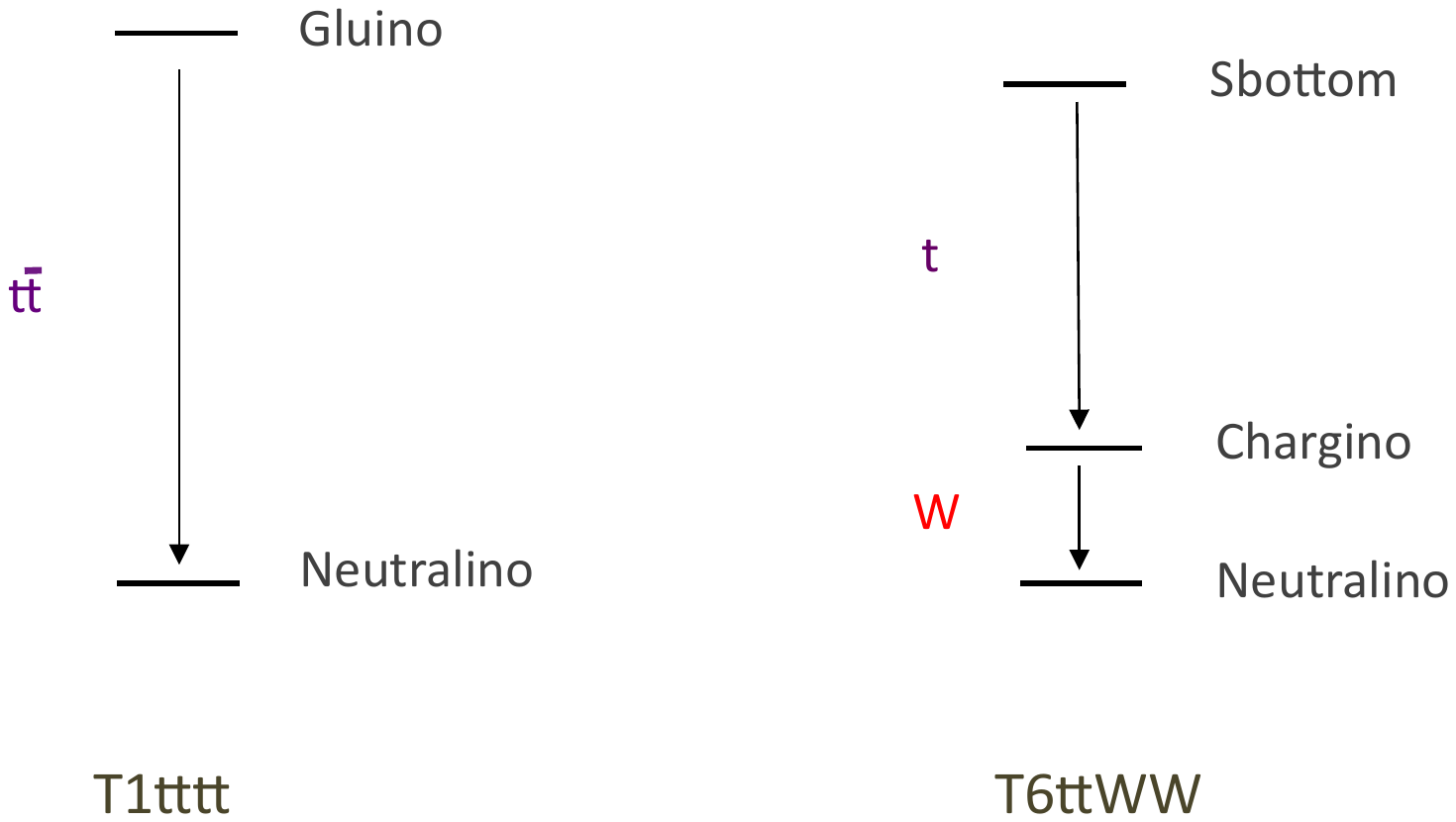}
\caption{Event diagram and a schematic superpartner mass spectrum for the SMS T6ttWW scenario.}
\label{fig:fd_spec_T6}
\end{figure*}

In the T6ttWW SMS scenario, we search for SUSY signals with
direct bottom-squark pair production~\cite{Lee:2012sy,Chatrchyan:2013sza}. An event diagram
and schematic mass spectrum are shown in Fig.~\ref{fig:fd_spec_T6}.
The bottom squark decays as $\sBot \to \cPqt \conem$,
while the chargino decays as $\conem \to \PWm \none$.
This scenario populates channels with tagged $\cPqb$ jets.

For simplicity, we consider on-shell charginos. The $\PW$ boson from the
chargino decay can be either on- or off-shell. Signal events are generated
using \MADGRAPH with normalization of the cross section performed to
NLO+NLL~\cite{Beenakker:1996ch,Kulesza:2008jb,Kulesza:2009kq,Beenakker:2009ha,Beenakker:2011fu}.
The uncertainty of the cross section calculation is 30\%~\cite{Kramer:2012bx}.

Figure~\ref{fig:SMSExclusion} (bottom) shows the exclusion limits for the T6ttWW scenario
in the chargino versus bottom-squark mass plane. The mass of the $\none$ is
assumed to be 50\GeV. We exclude bottom squarks with mass values less
than 550\GeV. This result complements our study of this same scenario performed
using same-sign dilepton events and obtains similar conclusions~\cite{Chatrchyan:2013fea}.

\section{Rare decay \texorpdfstring{$\cPqt \to \cPqc \PH$}{t -> cH}}
\label{tch}

Beyond the SUSY scenarios examined in Sec.~\ref{models}, we interpret our
results in the context of the flavor-changing decay of a top quark to
a Higgs boson and a charm quark. Although not forbidden in the SM, the SM branching fraction
is predicted to be extremely small ($10^{-13}$--$10^{-15}$~\cite{Craig:2012vj, Chen:2013qta}),
due to suppression both by the Glashow--Iliopoulos--Maiani mechanism~\cite{PhysRevD.2.1285}
and by the Cabibbo--Kobayashi--Maskawa quark-mixing matrix~\cite{Kobayashi:1973fv} factor.
Observation of the $\cPqt \to \cPqc \PH$ transition can therefore provide evidence
for BSM physics, \ie, for non-SM particles produced virtually in loops.
In this sense the $\cPqt \to \cPqc \PH$ transition plays a complementary role
to SUSY searches compared to the direct superpartner production scenarios considered in Sec.~\ref{models}.

In addition, the $\cPqt \to \cPqc \PH$ decay directly probes the
flavor-violating couplings of the Higgs boson to the top quark.
Since up-type quark-flavor violation is less constrained than down-type
quark-flavor violation~\cite{Antonelli:2009ws}, exploration of this issue is of general
interest.

The production of a $\ttbar$ pair followed by the decay of one top quark to a
$\cPqc \PH$ state and the other to a $\cPqb \PW$ state can yield a multilepton signature,
especially if the Higgs boson decays through one of the following channels:

\begin{itemize}
\item $\PH \to \PW \PW^* \to \ell \nu \ell \nu$,
\item $\PH \to \Pgt \Pgt$, or
\item $\PH \to \cPZ \cPZ^* \to jj \ell \ell, \nu \nu \ell \ell, \ell \ell \ell \ell$,
\end{itemize}

where $j$ refers to a jet. If the $\cPqt \to \cPqb \PW$  decay also produces a lepton,
there can be up to five leptons in an event.

To simulate signal events, we generate a $\ttbar$ sample in which one top quark decays
to $\cPqc \PH$ and the other to $\cPqb \PW$. We assume $m_{\PH} = 126\GeV$~\cite{Chatrchyan:2013mxa}
and that the Higgs boson has SM branching fractions. We only consider the decay
modes listed above because the contributions of other Higgs boson decay modes to the
multilepton final state are found to be negligible. Signal events are generated
using \MADGRAPH, with normalization performed at the next-to-next-to-leading
order~\cite{Czakon:2013goa}.

The signal events predominantly populate channels with three leptons, a tagged $\cPqb$ jet,
no $\tauh$-lepton candidate, and an OSSF off-$\cPZ$ pair or no OSSF pair. The most
sensitive channels are listed in Table~\ref{tab:tchChannels}. The main source
of SM background arises from $\ttbar$ production. The observed numbers of
events are seen to be in agreement with the SM expectations to within the uncertainties.

\begin{table*}[h!t]
\centering
\topcaption{The ten most sensitive signal regions for the $\cPqt \to \cPqc \PH$ process,
along with the number of observed (Obs.), background (Exp.), and expected signal (Sig.) events,
assuming $\mathcal{B}(\cPqt \to \cPqc \PH) = 1\%$, ordered by sensitivity.
All signal regions shown have exactly three selected leptons. The results are binned in
$\MET$, $\HT$, number of tagged $\cPqb$ jets or $\tauh$ candidates, and, if an OSSF pair is present, its invariant mass
with respect to the $\cPZ$-boson mass window.
}
\label{tab:tchChannels}
\begin{scotch}{cccccccc}
OSSF pair & ${N}_{\tauh}$ & $\MET$ (\GeVns) & $\HT$ (\GeVns) & ${N}_{\cPqb}$ & Obs. & Exp. & Sig. \\
\hline
Below-$\cPZ$ & 0 & 50--100 & 0--200 & $\geq$1 & 48 & 48 $\pm$ 23 & 9.5 $\pm$ 2.3\\
\NA & 0 & 50--100 & 0--200 & $\geq$1 & 29 & 26 $\pm$ 13 & 5.9 $\pm$ 1.3\\
Below-$\cPZ$ & 0 & 0--50 & 0--200 & $\geq$1 & 34 & 42 $\pm$ 11 & 5.9 $\pm$ 1.2\\
\NA & 0 & 0--50 & 0--200 & $\geq$1 & 29 & 23 $\pm$ 10 & 4.3 $\pm$ 1.1\\
Below-$\cPZ$ & 0 & 50--100 & $> 200$ & $\geq$1 & 10 & 9.9 $\pm$ 3.7 & 3.0 $\pm$ 1.1\\
Below-$\cPZ$ & 0 & 0--50 & $> 200$ & $\geq$1 & 5 & 10 $\pm$ 2.5 & 2.8 $\pm$ 0.8\\
Below-$\cPZ$ & 0 & 50--100 & 0--200 & 0 & 142 & 125 $\pm$ 27 & 9.7 $\pm$ 2.1\\
\NA & 1 & 0--50 & 0--200 & $\geq$1 & 237 & 240 $\pm$ 113 & 13.1 $\pm$ 2.6\\
\NA & 0 & 50--100 & 0--200 & 0 & 35 & 38 $\pm$ 15 & 4.3 $\pm$ 1.1\\
Above-$\cPZ$ & 0 & 0--50 & 0--200 & $\geq$1 & 17 & 18 $\pm$ 6.7 & 2.8 $\pm$ 0.8\\
\end{scotch}
\end{table*}

Using the same limit-setting procedure as in Sec.~\ref{models}, we
obtain a 95\% \CL upper limit on the branching fraction of $\mathcal{B}(\cPqt \to \cPqc \PH) < 1.3\%$.
The measured branching fraction is $(1.2^{+0.5}_{-0.3})\%$. The uncertainties include both the statistical and systematic terms.
The observed limit corresponds to a bound on the left- and right-handed top-charm flavor-violating Higgs
Yukawa couplings, $\lambda_{\cPqt \cPqc}^{\PH}$ and $\lambda_{\cPqc \cPqt}^{\PH}$, respectively, of
$\sqrt{|\lambda_{\cPqt \cPqc}^{\PH}|^2 + |\lambda_{\cPqc \cPqt}^{\PH}|^2} < 0.21$.
This result represents a significant improvement compared with the inferred
result $\sqrt{|\lambda_{\cPqt \cPqc}^{\PH}|^2 + |\lambda_{\cPqc \cPqt}^{\PH}|^2} < 0.31$
from Ref.~\cite{Craig:2012vj}, which is based on our 7\TeV results~\cite{Chatrchyan:2012mea}. 
Reference~\cite{Aad:2014dya} presents recent results from the ATLAS Collaboration. 
Table~\ref{tab:tch} presents upper limits from individual Higgs boson decay modes.
All other decay modes are ignored when calculating these limits. It is seen that
the $\PH \to \PW \PW^* \to \ell \nu \ell \nu$ mode dominates the overall result.

\begin{table}[h!t]
\centering
\topcaption{Comparison of the observed (Obs.) and median expected (Exp.) 95\% \CL upper limits on
$\mathcal{B}(\cPqt \to \cPqc \PH)$ from individual Higgs boson
decay modes, along with their one standard deviation ($\sigma$) uncertainties.
The uncertainties include both statistical and systematic terms.}
\label{tab:tch}
\begin{scotch}{ll|ccc}
\multicolumn{2}{c|}{Higgs boson decay mode} & \multicolumn{3}{c}{Upper limits on $\mathcal{B}(\cPqt \to \cPqc \PH)$} \\
& & Obs. & Exp. & $1\sigma$ range \\
\hline
$\mathcal{B}(\PH \to \PW \PW^*)$ & $= 23.1\%$ & 1.6 \% & 1.6\% & (1.0--2.2)\% \\
$\mathcal{B}(\PH \to \Pgt \Pgt)$ & $= 6.2\%$ & 7.01\% & 5.0 \% & (3.5--7.7)\% \\
$\mathcal{B}(\PH \to \cPZ \cPZ^*)$ & $= 2.9\%$ & 5.3\% & 4.11\% & (2.9--6.5)\% \\
\hline
\multicolumn{2}{c|}{Combined} & 1.3\% & 1.2\% & (0.9--1.7)\% \\
\end{scotch}
\end{table}

\section{Summary}
\label{summary}

We have performed a search for physics beyond the standard model based
on events with three or more leptons, where one of these leptons can be a
hadronically decaying $\tau$ lepton. We search in channels with $\Pep \Pem$
or $\Pgmp \Pgmm$ pairs that are either consistent or inconsistent with $\cPZ$ boson
decay, in channels without such a pair, in channels with or without a
hadronically decaying $\tau$-lepton candidate, in channels with and without
a tagged bottom-quark jet, in events with and without a large level of jet activity
(measured with the scalar sum of jet $\pt$ values), and in different bins of
missing transverse energy. We find no significant excesses compared to the
expectations from standard model processes. The search is performed
separately for events with exactly three leptons and with four or more leptons.

We examine a broad class of supersymmetric scenarios that, taken together, populate
a broad spectrum of multilepton final states. Compared to previous results, we probe
new regions of the parameter space for the natural higgsino next-to-lightest
supersymmetric particle (NLSP), slepton co-NLSP, and stau-(N)NLSP scenarios,
where (N)NLSP denotes the (next-to-)next-to-lightest-supersymmetric particle. In addition,
we investigate scenarios with gluino pair production followed by gluino decay to a top-antitop
pair and the lightest supersymmetric particle, and direct bottom-squark pair production.
Cross section upper limits at 95\% confidence level are presented for all these scenarios.

We further explore rare transitions of the top quark to a charm quark and a Higgs boson, $\cPqt \to \cPqc \PH$.
We set a 95\% confidence level upper limit of 1.3\% on the branching fraction of this decay,
which corresponds to an upper bound $\sqrt{\abs{\lambda_{\cPqt \cPqc}^{\PH}}^2 + \abs{\lambda_{\cPqc \cPqt}^{\PH}}^2} < 0.21$
on the flavor-violating couplings of a Higgs boson to a $\cPqt \cPqc$ quark combination.

\section*{Acknowledgements}
\label{ack}

\hyphenation{Bundes-ministerium Forschungs-gemeinschaft Forschungs-zentren} We congratulate our colleagues in the CERN accelerator departments for the excellent performance of the LHC and thank the technical and administrative staffs at CERN and at other CMS institutes for their contributions to the success of the CMS effort. In addition, we gratefully acknowledge the computing centers and personnel of the Worldwide LHC Computing Grid for delivering so effectively the computing infrastructure essential to our analyses. Finally, we acknowledge the enduring support for the construction and operation of the LHC and the CMS detector provided by the following funding agencies: the Austrian Federal Ministry of Science, Research and Economy and the Austrian Science Fund; the Belgian Fonds de la Recherche Scientifique, and Fonds voor Wetenschappelijk Onderzoek; the Brazilian Funding Agencies (CNPq, CAPES, FAPERJ, and FAPESP); the Bulgarian Ministry of Education and Science; CERN; the Chinese Academy of Sciences, Ministry of Science and Technology, and National Natural Science Foundation of China; the Colombian Funding Agency (COLCIENCIAS); the Croatian Ministry of Science, Education and Sport, and the Croatian Science Foundation; the Research Promotion Foundation, Cyprus; the Ministry of Education and Research, Recurrent Financing Contract No. SF0690030s09 and European Regional Development Fund, Estonia; the Academy of Finland, Finnish Ministry of Education and Culture, and Helsinki Institute of Physics; the Institut National de Physique Nucl\'eaire et de Physique des Particules~/~CNRS, and Commissariat \`a l'\'Energie Atomique et aux \'Energies Alternatives~/~CEA, France; the Bundesministerium f\"ur Bildung und Forschung, Deutsche Forschungsgemeinschaft, and Helmholtz-Gemeinschaft Deutscher Forschungszentren, Germany; the General Secretariat for Research and Technology, Greece; the National Scientific Research Foundation, and National Innovation Office, Hungary; the Department of Atomic Energy and the Department of Science and Technology, India; the Institute for Studies in Theoretical Physics and Mathematics, Iran; the Science Foundation, Ireland; the Istituto Nazionale di Fisica Nucleare, Italy; the Korean Ministry of Education, Science and Technology and the World Class University program of NRF, Republic of Korea; the Lithuanian Academy of Sciences; the Ministry of Education, and University of Malaya (Malaysia); the Mexican Funding Agencies (CINVESTAV, CONACYT, SEP, and UASLP-FAI); the Ministry of Business, Innovation and Employment, New Zealand; the Pakistan Atomic Energy Commission; the Ministry of Science and Higher Education and the National Science Centre, Poland; the Funda\c{c}\~ao para a Ci\^encia e a Tecnologia, Portugal; JINR, Dubna; the Ministry of Education and Science of the Russian Federation, the Federal Agency of Atomic Energy of the Russian Federation, Russian Academy of Sciences, and the Russian Foundation for Basic Research; the Ministry of Education, Science and Technological Development of Serbia; the Secretar\'{\i}a de Estado de Investigaci\'on, Desarrollo e Innovaci\'on and Programa Consolider-Ingenio 2010, Spain; the Swiss Funding Agencies (ETH Board, ETH Zurich, PSI, SNF, UniZH, Canton Zurich, and SER); the Ministry of Science and Technology, Taipei; the Thailand Center of Excellence in Physics, the Institute for the Promotion of Teaching Science and Technology of Thailand, Special Task Force for Activating Research and the National Science and Technology Development Agency of Thailand; the Scientific and Technical Research Council of Turkey, and Turkish Atomic Energy Authority; the National Academy of Sciences of Ukraine, and State Fund for Fundamental Researches, Ukraine; the Science and Technology Facilities Council, UK; the U.S. Department of Energy, and the U.S. National Science Foundation.

Individuals have received support from the Marie-Curie program and the European Research Council and EPLANET (European Union); the Leventis Foundation; the A. P. Sloan Foundation; the Alexander von Humboldt Foundation; the Belgian Federal Science Policy Office; the Fonds pour la Formation \`a la Recherche dans l'Industrie et dans l'Agriculture (FRIA-Belgium); the Agentschap voor Innovatie door Wetenschap en Technologie (IWT-Belgium); the Ministry of Education, Youth and Sports (MEYS) of Czech Republic; the Council of Science and Industrial Research, India; the Compagnia di San Paolo (Torino); the HOMING PLUS program of Foundation for Polish Science, cofinanced by EU, Regional Development Fund; and the Thalis and Aristeia programs cofinanced by EU-ESF and the Greek NSRF.

\bibliography{auto_generated}   % will be created by the tdr script.

\cleardoublepage \appendix\section{The CMS Collaboration \label{app:collab}}\begin{sloppypar}\hyphenpenalty=5000\widowpenalty=500\clubpenalty=5000\input{SUS-13-002-authorlist.tex}\end{sloppypar}
\end{document}

%% file: SUS-13-002-authorlist.tex
\textbf{Yerevan Physics Institute,  Yerevan,  Armenia}\\*[0pt]
S.~Chatrchyan, V.~Khachatryan, A.M.~Sirunyan, A.~Tumasyan
\vskip\cmsinstskip
\textbf{Institut f\"{u}r Hochenergiephysik der OeAW,  Wien,  Austria}\\*[0pt]
W.~Adam, T.~Bergauer, M.~Dragicevic, J.~Er\"{o}, C.~Fabjan\cmsAuthorMark{1}, M.~Friedl, R.~Fr\"{u}hwirth\cmsAuthorMark{1}, V.M.~Ghete, C.~Hartl, N.~H\"{o}rmann, J.~Hrubec, M.~Jeitler\cmsAuthorMark{1}, W.~Kiesenhofer, V.~Kn\"{u}nz, M.~Krammer\cmsAuthorMark{1}, I.~Kr\"{a}tschmer, D.~Liko, I.~Mikulec, D.~Rabady\cmsAuthorMark{2}, B.~Rahbaran, H.~Rohringer, R.~Sch\"{o}fbeck, J.~Strauss, A.~Taurok, W.~Treberer-Treberspurg, W.~Waltenberger, C.-E.~Wulz\cmsAuthorMark{1}
% ...
\vskip\cmsinstskip
% ...
1:~~Also at Vienna University of Technology, Vienna, Austria\\
2:~~Also at CERN, European Organization for Nuclear Research, Geneva, Switzerland\\
% ...


%% file: SUS-13-002_temp.bbl
\providecommand{\href}[2]{#2}\begingroup\raggedright\begin{thebibliography}{10}%
\makeatletter
\providecommand{\hrefCMSnoop }[0]{\@secondoftwo}%
\makeatother
\providecommand{\doi}{\texttt{doi:}\begingroup \urlstyle{tt}\Url}

\bibitem{Aad:2012tfa}
\hrefCMSnoop {} {{ ATLAS} Collaboration, ``{Observation of a new particle in
  the search for the Standard Model Higgs boson with the ATLAS detector at the
  LHC}'',} \textit{ Phys. Lett. B} \textbf{ 716} (2012) 1,
  \href{http://dx.doi.org/10.1016/j.physletb.2012.08.020}{\doi{10.1016/j.physletb.2012.08.020}},
\href{http://www.arXiv.org/abs/1207.7214}{\texttt{ arXiv:1207.7214}}.
%%CITATION = ARXIV:1207.7214;%%.

\bibitem{Chatrchyan:2012ufa}
\hrefCMSnoop {} {{ CMS} Collaboration, ``{Observation of a new boson at a mass
  of 125 GeV with the CMS experiment at the LHC}'',} \textit{ Phys. Lett. B}
  \textbf{ 716} (2012) 30,
  \href{http://dx.doi.org/10.1016/j.physletb.2012.08.021}{\doi{10.1016/j.physletb.2012.08.021}},
\href{http://www.arXiv.org/abs/1207.7235}{\texttt{ arXiv:1207.7235}}.
%%CITATION = ARXIV:1207.7235;%%.

\bibitem{Chatrchyan:2013lba}
\hrefCMSnoop {} {{ CMS} Collaboration, ``Observation of a new boson with mass
  near 125 GeV in $\textrm{pp}$ collisions at $\sqrt{s}$ = 7 and 8 TeV'',}
  \textit{ JHEP} \textbf{ 06} (2013) 081,
  \href{http://dx.doi.org/10.1007/JHEP06(2013)081}{\doi{10.1007/JHEP06(2013)081}},
\href{http://www.arXiv.org/abs/1303.4571}{\texttt{ arXiv:1303.4571}}.
%%CITATION = ARXIV:1303.4571;%%.

\bibitem{Nilles:1983ge}
\hrefCMSnoop {} {H.~P. Nilles, ``Supersymmetry, Supergravity and Particle
  Physics'',} \textit{ Phys. Rept.} \textbf{ 110} (1984) 1,
\href{http://dx.doi.org/10.1016/0370-1573(84)90008-5}{\doi{10.1016/0370-1573(84)90008-5}}.
%%CITATION = PRPLC,110,1;%%.

\bibitem{Haber:1984rc}
\hrefCMSnoop {} {H.~E. Haber and G.~L. Kane, ``The Search for Supersymmetry:
  Probing Physics Beyond the Standard Model'',} \textit{ Phys. Rept.} \textbf{
  117} (1985) 75,
\href{http://dx.doi.org/10.1016/0370-1573(85)90051-1}{\doi{10.1016/0370-1573(85)90051-1}}.
%%CITATION = PRPLC,117,75;%%.

\bibitem{deBoer:1994dg}
\hrefCMSnoop {} {W.~de~Boer, ``Grand unified theories and supersymmetry in
  particle physics and cosmology'',} \textit{ Prog. Part. Nucl. Phys.} \textbf{
  33} (1994) 201,
\href{http://dx.doi.org/10.1016/0146-6410(94)90045-0}{\doi{10.1016/0146-6410(94)90045-0}}.
%%CITATION = HEP-PH/9402266;%%.

\bibitem{Farrar:1978xj}
\hrefCMSnoop {} {G.~R. Farrar and P.~Fayet, ``Phenomenology of the production,
  decay, and detecion of new hadronic states associated with supersymmetry'',}
  \textit{ Phys. Lett. B} \textbf{ 76} (1978) 575,
\href{http://dx.doi.org/0.1016/0370-2693(78)90858-4}{\doi{0.1016/0370-2693(78)90858-4}}.
%%CITATION = PHLTA,B76,575;%%.

\bibitem{Brooijmans:2010tn}
\hrefCMSnoop {} {G.~Brooijmans {et~al.}, ``New Physics at the {LHC}: A {Les
  Houches} Report'',} (2010).
\href{http://www.arXiv.org/abs/1005.1229}{\texttt{ arXiv:1005.1229}}.
%%CITATION = ARXIV:1005.1229;%%.

\bibitem{Craig:2012vj}
N.~Craig\hrefCMSnoop {} { {et~al.}, ``{Searching for $\cPqt \rightarrow \cPqc
  \PH$ with Multi-Leptons}'',} \textit{ Phys. Rev. D} \textbf{ 86} (2012)
  075002,
  \href{http://dx.doi.org/10.1103/PhysRevD.86.075002}{\doi{10.1103/PhysRevD.86.075002}},
\href{http://www.arXiv.org/abs/1207.6794v2}{\texttt{ arXiv:1207.6794v2}}.
%%CITATION = ARXIV:1207.6794;%%.

\bibitem{Chen:2013qta}
\hrefCMSnoop {} {K.-F. Chen, W.-S. Hou, C.~Kao, and M.~Kohda, ``{When the Higgs
  meets the Top: Search for $\cPqt \rightarrow \cPqc \PH$ at the LHC}'',}
  \textit{ Phys. Lett. B} \textbf{ 725} (2013) 378,
  \href{http://dx.doi.org/10.1016/j.physletb.2013.07.060}{\doi{10.1016/j.physletb.2013.07.060}},
\href{http://www.arXiv.org/abs/1304.8037v2}{\texttt{ arXiv:1304.8037v2}}.
%%CITATION = ARXIV:1304.8037;%%.

\bibitem{Chatrchyan:2012mea}
\hrefCMSnoop {} {{ CMS} Collaboration, ``{Search for anomalous production of
  multilepton events in $\textrm{pp}$ collisions at $\sqrt{s}=7$ TeV}'',}
  \textit{ JHEP} \textbf{ 06} (2012) 169,
  \href{http://dx.doi.org/10.1007/JHEP06(2012)169}{\doi{10.1007/JHEP06(2012)169}},
\href{http://www.arXiv.org/abs/1204.5341}{\texttt{ arXiv:1204.5341}}.
%%CITATION = ARXIV:1204.5341;%%.

\bibitem{Chatrchyan:2013xsw}
\hrefCMSnoop {} {{ CMS} Collaboration, ``{Search for top squarks in
  R-parity-violating supersymmetry using three or more leptons and b-tagged
  jets}'',} \textit{ Phys. Rev. Lett.} \textbf{ 111} (2013) 221801,
  \href{http://dx.doi.org/10.1103/PhysRevLett.111.221801}{\doi{10.1103/PhysRevLett.111.221801}},
\href{http://www.arXiv.org/abs/1306.6643v2}{\texttt{ arXiv:1306.6643v2}}.
%%CITATION = ARXIV:1306.6643;%%.

\bibitem{ATLAS:2012kr}
\hrefCMSnoop {} {{ ATLAS} Collaboration, ``{Search for R-parity-violating
  supersymmetry in events with four or more leptons in $\sqrt{s} = 7$ TeV
  $\textrm{pp}$ collisions with the ATLAS detector}'',} \textit{ JHEP} \textbf{
  12} (2012) 124,
  \href{http://dx.doi.org/10.1007/JHEP12(2012)124}{\doi{10.1007/JHEP12(2012)124}},
\href{http://www.arXiv.org/abs/1210.4457}{\texttt{ arXiv:1210.4457}}.
%%CITATION = ARXIV:1210.4457;%%.

\bibitem{Aad:2014dya}
\hrefCMSnoop {} {{ATLAS Collaboration}, ``{Search for top quark decays $\cPqt
  \rightarrow \textrm{q} \PH$ with $\PH \rightarrow \gamma \gamma$ using the
  ATLAS detector}'',} (2014).
\href{http://www.arXiv.org/abs/1403.6293v1}{\texttt{ arXiv:1403.6293v1}}.
%%CITATION = ARXIV:1403.6293;%%.

\bibitem{Aad:2014pda}
\hrefCMSnoop {} {{ATLAS Collaboration}, ``{Search for supersymmetry at
  $\sqrt{s}$ = 8 TeV in final states with jets and two same-sign leptons or
  three leptons with the ATLAS detector}'',} (2014).
\href{http://www.arXiv.org/abs/1404.2500v1}{\texttt{ arXiv:1404.2500v1}}.
%%CITATION = ARXIV:1404.2500;%%.

\bibitem{Aad:2014mha}
\hrefCMSnoop {} {{ATLAS Collaboration}, ``{Search for direct top squark pair
  production in events with a $\cPZ$ boson, b-jets and missing transverse
  momentum in $\sqrt{s}$ = 8 TeV $\Pp \Pp$ collisions with the ATLAS
  detector}'',} (2014).
\href{http://www.arXiv.org/abs/1403.5222v1}{\texttt{ arXiv:1403.5222v1}}.
%%CITATION = ARXIV:1403.5222;%%.

\bibitem{Aad:2014nua}
\hrefCMSnoop {} {{ ATLAS} Collaboration, ``{Search for direct production of
  charginos and neutralinos in events with three leptons and missing transverse
  momentum in $\sqrt{s}$ = 8TeV $\Pp \Pp$ collisions with the ATLAS
  detector}'',} \textit{ JHEP} \textbf{ 04} (2014) 169,
  \href{http://dx.doi.org/10.1007/JHEP04(2014)169}{\doi{10.1007/JHEP04(2014)169}},
\href{http://www.arXiv.org/abs/1402.7029v3}{\texttt{ arXiv:1402.7029v3}}.
%%CITATION = ARXIV:1402.7029;%%.

\bibitem{Khachatryan:2014qwa}
\hrefCMSnoop {} {{CMS Collaboration}, ``{Searches for electroweak production of
  charginos, neutralinos, and sleptons decaying to leptons and $\PW$, $\cPZ$,
  and Higgs bosons in $\Pp \Pp$ collisions at 8 TeV}'',} (2014).
\href{http://www.arXiv.org/abs/1405.7570v1}{\texttt{ arXiv:1405.7570v1}}.
%%CITATION = ARXIV:1405.7570;%%.

\bibitem{Khachatryan:2014doa}
\hrefCMSnoop {} {{CMS Collaboration}, ``{Search for top-squark pairs decaying
  into Higgs or $\cPZ$ bosons in $\Pp \Pp$ collisions at $\sqrt{s}$ = 8
  TeV}'',} (2014).
\href{http://www.arXiv.org/abs/1405.3886v1}{\texttt{ arXiv:1405.3886v1}}.
%%CITATION = ARXIV:1405.3886;%%.

\bibitem{Chatrchyan:2014lfa}
\hrefCMSnoop {} {{CMS Collaboration}, ``{Search for new physics in the multijet
  and missing transverse momentum final state in proton-proton collisions at
  $\sqrt{s}$ = 8 TeV}'',} (2014).
\href{http://www.arXiv.org/abs/1402.4770v1}{\texttt{ arXiv:1402.4770v1}}.
%%CITATION = ARXIV:1402.4770;%%.

\bibitem{Chatrchyan:2013mya}
\hrefCMSnoop {} {{ CMS} Collaboration, ``{Search for top squark and higgsino
  production using diphoton Higgs boson decays}'',} \textit{ Phys. Rev. Lett.}
  \textbf{ 112} (2014) 161802,
  \href{http://dx.doi.org/10.1103/PhysRevLett.112.161802}{\doi{10.1103/PhysRevLett.112.161802}},
\href{http://www.arXiv.org/abs/1312.3310v2}{\texttt{ arXiv:1312.3310v2}}.
%%CITATION = ARXIV:1312.3310;%%.

\bibitem{Chatrchyan:2013fea}
\hrefCMSnoop {} {{ CMS} Collaboration, ``{Search for new physics in events with
  same-sign dileptons and jets in $\Pp \Pp$ collisions at $\sqrt{s}$ = 8
  TeV}'',} \textit{ JHEP} \textbf{ 01} (2014) 163,
  \href{http://dx.doi.org/10.1007/JHEP01(2014)163}{\doi{10.1007/JHEP01(2014)163}},
\href{http://www.arXiv.org/abs/1311.6736v2}{\texttt{ arXiv:1311.6736v2}}.
%%CITATION = ARXIV:1311.6736;%%.

\bibitem{Chatrchyan:2013xna}
\hrefCMSnoop {} {{ CMS} Collaboration, ``{Search for top-squark pair production
  in the single-lepton final state in $\Pp \Pp$ collisions at $\sqrt{s}$ = 8
  TeV}'',} \textit{ Eur. Phys. J. C} \textbf{ 73} (2013) 2677,
  \href{http://dx.doi.org/10.1140/epjc/s10052-013-2677-2}{\doi{10.1140/epjc/s10052-013-2677-2}},
\href{http://www.arXiv.org/abs/1308.1586v2}{\texttt{ arXiv:1308.1586v2}}.
%%CITATION = ARXIV:1308.1586;%%.

\bibitem{Chatrchyan:2013wxa}
\hrefCMSnoop {} {{ CMS} Collaboration, ``{Search for gluino mediated bottom-
  and top-squark production in multijet final states in $\Pp \Pp$ collisions at
  8 TeV}'',} \textit{ Phys. Lett. B} \textbf{ 725} (2013) 243,
  \href{http://dx.doi.org/10.1016/j.physletb.2013.06.058}{\doi{10.1016/j.physletb.2013.06.058}},
\href{http://www.arXiv.org/abs/1305.2390v2}{\texttt{ arXiv:1305.2390v2}}.
%%CITATION = ARXIV:1305.2390;%%.

\bibitem{Chatrchyan:2013lya}
\hrefCMSnoop {} {{ CMS} Collaboration, ``{Search for supersymmetry in hadronic
  final states with missing transverse energy using the variables $\alpha_{T}$
  and b-quark multiplicity in $\Pp \Pp$ collisions at $\sqrt{s}$ = 8 TeV}'',}
  \textit{ Eur. Phys. J. C} \textbf{ 73} (2013) 2568,
  \href{http://dx.doi.org/10.1140/epjc/s10052-013-2568-6}{\doi{10.1140/epjc/s10052-013-2568-6}},
\href{http://www.arXiv.org/abs/1303.2985v2}{\texttt{ arXiv:1303.2985v2}}.
%%CITATION = ARXIV:1303.2985;%%.

\bibitem{Chatrchyan:2012paa}
\hrefCMSnoop {} {{ CMS} Collaboration, ``{Search for new physics in events with
  same-sign dileptons and b jets in $\Pp \Pp$ collisions at $\sqrt{s}$ = 8
  TeV}'',} \textit{ JHEP} \textbf{ 03} (2013) 037,
  \href{http://dx.doi.org/10.1007/JHEP03(2013)037}{\doi{10.1007/JHEP03(2013)037}},
  \href{http://www.arXiv.org/abs/1212.6194v3}{\texttt{ arXiv:1212.6194v3}}.
[Erratum: \DOI{10.1007/JHEP07(2013)041}].
%%CITATION = ARXIV:1212.6194;%%.

\bibitem{Chatrchyan:2013iqa}
\hrefCMSnoop {} {{ CMS} Collaboration, ``{Search for supersymmetry in $\Pp \Pp$
  collisions at $\sqrt{s}$ = 8 TeV in events with a single lepton, large jet
  multiplicity, and multiple b jets}'',} \textit{ Phys. Lett. B} \textbf{ 733}
  (2014) 328,
  \href{http://dx.doi.org/10.1016/j.physletb.2014.04.023}{\doi{10.1016/j.physletb.2014.04.023}},
\href{http://www.arXiv.org/abs/1311.4937v1}{\texttt{ arXiv:1311.4937v1}}.
%%CITATION = ARXIV:1311.4937;%%.

\bibitem{CMS:2008zzk}
\hrefCMSnoop {} {{ CMS} Collaboration, ``The {CMS} experiment at the {CERN}
  {LHC}'',} \textit{ JINST} \textbf{ 3} (2008) S08004,
\href{http://dx.doi.org/10.1088/1748-0221/3/08/S08004}{\doi{10.1088/1748-0221/3/08/S08004}}.
%%CITATION = JINST,3,S08004;%%.

\bibitem{PFT-10-002}
\href {http://cdsweb.cern.ch/record/1279341} {{ CMS} Collaboration,
  ``Commissioning of the Particle-Flow Reconstruction in Minimum-Bias and Jet
  Events from {\Pp\Pp} Collisions at 7 {TeV}'',} CMS Physics Analysis Summary
  CMS-PAS-PFT-10-002, 2010.

\bibitem{CMS-PAS-PFT-09-001}
\href {http://cdsweb.cern.ch/record/1194487} {{ CMS} Collaboration,
  ``Particle--Flow Event Reconstruction in {CMS} and Performance for Jets,
  Taus, and {\MET}'',} CMS Physics Analysis Summary CMS-PAS-PFT-09-001, 2009.

\bibitem{EGM-10-004}
\href {http://cdsweb.cern.ch/record/1299116} {{ CMS} Collaboration, ``Electron
  Reconstruction and Identification at $\sqrt{s} = 7$ {TeV}'',} CMS Physics
  Analysis Summary CMS-PAS-EGM-10-004, 2010.

\bibitem{JME-10-005}
\href {http://cdsweb.cern.ch/record/1294501} {{ CMS} Collaboration, ``CMS MET
  Performance in Events Containing Electroweak Bosons from pp Collisions at
  $\sqrt{s}=7$ {TeV}'',} CMS Physics Analysis Summary CMS-PAS-JME-10-005, 2010.

\bibitem{CMS-PAPERS-JME-10-009}
\hrefCMSnoop {} {{ CMS} Collaboration, ``Missing transverse energy performance
  of the {CMS} detector'',} \textit{ JINST} \textbf{ 6} (2011) P09001,
  \href{http://dx.doi.org/10.1088/1748-0221/6/09/P09001}{\doi{10.1088/1748-0221/6/09/P09001}}.

\bibitem{Cacciari:2008gp}
\hrefCMSnoop {} {M.~Cacciari, G.~P. Salam, and G.~Soyez, ``{The anti-$\kt$ jet
  clustering algorithm}'',} \textit{ JHEP} \textbf{ 04} (2008) 063,
  \href{http://dx.doi.org/10.1088/1126-6708/2008/04/063}{\doi{10.1088/1126-6708/2008/04/063}},
\href{http://www.arXiv.org/abs/0802.1189v2}{\texttt{ arXiv:0802.1189v2}}.
%%CITATION = ARXIV:0802.1189;%%.

\bibitem{Cacciari:2011ma}
\hrefCMSnoop {} {M.~Cacciari, G.~P. Salam, and G.~Soyez, ``{FastJet User
  Manual}'',} \textit{ Eur. Phys. J. C} \textbf{ 72} (2012) 1896,
  \href{http://dx.doi.org/10.1140/epjc/s10052-012-1896-2}{\doi{10.1140/epjc/s10052-012-1896-2}},
\href{http://www.arXiv.org/abs/1111.6097v1}{\texttt{ arXiv:1111.6097v1}}.
%%CITATION = ARXIV:1111.6097;%%.

\bibitem{CMS-PAPERS-JME-10-011}
\hrefCMSnoop {} {{ CMS} Collaboration, ``Determination of jet energy
  calibration and transverse momentum resolution in {CMS}'',} \textit{ JINST}
  \textbf{ 6} (2011) P11002,
  \href{http://dx.doi.org/10.1088/1748-0221/6/11/P11002}{\doi{10.1088/1748-0221/6/11/P11002}}.

\bibitem{Cacciari:2007fd}
\hrefCMSnoop {} {M.~Cacciari and G.~P. Salam, ``{Pileup subtraction using jet
  areas}'',} \textit{ Phys. Lett. B} \textbf{ 659} (2008) 119,
  \href{http://dx.doi.org/10.1016/j.physletb.2007.09.077}{\doi{10.1016/j.physletb.2007.09.077}},
\href{http://www.arXiv.org/abs/0707.1378v2}{\texttt{ arXiv:0707.1378v2}}.
%%CITATION = ARXIV:0707.1378;%%.

\bibitem{Chatrchyan:2012jua}
\hrefCMSnoop {} {{ CMS} Collaboration, ``{Identification of b-quark jets with
  the CMS experiment}'',} \textit{ JINST} \textbf{ 8} (2013) P04013,
  \href{http://dx.doi.org/10.1088/1748-0221/8/04/P04013}{\doi{10.1088/1748-0221/8/04/P04013}},
\href{http://www.arXiv.org/abs/1211.4462}{\texttt{ arXiv:1211.4462}}.
%%CITATION = ARXIV:1211.4462;%%.

\bibitem{Maltoni:2002qb}
\hrefCMSnoop {} {F.~Maltoni and T.~Stelzer, ``{MadEvent}: Automatic event
  generation with {MadGraph}'',} \textit{ JHEP} \textbf{ 02} (2003) 027,
  \href{http://dx.doi.org/10.1088/1126-6708/2003/02/027}{\doi{10.1088/1126-6708/2003/02/027}},
\href{http://www.arXiv.org/abs/hep-ph/0208156v1}{\texttt{
  arXiv:hep-ph/0208156v1}}.
%%CITATION = HEP-PH/0208156;%%.

\bibitem{Nadolsky:2008zw}
P.~M. Nadolsky\hrefCMSnoop {} { {et~al.}, ``{Implications of CTEQ global
  analysis for collider observables}'',} \textit{ Phys. Rev. D} \textbf{ 78}
  (2008) 013004,
  \href{http://dx.doi.org/10.1103/PhysRevD.78.013004}{\doi{10.1103/PhysRevD.78.013004}},
\href{http://www.arXiv.org/abs/0802.0007}{\texttt{ arXiv:0802.0007}}.
%%CITATION = ARXIV:0802.0007;%%.

\bibitem{Agostinelli:2002hh}
\hrefCMSnoop {} {{ GEANT4} Collaboration, ``{GEANT4}---a simulation toolkit'',}
  \textit{ Nucl. Instrum. Meth. A} \textbf{ 506} (2003) 250,
\href{http://dx.doi.org/10.1016/S0168-9002(03)01368-8}{\doi{10.1016/S0168-9002(03)01368-8}}.
%%CITATION = NUIMA,A506,250;%%.

\bibitem{MCFM}
\hrefCMSnoop {} {J.~Campbell, R.~K. Ellis, and C.~Williams, ``Vector boson pair
  production at the LHC'',} \textit{ JHEP} \textbf{ 07} (2011) 018,
  \href{http://dx.doi.org/10.1007/JHEP07(2011)018}{\doi{10.1007/JHEP07(2011)018}},
\href{http://www.arXiv.org/abs/1105.0020}{\texttt{ arXiv:1105.0020}}.
%%CITATION = 0802.0007;%%.

\bibitem{Campbell:2012dh}
\hrefCMSnoop {} {J.~M. Campbell and R.~K. Ellis, ``{$\ttbar \PW^{\pm}$
  production and decay at NLO}'',} \textit{ JHEP} \textbf{ 07} (2012) 052,
  \href{http://dx.doi.org/10.1007/JHEP07(2012)052}{\doi{10.1007/JHEP07(2012)052}},
\href{http://www.arXiv.org/abs/1204.5678v1}{\texttt{ arXiv:1204.5678v1}}.
%%CITATION = ARXIV:1204.5678;%%.

\bibitem{Garzelli:2012bn}
\hrefCMSnoop {} {M.~V. Garzelli, A.~Kardos, C.~G. Papadopoulos, and
  Z.~Trocsanyi, ``{$\ttbar \PW^{\pm}$ and $\ttbar \cPZ$ Hadroproduction at NLO
  accuracy in QCD with Parton Shower and Hadronization effects}'',} \textit{
  JHEP} \textbf{ 11} (2012) 056,
  \href{http://dx.doi.org/10.1007/JHEP11(2012)056}{\doi{10.1007/JHEP11(2012)056}},
\href{http://www.arXiv.org/abs/1208.2665v1}{\texttt{ arXiv:1208.2665v1}}.
%%CITATION = ARXIV:1208.2665;%%.

\bibitem{Sjostrand:2007gs}
\hrefCMSnoop {} {T.~Sj{\"o}strand, S.~Mrenna, and P.~Z. Skands, ``A brief
  introduction to {PYTHIA} 8.1'',} \textit{ Comput. Phys. Commun.} \textbf{
  178} (2008) 852,
  \href{http://dx.doi.org/10.1016/j.cpc.2008.01.036}{\doi{10.1016/j.cpc.2008.01.036}},
\href{http://www.arXiv.org/abs/0710.3820v1}{\texttt{ arXiv:0710.3820v1}}.
%%CITATION = 0710.3820;%%.

\bibitem{Orbaker:2010zz}
\hrefCMSnoop {} {{ CMS} Collaboration, ``{Fast simulation of the CMS
  detector}'',} \textit{ J. Phys. Conf. Ser.} \textbf{ 219} (2010) 032053,
\href{http://dx.doi.org/10.1088/1742-6596/219/3/032053}{\doi{10.1088/1742-6596/219/3/032053}}.
%%CITATION = 00462,219,032053;%%.

\bibitem{CMS-PAPERS-MUO-10-004}
\hrefCMSnoop {} {{ CMS} Collaboration, ``Performance of {CMS} muon
  reconstruction in pp collision events at {$\sqrt{s} = 7$\TeV}'',} \textit{
  JINST} \textbf{ 7} (2012) P10002,
  \href{http://dx.doi.org/10.1088/1748-0221/7/10/P10002}{\doi{10.1088/1748-0221/7/10/P10002}}.

\bibitem{CMS:2012ra}
\hrefCMSnoop {} {{ CMS} Collaboration, ``{Search for heavy lepton partners of
  neutrinos in proton-proton collisions in the context of the type III seesaw
  mechanism}'',} \textit{ Phys. Lett. B} \textbf{ 718} (2012) 348,
  \href{http://dx.doi.org/10.1016/j.physletb.2012.10.070}{\doi{10.1016/j.physletb.2012.10.070}},
\href{http://www.arXiv.org/abs/1210.1797}{\texttt{ arXiv:1210.1797}}.
%%CITATION = ARXIV:1210.1797;%%.

\bibitem{CMS-PAPERS-TAU-11-001}
\hrefCMSnoop {} {{ CMS} Collaboration, ``Performance of $\tau$-lepton
  reconstruction and identification in {CMS}'',} \textit{ JINST} \textbf{ 7}
  (2012) P01001,
  \href{http://dx.doi.org/10.1088/1748-0221/7/01/P01001}{\doi{10.1088/1748-0221/7/01/P01001}}.

\bibitem{CMS-PAS-LUM-13-001}
\href {http://cdsweb.cern.ch/record/1598864} {{ CMS} Collaboration, ``CMS
  Luminosity Based on Pixel Cluster Counting - Summer 2013 Update'',} CMS
  Physics Analysis Summary CMS-PAS-LUM-13-001, 2013.

\bibitem{ATLAS:1379837}
\href {http://cdsweb.cern.ch/record/1379837} {{ATLAS and CMS Collaborations},
  ``Procedure for the LHC Higgs boson search combination in Summer 2011'',}
  Technical Report CMS-NOTE-2011-005, ATLAS/CMS, Geneva, 2011.

\bibitem{Junk:1999kv}
\hrefCMSnoop {} {T.~Junk, ``{Confidence Level Computation for Combining
  Searches with Small Statistics}'',} \textit{ Nucl. Instrum. Meth. A} \textbf{
  434} (1999) 435,
  \href{http://dx.doi.org/10.1016/S0168-9002(99)00498-2}{\doi{10.1016/S0168-9002(99)00498-2}},
\href{http://www.arXiv.org/abs/hep-ex/9902006}{\texttt{ arXiv:hep-ex/9902006}}.
%%CITATION = HEP-EX/9902006;%%.

\bibitem{Read:2002hq}
\hrefCMSnoop {} {A.~L. Read, ``{Presentation of search results: The CL(s)
  technique}'',} \textit{ J. Phys. G} \textbf{ 28} (2002) 2693,
\href{http://dx.doi.org/10.1088/0954-3899/28/10/313}{\doi{10.1088/0954-3899/28/10/313}}.
%%CITATION = JPHGB,G28,2693;%%.

\bibitem{PhysRevD.62.077702}
\hrefCMSnoop {} {K.~T. Matchev and S.~Thomas, ``Higgs and Z-boson signatures of
  supersymmetry'',} \textit{ Phys. Rev. D} \textbf{ 62} (2000) 077702,
  \href{http://dx.doi.org/10.1103/PhysRevD.62.077702}{\doi{10.1103/PhysRevD.62.077702}},
  \href{http://www.arXiv.org/abs/hep-ph/9908482}{\texttt{
  arXiv:hep-ph/9908482}}.

\bibitem{Dimopoulos:1996va}
\hrefCMSnoop {} {S.~Dimopoulos, S.~D. Thomas, and J.~D. Wells, ``Implications
  of low energy supersymmetry breaking at the {F}ermilab Tevatron'',} \textit{
  Phys. Rev. D} \textbf{ 54} (1996) 3283,
  \href{http://dx.doi.org/10.1103/PhysRevD.54.3283}{\doi{10.1103/PhysRevD.54.3283}},
\href{http://www.arXiv.org/abs/hep-ph/9604452v3}{\texttt{
  arXiv:hep-ph/9604452v3}}.
%%CITATION = HEP-PH/9604452;%%.

\bibitem{Culbertson:2000am}
\hrefCMSnoop {} {{ SUSY Working Group} Collaboration, ``Low scale and gauge
  mediated supersymmetry breaking at the {Fermilab Tevatron Run II}'',} (2000).
\href{http://www.arXiv.org/abs/hep-ph/0008070v2}{\texttt{
  arXiv:hep-ph/0008070v2}}.
%%CITATION = HEP-PH/0008070;%%.

\bibitem{Ruderman:2010kj}
\hrefCMSnoop {} {J.~T. Ruderman and D.~Shih, ``{Slepton co-NLSPs at the
  Tevatron}'',} \textit{ JHEP} \textbf{ 11} (2010) 046,
  \href{http://dx.doi.org/10.1007/JHEP11(2010)046}{\doi{10.1007/JHEP11(2010)046}},
  \href{http://www.arXiv.org/abs/1009.1665}{\texttt{ arXiv:1009.1665}}.

\bibitem{Alves:2011wf}
\hrefCMSnoop {} {D.~Alves {et~al.}, ``Simplified Models for {LHC} New Physics
  Searches'',} \textit{ J. Phys. G} \textbf{ 39} (2012) 105005,
  \href{http://dx.doi.org/10.1088/0954-3899/39/10/105005}{\doi{10.1088/0954-3899/39/10/105005}},
\href{http://www.arXiv.org/abs/1105.2838}{\texttt{ arXiv:1105.2838}}.
%%CITATION = ARXIV:1105.2838;%%.

\bibitem{Beenakker:1996ch}
\hrefCMSnoop {} {W.~Beenakker, R.~H{\"o}pker, M.~Spira, and P.~M. Zerwas,
  ``{Squark and gluino production at hadron colliders}'',} \textit{ Nucl. Phys.
  B} \textbf{ 492} (1997) 51,
  \href{http://dx.doi.org/10.1016/S0550-3213(97)80027-2}{\doi{10.1016/S0550-3213(97)80027-2}},
\href{http://www.arXiv.org/abs/hep-ph/9610490v1}{\texttt{
  arXiv:hep-ph/9610490v1}}.
%%CITATION = HEP-PH/9610490;%%.

\bibitem{Beenakker:1996ed}
W.~Beenakker, R.~H{\"o}pker, and M.~Spira, ``PROSPINO: A program for the
  production of supersymmetric particles in next-to-leading order QCD'', 1996,
  \href{http://www.arXiv.org/abs/hep-ph/9611232v1}{\texttt{
  arXiv:hep-ph/9611232v1}}.

\bibitem{Essig:2011qg}
\hrefCMSnoop {} {R.~Essig, E.~Izaguirre, J.~Kaplan, and J.~G. Wacker, ``{Heavy
  Flavor Simplified Models at the LHC}'',} \textit{ JHEP} \textbf{ 01} (2012)
  074,
  \href{http://dx.doi.org/10.1007/JHEP01(2012)074}{\doi{10.1007/JHEP01(2012)074}},
\href{http://www.arXiv.org/abs/1110.6443v1}{\texttt{ arXiv:1110.6443v1}}.
%%CITATION = ARXIV:1110.6443;%%.

\bibitem{Chatrchyan:2013sza}
\hrefCMSnoop {} {{ CMS} Collaboration, ``{Interpretation of Searches for
  Supersymmetry with Simplified Models}'',} \textit{ Phys. Rev. D} \textbf{ 88}
  (2013) 052017,
  \href{http://dx.doi.org/10.1103/PhysRevD.88.052017}{\doi{10.1103/PhysRevD.88.052017}},
\href{http://www.arXiv.org/abs/1301.2175v2}{\texttt{ arXiv:1301.2175v2}}.
%%CITATION = ARXIV:1301.2175;%%.

\bibitem{Kulesza:2008jb}
\hrefCMSnoop {} {A.~Kulesza and L.~Motyka, ``{Threshold resummation for
  squark-antisquark and gluino-pair production at the LHC}'',} \textit{ Phys.
  Rev. Lett.} \textbf{ 102} (2009) 111802,
  \href{http://dx.doi.org/10.1103/PhysRevLett.102.111802}{\doi{10.1103/PhysRevLett.102.111802}},
\href{http://www.arXiv.org/abs/0807.2405v1}{\texttt{ arXiv:0807.2405v1}}.
%%CITATION = ARXIV:0807.2405;%%.

\bibitem{Kulesza:2009kq}
\hrefCMSnoop {} {A.~Kulesza and L.~Motyka, ``{Soft gluon resummation for the
  production of gluino-gluino and squark-antisquark pairs at the LHC}'',}
  \textit{ Phys. Rev. D} \textbf{ 80} (2009) 095004,
  \href{http://dx.doi.org/10.1103/PhysRevD.80.095004}{\doi{10.1103/PhysRevD.80.095004}},
\href{http://www.arXiv.org/abs/0905.4749v1}{\texttt{ arXiv:0905.4749v1}}.
%%CITATION = ARXIV:0905.4749;%%.

\bibitem{Beenakker:2009ha}
W.~Beenakker\hrefCMSnoop {} { {et~al.}, ``Soft-gluon resummation for squark and
  gluino hadroproduction'',} \textit{ JHEP} \textbf{ 12} (2009) 041,
  \href{http://dx.doi.org/10.1088/1126-6708/2009/12/041}{\doi{10.1088/1126-6708/2009/12/041}},
\href{http://www.arXiv.org/abs/0909.4418v1}{\texttt{ arXiv:0909.4418v1}}.
%%CITATION = ARXIV:0909.4418;%%.

\bibitem{Beenakker:2011fu}
W.~Beenakker\hrefCMSnoop {} { {et~al.}, ``Squark and gluino hadroproduction'',}
  \textit{ Int. J. Mod. Phys. A} \textbf{ 26} (2011) 2637,
  \href{http://dx.doi.org/10.1142/S0217751X11053560}{\doi{10.1142/S0217751X11053560}},
\href{http://www.arXiv.org/abs/1105.1110v1}{\texttt{ arXiv:1105.1110v1}}.
%%CITATION = ARXIV:1105.1110;%%.

\bibitem{Kramer:2012bx}
M.~Kr{\"a}mer\hrefCMSnoop {} { {et~al.}, ``{Supersymmetry production cross
  sections in $\Pp \Pp$ collisions at $\sqrt{s} = 7\,\TeV$}'',} (2012).
\href{http://www.arXiv.org/abs/1206.2892v1}{\texttt{ arXiv:1206.2892v1}}.
%%CITATION = ARXIV:1206.2892;%%.

\bibitem{Lee:2012sy}
\hrefCMSnoop {} {H.~M. Lee, V.~Sanz, and M.~Trott, ``{Hitting sbottom in
  natural SUSY}'',} \textit{ JHEP} \textbf{ 05} (2012) 139,
  \href{http://dx.doi.org/10.1007/JHEP05(2012)139}{\doi{10.1007/JHEP05(2012)139}},
\href{http://www.arXiv.org/abs/1204.0802}{\texttt{ arXiv:1204.0802}}.
%%CITATION = ARXIV:1204.0802;%%.

\bibitem{PhysRevD.2.1285}
\hrefCMSnoop {} {S.~L. Glashow, J.~Iliopoulos, and L.~Maiani, ``{Weak
  Interactions with Lepton-Hadron Symmetry}'',} \textit{ Phys. Rev. D} \textbf{
  2} (1970) 1285,
  \href{http://dx.doi.org/10.1103/PhysRevD.2.1285}{\doi{10.1103/PhysRevD.2.1285}}.

\bibitem{Kobayashi:1973fv}
\hrefCMSnoop {} {M.~Kobayashi and T.~Maskawa, ``{CP Violation in the
  Renormalizable Theory of Weak Interaction}'',} \textit{ Prog. Theor. Phys.}
  \textbf{ 49} (1973) 652,
\href{http://dx.doi.org/10.1143/PTP.49.652}{\doi{10.1143/PTP.49.652}}.
%%CITATION = PTPKA,49,652;%%.

\bibitem{Antonelli:2009ws}
\hrefCMSnoop {} {M.~Antonelli {et~al.}, ``Flavor physics in the quark
  sector'',} \textit{ Phys. Rept.} \textbf{ 494} (2010) 197,
  \href{http://dx.doi.org/10.1016/j.physrep.2010.05.003}{\doi{10.1016/j.physrep.2010.05.003}},
\href{http://www.arXiv.org/abs/0907.5386v2}{\texttt{ arXiv:0907.5386v2}}.
%%CITATION = ARXIV:0907.5386;%%.

\bibitem{Chatrchyan:2013mxa}
\hrefCMSnoop {} {{ CMS} Collaboration, ``{Measurement of the properties of a
  Higgs boson in the four-lepton final state}'',} \textit{ Phys. Rev. D}
  \textbf{ 89} (2014) 092007,
  \href{http://dx.doi.org/10.1103/PhysRevD.89.092007}{\doi{10.1103/PhysRevD.89.092007}},
\href{http://www.arXiv.org/abs/1312.5353v3}{\texttt{ arXiv:1312.5353v3}}.
%%CITATION = ARXIV:1312.5353;%%.

\bibitem{Czakon:2013goa}
\hrefCMSnoop {} {M.~Czakon, P.~Fiedler, and A.~Mitov, ``{The total top quark
  pair production cross-section at hadron colliders through
  $O(\alpha_{S}^{4})$}'',} \textit{ Phys. Rev. Lett.} \textbf{ 110} (2013)
  252004,
  \href{http://dx.doi.org/10.1103/PhysRevLett.110.252004}{\doi{10.1103/PhysRevLett.110.252004}},
\href{http://www.arXiv.org/abs/1303.6254}{\texttt{ arXiv:1303.6254}}.
%%CITATION = ARXIV:1303.6254;%%.

\end{thebibliography}\endgroup
